\documentclass{emulateapj}
\usepackage{longtable}

\newcommand{\hii}{H$\,${\sc ii}}
\newcommand{\hi}{H$\,${\sc i}}

\newcommand{\sfr}{SFR$_{0.1}$}
\newcommand{\sfg}{SFR$_{1}$}
\newcommand{\sfl}{SFR$_{13.7}$}

\shorttitle{Outside-in Shrinking of the Star-forming Disk of Dwarf Irregular Galaxies}
\shortauthors{Zhang et al.}

\begin{document}

\title{Outside-in Shrinking of the Star-forming Disk of Dwarf Irregular Galaxies
  \footnote{
  Based on data from the LITTLE THINGS Survey (Hunter et al., in preparation),
  funded in part by the National Science Foundation through grants AST-0707563, AST-0707426, 
  AST-0707468, and AST-0707835 to US-based LITTLE THINGS team members and with generous 
  support from the National Radio Astronomy Observatory.}}

\author{
Hong-Xin Zhang\altaffilmark{1,2,3,4}, Deidre A. Hunter\altaffilmark{1}, Bruce G. Elmegreen\altaffilmark{5}, 
Yu Gao\altaffilmark{2,3}, Andreas Schruba\altaffilmark{6}
}

\altaffiltext{1}{Lowell Observatory, 1400 West Mars Hill Road, Flagstaff, Arizona 86001 USA; hxzhang@lowell.edu; dah@lowell.edu}
\altaffiltext{2}{Purple Mountain Observatory, Chinese Academy of Sciences, 2 West Beijing Road, Nanjing 210008 China; yugao@pmo.ac.cn}
\altaffiltext{3}{Key Laboratory of Radio Astronomy, Chinese Academy of Sciences, Nanjing 210008 China}
\altaffiltext{4}{Graduate School of the Chinese Academy of Sciences, Beijing 100080 China}
\altaffiltext{5}{IBM T. J. Watson Research Center, PO Box 218, Yorktown Heights, New York 10598 USA; bge@us.ibm.com}
\altaffiltext{6}{Max-Planck-Institut f\"ur Astronomie, K\"onigstuhl 17, 69117 Heidelberg Germany; schruba@mpia-hd.mpg.de}

\begin{abstract}
We have studied multi-band surface brightness profiles of a representative sample of 34 nearby dwarf irregular galaxies.\
Our data include $Galaxy$ $Evolution$ $Explorer$ ({\it GALEX}) FUV/NUV, $UBV$, H$\alpha$ and {\it Spitzer} 3.6 $\mu$m images.\ These galaxies constitute the majority of the LITTLE THINGS survey\footnote{Local Irregulars That Trace Luminosity Extremes -- The \hi\ Nearby Galaxy Survey, http://www.lowell.edu/users/dah/littlethings/index.html}.\  
By modeling the azimuthal averages of the spectral energy distributions with a complete library of star formation histories, 
we derived the stellar mass surface density distributions and the star formation rate averaged over three different timescales: 
the recent 0.1 Gyr, 1 Gyr and a Hubble time.

We find that, for $\sim$ 80\% (27 galaxies) of our sample galaxies, radial profiles (at least in the outer part) at shorter wavelengths, corresponding to younger stellar populations, have shorter disk scale lengths than those at longer wavelengths, corresponding to older stellar populations.\ This indicates that the star-forming disk has been shrinking.\  
In addition, the radial distributions of the stellar mass surface density are well described as piece-wise exponential profiles,
and $\sim$ 80\% of the galaxies have steeper mass profiles in the outer disk than in the inner region.\ 
The steep radial decline of star formation rate in the outer parts compared to that in the inner disks gives a natural explanation 
for the down-bending stellar mass surface density profiles.\ Within the inner disks, our sample galaxies on average 
have constant ratios of recent star formation rate to stellar mass with radius.\ 
Nevertheless, $\sim$ 35\% (12 galaxies, among which 7 have baryonic mass $\lesssim$ 10$^{8}$ M$_{\odot}$) of 
the sample exhibit negative slopes across the observed disk, which is in contrast with the so-called ``inside-out" disk 
growth scenario suggested for luminous spiral galaxies.\ 
The tendency of star formation to become concentrated toward the inner disks in low mass dwarf irregular galaxies is interpreted 
as a result of their susceptibility to environmental effects and regulation through stellar feedback.\
\end{abstract}


\section{Introduction}
Dwarf irregular (dIrr) galaxies numerically dominate actively star-forming galaxies in the local universe (van den Bergh 1966; Gallagher \& Hunter 1984; Dale et al.\ 2009). Compared to local spiral galaxies, dIrr galaxies are generally smaller, bluer, less luminous, and gas-rich.\  dIrr galaxies tend to have low metallicity and a flat metallicity gradient within measurement uncertainties in the interstellar medium (ISM; e.g.\ Kobulnicky \& Skillman 1996; Hunter \& Hoffmann 1999; van Zee \& Haynes 2006; Croxall et al.\ 2009).\

The star formation (SF) process in seemingly simple dIrr galaxies is unclear.\ The classical Toomre gravitational instability model (Toomre 1964; Quirk 1972), which seems to explain the SF threshold successfully in spiral galaxies (Kennicutt 1989; Martin \& Kennicutt 2001) and elliptical galaxies (e.g.\ Vader \& Vigroux 1991), predicts that the gas surface density of dIrr  galaxies, both at intermediate radii and outer radii, is usually well below the threshold of large-scale SF  for a single-fluid gas disk (Hunter, Elmegreen \& Baker 1998, hereafter H98; Leroy et al.\ 2008).\ Even though recent studies (e.g.\ Yang, Gruendl \& Chu 2007) suggest that incorporation of the gravitational contribution of the stars may make the gas disk marginally stable, the uncertainties (e.g.\ the gas/stellar velocity dispersion, the disk scale height) included in the stability analysis are substantial.\ 
Nevertheless, widely distributed SF, although apparently inefficient, has been observed in dIrr galaxies, which implies that some other internal or local process must play an important role (e.g.\ Elmegreen \& Hunter 2006).\ 

H98 found that, for most dIrr galaxies, the current SF rate (SFR, traced by H$\alpha$) correlates with the optical (e.g.\ $V$) starlight profiles better than with the usually mass-dominant atomic gas surface density.\ The correlation with older stars indicates that SF may be a self-regulating process and/or the SF may be affected by the existing stars to a certain degree.\ 
By analyzing optical and near-IR (NIR) broadband data, Bell \& de Jong (2000) and Bell et al.\ (2000) found that the SF histories (SFHs, described as luminosity-weighted ages there) primarily correlate with the $K$-band surface brightness in both spiral galaxies and low surface brightness (LSB) galaxies. Recently, Leroy et al.\ (2008) found that the SF efficiency (SFE $\sim$ SFR/M$_{atomic~gas}$) shows a strong relationship with the stellar mass surface density, rather than with the gas, in atomic gas-dominated regions, such as dIrr galaxies and the outer disks of spiral galaxies.\ Shi et al.\ (2011) proposed an ``extended Schmidt law" with explicit dependence of the SFE on the stellar mass surface density.\ Therefore, an important question to be addressed is how do the SF, SFH, and stellar mass surface density relate to each other? 

Generally speaking, the SFR, SFH, and stellar mass density could be derived directly from analysis of resolved stellar color-magnitude diagrams (CMD).\ However, besides the problem of stellar crowding in intense star-forming regions, currently full-scale---in terms of both the observational depth and spatial coverage---CMD analysis of dIrr galaxies is still rare (e.g.\ Cole et al.\ 2007; Hidalgo et al.\ 2011). Obtaining deep CMDs which reach the old main sequence (MS) turnoffs is crucial to reconstructing the whole SFH of a galaxy.\ As pointed out by Gallart, Zoccali, \& Aparicio (2005), beyond $\sim$2 Gyr ago, shallow (e.g.\  completeness limit $M_{I}$ $<$ 2) CMDs at best can reliably constrain the {\it integrated} SF.\ 
Therefore, analyzing spatially integrated multi-wavelength broadband data is still an efficient and helpful way to constrain the SFHs for a large sample of dIrr galaxies.\ 

Specifically, ultraviolet (UV) emission is mainly sensitive to the recent ($\lesssim$ 100 Myr, e.g.\ Kennicutt 1998) SF; the optical broadband emission has long been shown to be able to constrain the SF during the past $\sim$ Gyr (e.g.\ Gallagher, Hunter, \& Tutukov 1984); the NIR bands have the least sensitivity to recent SF and dust extinction compared to shorter wavelengths, and thus give the tightest constraint on the stellar mass (e.g.\ Elmegreen \& Elmegreen 1984; Kendall et al.\ 2008).\
Even more, the combination of optical and NIR data is able to largely break the notorious age-metallicity degeneracy (Worthey 1994; Anders \& Fritze-v.\ Alvensleben 2003; Lee et al.\ 2007). Cardiel et al.\ (2003) found that the inclusion of an NIR band improves the constraints on the stellar populations by $\sim$ 30 times over using optical colors alone.\ Bell \& de Jong (2000) found that there are significant variations in stellar mass-to-light ratios even in NIR bands, which is consistent with recent modeling results (Dutton 2009).\ Above all, by fitting the whole spectral energy distribution (SED) from the UV to the NIR, the stellar mass distribution and recent ($\lesssim$ Gyr) SFHs can be reasonably constrained.\ 

With the advent of {\it GALEX} (Martin et al.\ 2005) and {\it Spitzer} (Werner et al.\ 2004), acquiring multi-band data from the far-UV (FUV) to the NIR for large samples of dIrr galaxies has become possible.\ 
The {\it Spitzer} Infrared Array Camera (IRAC, Fazio et al.\ 2004) 3.6 $\mu$m passband is primarily stellar emission from the mass-dominant older populations (e.g.\ Helou et al.\ 2004), even though there could be a significant contribution from hot dust emission in intense star-forming regions (e.g.\ Zhang, Gao, \& Kong 2010).\ Furthermore, in low-metallicity dIrr galaxies, the contribution of polycyclic aromatic hydrocarbon (PAH) dust emission to the 3.6 $\mu$m passband is substantially weaker compared to that in high-metallicity spiral galaxies (Engelbracht et al.\ 2008), which suggests that the IRAC 3.6 $\mu$m band is a reliable tracer of underlying older stellar populations in dIrr galaxies.\  

In an attempt to probe the relationship between the stellar mass surface density, SFH, and recent SF, this work derives stellar mass surface density profiles and rough SFH variations as a function of radius by synthesizing multi-wavelength data from the {\it GALEX} FUV to {\it Spitzer} 3.6 $\mu$m of a representative sample of 34 dIrr galaxies.\ 

This paper is organized as follows: 
In \S~2, we describe the sample and introduce the data used in this work.\ The multi-band surface brightness profiles are presented in \S~3.\ Then the stellar population synthesis method used in this work follows in \S~4.\  In \S~5, we present the results of the population modeling, including the stellar mass surface density profiles and the SFH variations with radius.\ The discussion and the summary are provided in \S~6 and \S~7, respectively.\ We also test the reliability of our SED modeling in Appendix A.\ 

\section{Data and Photometry}
\subsection{Data}
The 34 galaxies included in this work (listed in Table 1) are drawn from the LITTLE THINGS sample, which is a representative sample of 41 nearby dIrr galaxies spanning a large range in galactic parameters: integrated luminosity ($M_{B}$ of $-9$ to $-19$), surface SFR density (0 - 1.3 $M_{\odot}$yr$^{-1}$kpc$^{-2}$), (atomic) gas richness (0.02 - 5 $M_{\odot}$/$L_{B}$), and central surface brightness $\mu$$_{0}$$^{V}$ (18.5 - 25.5 mag arcsec$^{-2}$).\ The sample galaxies were selected to be relatively isolated.\ LITTLE THINGS is a large VLA project that was granted nearly 376 hours of VLA time in the B, C, and D array configurations to obtain deep \hi-line emission maps of dIrr galaxies with high angular and velocity resolutions.\ The subsample of 34 galaxies studied here was mostly chosen because these galaxies have the full complement of data available, including {\it GALEX} FUV/NUV, {\it UBV}, H$\alpha$ and {\it Spitzer} 3.6 $\mu$m.\ 
The Two Micron All Sky Survey (2MASS) $JHKs$ images are too shallow to be used in this work.\ 
However, UGC 8508, which cannot be observed by $GALEX$ due to a nearby bright star, and DDO 165 which does not have FUV observations, are also included in the current sample.\  
Another three galaxies without an H$\alpha$ detection (LGS 3, M81dwA, DDO 210) were also included.\

The acquisition and reduction of the $UBV$ and H$\alpha$ narrowband images of our sample were described by Hunter \& Elmegreen (2004, 2006; hereafter HE04, HE06).\ The {\it GALEX} UV data for 21 galaxies included in the current sample were presented by Hunter, Elmegreen \& Ludka (2010).\ 

There have been significant improvements in the flat fields and absolute calibration of the {\it GALEX} imaging data from the GR2/3 pipeline to the GR4/5 pipeline.\ The work of Hunter et al.\  (2010) used the GR2/3 pipeline data for some galaxies.\ 
Therefore, we obtained the GR4/5 imaging data of all of our galaxies from the {\it GALEX} archive.\ 
Most of the {\it Spitzer} 3.6 $\mu$m images were obtained and reduced by the Local Volume Legacy survey (LVL, Dale et al.\ 2009) and the {\it Spitzer} Infrared Nearby Galaxy Survey (SINGS, Kennicutt et al.\ 2003).\ Besides the standard post-pipeline processing, additional processing, including distortion corrections, rotation of the individual frames, bias structure and bias drift corrections, has been done by the teams of these two surveys.\   

\subsection{Surface Photometry}
We follow the procedures as laid out in HE04 for the surface photometry.\
First, 
the background-subtracted images were geometrically transformed and aligned with the {\it V}-band image.\  Background-and-foreground sources not belonging to the galaxy were masked.\ Then the surface photometry was carried out
with the Image Reduction and Analysis Facility (IRAF\footnote{IRAF is distributed by the
National Optical Astronomy Observatory, which is operated by the Association of Universities for Research in Astronomy
(AURA) under cooperative agreement with the National Science Foundation.}) 
task ELLIPSE, adopting the same geometrical parameters as derived by HE04.\ Briefly, the center of the galaxy, position angle, and the ellipticity were determined from a contour in the outer half of the {\it V}-band image that was block-averaged by factors of a few to increase the signal-to-noise.\ The center was fixed as the geometrical center of this isophote, and the major axis is the longest bisector that passes through the center that, as much as possible, symmetrically divides the galaxy.\ 
The surface photometry of the {\it UBV} and H$\alpha$ images has been described in HE04 and HE06.

To perform photometry on the {\it GALEX} images, as described in Hunter et al.\ (2010), we masked foreground stars and background galaxies identified mainly in the higher resolution optical band images.\ For the removal of sky background, because the background was quite flat, we measured the sky level in tens of square regions (10 pixel by 10 pixel) sampled along an ellipse around the galaxy, but far enough from the galactic center to avoid emission from the target galaxy.\ The final sky level was determined as the average of the sky values in all sampled subregions.\ 
As to the uncertainties of the background removal, the standard deviation of the mean sky values $\sigma_{\langle sky \rangle}$ among the surrounding sky regions is usually an order of magnitude smaller than the mean of the standard deviation in individual regions $\langle \sigma_{sky} \rangle$.\ Therefore, only the $\langle \sigma_{sky} \rangle$ were considered.\

For the {\it Spitzer} 3.6 $\mu$m images, 
we first applied the same photometry mask as was used for the optical band images.\ 
Background galaxies can significantly affect the photometry on 3.6 $\mu$m images of faint dwarf galaxies.\ Whenever available, high resolution archival Hubble Space Telescope ({\it HST}) imaging data were inspected to further remove the background galaxies based on morphology.\ 

Finally, 
by dividing the luminosity differences between adjacent ellipses by the area of the annulus, we obtained azimuthally-averaged surface brightness profiles.\ 
A value of 0.15 mag was taken as the calibration uncertainty for the {\it GALEX} UV data, and 0.1 mag for {\it Spitzer} 3.6 $\mu$m.\  
The calibration uncertainties of the {\it UBV} data for each galaxy were given in HE06.\ We adopt a 20\% calibration uncertainty for all of the H$\alpha$ photometry.\ 

Routinely one uses the average Milky Way (MW) extinction curve ($R_{V}$=$A_{V}$/$E_{B-V}$=3.1, Cardelli, Clayton \& Mathis 1989) to correct the photometry for the Galactic extinction.\ 
However, there is considerable scatter of the extinction curve shape from sightline to sightline through the Galaxy (Fitzpatrick 1999), which could significantly influence de-reddening of the short-wavelength data (e.g.\ UV, $U$) when only an average extinction curve was adopted.\ 
Fitzpatrick (1999) gave the typical uncertainties ($\sigma_{E(\lambda-V)/E(B-V)}$, Table 1 in Fitzpatrick (1999)) when correcting the data with the average extinction curve.\  
In this work we included the uncertainty ($E(B-V)$$\times$$\sigma_{E(\lambda-V)/E(B-V)}$) of adopting the average MW extinction curve.\ 
The final uncertainties assigned to each annulus are a quadratic sum of four contributions: $\langle \sigma_{sky} \rangle$, photometric calibration uncertainties, Poisson noise and $E(B-V)$$\times$$\sigma_{E(\lambda-V)/E(B-V)}$.\ 
The flux calibration uncertainties usually dominate all other contributions except in the outer disks.\ 

\section{Multi-band Surface Brightness Profiles}

The surface brightness profiles of H$\alpha$, FUV, {\it B}, and 3.6 $\mu$m for all galaxies are presented in the first column of Figure \ref{fig1}.\ 
The multi-band surface brightness profiles broadly follow each other.\
Nevertheless, for about half of our galaxies, 
the surface brightness profiles of H$\alpha$ and FUV peak in the circumnuclear regions or even in the intermediate radii, rather than the center.\ 
The recent SF indicator H$\alpha$ (and also FUV) usually follows the shorter wavelength stellar emission better than the longer wavelength emission, as is quantified below (Table 2; Figure \ref{fig2}).\
Therefore, the correlation between H$\alpha$ and the optical bands seen by others was partly caused by a dominant contribution of recent SF to the optical passbands.\ 
The three blue compact dwarf (BCD) galaxies (Haro 29, VIIZw 403, Mrk 178) included in our sample show obvious central enhancements even at 3.6 $\mu$m, which suggests that the central regions have sustained the central starburst that characterizes BCDs for an extended period of time (e.g.\ Zhao, Gu \& Gao 2011).\ 
 
To quantify the radial fall-off of the multi-wavelength emission, 
we obtained scale lengths of the exponential disk, $R_{D}$, of the FUV, NUV, {\it U}, {\it B}, {\it V} and 3.6 $\mu$m surface brightness profiles by fitting a single or multi-piece exponential to the profiles with the least squares method.\ The results are given in Table 2.\
Negative values of the scale lengths in Table 2 indicate that the relevant surface brightness increases toward larger radii.\ 
To visualize the difference of the scale lengths at different wavelengths, $R_{D}^{FUV}$/$R_{D}^{3.6 \mu m}$ is plotted against $R_{D}^{B}$/$R_{D}^{3.6 \mu m}$ for the galaxies (Figure \ref{fig2}).\ For the double exponential profiles, we only plot the disk scale length measured in the outer part.\ As is shown, the disk scale lengths measured at shorter wavelengths are smaller than those measured at longer wavelengths for $\sim$ 80\% (27) of our sample galaxies, at least in the outer disks.\ From Table 2.\  $\sim$ 65\% (22) of the galaxies have shorter disk scale lengths measured at shorter wavelengths in the inner disk, and 6 galaxies with broken surface brightness profiles have shorter scale lengths measured at shorter wavelengths in both the inner and outer disk.

FUV traces SF over the past $\sim$ 100 Myr, {\it B}-band emission is dominated by stellar populations younger than a few Gyr, and 3.6 $\mu$m is a very good proxy for SF over the whole lifetime of the galaxy (see references in the Introduction).\ 
The low metallicity of dIrr galaxies, combined with the lack of radial metallicity gradients, indicates that the multi-band surface brightness profiles should be predominantly determined by the distribution of the different stellar populations, with negligible effects due to changing internal extinction.\ 
Therefore, the fact that shorter wavelengths have shorter scale lengths most probably implies that the star-forming disks are shrinking towards the inner regions for most of the dIrr galaxies in our sample.\ Below, we will come back to this point with the help of multi-band SED modeling.\ The trends we see here in dIrr galaxies are in striking contrast with the findings in the luminous spiral galaxies (e.g.\ Ryder \& Dopita 1994), for which the inside-out disk growth scenario has been suggested and the disk scale length decreases with age.\ 
 
We note that for more than 50\% of the sample galaxies the multi-band surface brightness profiles show obvious breaks, 
especially at shorter wavelengths.\
$R_{D(out)}^{FUV}$/$R_{D(in)}^{FUV}$ is plotted against $R_{D(out)}^{3.6 \mu m}$/$R_{D(in)}^{3.6 \mu m}$ for galaxies exhibiting obvious surface brightness breaks in Figure \ref{fig3}.\ As can be seen, most of the galaxies with broken profiles are characterized as down-bending profiles.\ Furthermore, for the down-bending profiles, there is a trend that the breaks in FUV profiles are stronger (smaller ratio of $R_{D(out)}$/$R_{D(in)}$ for a down-bending profile) than that in 3.6 $\mu$m, indicating that the recent SFHs may play an important role in shaping the broken profiles.\ We note that the smaller ratio of $R_{D(out)}^{FUV}$/$R_{D(in)}^{FUV}$ is primarily due to the much steeper radial decline of the FUV in the outer disk for most galaxies.\ A down-bending profile has been found in the majority of spiral (e.g.\ Pohlen et al.\ 2002; Pohlen \& Trujillo 2006) and dIrr (HE06) galaxies.\

\section{SED modeling technique}
\subsection{Creating a Library of Star Formation Histories}
Detailed studies of high-quality stellar CMDs for very nearby dIrr galaxies (e.g.\ Dolphin 2000; Skillman et al.\ 2003; Cole et al.\ 2007; Tolstoy, Hill \& Tosi 2009) suggest that low-mass dIrr galaxies exhibit a wide variety of SFHs over their lifetime.\ Therefore, rather than adopting the commonly used two-component form of SFHs for larger galaxies (an exponentially declining component superimposed by random bursts, e.g.\ Kauffmann et al.\ 2003; da Cunha, Charlot \& Elbaz  2008), we use multi-component population models to create our SFH library.\ Specifically, we divided logarithmically the lifetime from the present to 13.70 Gyr ago into six independent age bins (0 - 0.03 Gyr, 0.03 - 0.10 Gyr, 0.10 - 0.35 Gyr, 0.35 - 1.18 Gyr, 1.18 - 4.02 Gyr, 4.02 - 13.70 Gyr), where each age bin has a constant SFR.\ 
The way we choose these age bins reflects the fact that the separation between isochrones of different ages strongly decreases with age.\ 
Our choice of the first age bin (0 - 0.03 Gyr) is justified by the nearly constant UV colors in the first $\sim$ 0.03 Gyr of evolution (e.g.\  Leitherer et al.\ 1999).\  
Similarly, the last age bin (from $\sim$ 4 Gyr to 13.7 Gyr) is justified by the nearly un-evolving shape of the optical to NIR spectrum at ages older than $\sim$ 4 Gyr (e.g.\ Bruzual \& Charlot 2003).\ 
In Appendix A, we show that the commonly assumed forms of SFHs tend to systematically bias the most probable estimates of the physical parameters.\ We make no assumptions here.\
Similarly, a common way to generate the SFH library is by Monte Carlo realizations of different SFHs.\ Instead, our library consists of a uniform, multi-dimensional grid of models without any prior assumption of the relevant physical parameters.\ This is necessary for exploring the complex SFHs of low-mass dIrr galaxies.\ 
When creating the SFH library, we allow the relative SFR among different age bins to logarithmically vary by an order of 1.5.\ 

We use the Charlot \& Bruzual (in preparation) stellar population models, which have implemented several important improvements compared to the Bruzual \& Charlot (2003) models.\ In particular, the new models include the recently improved treatment (Marigo et al.\ 2008) of the thermally pulsing asymptotic giant branch (TP-AGB), which at maximum (from a few hundred Myr to $\sim$ 1 Gyr for a single stellar population) could dominate the NIR emission.\ 
We adopt the stellar initial mass function (IMF) parametrized by Chabrier (2003).\ 
We did not consider chemical evolution in the models, all the components of each model in the library have the same metallicity.\ The metallicities of the population models are allowed to vary among 0.02, 0.2, and 0.4 Z$_{\odot}$, which are adequate for our dIrr galaxies.\ 

To make use of the H$\alpha$ data,  
the nebular hydrogen emission lines from \hii\ regions are also obtained for each model in the library.\ 
Collisionless Case B recombination for a 10,000 K gas is assumed.\ 
Using the recombination rates given by Hummer \& Sorey (1987), 
the H$\alpha$ luminosity is calculated from the UV stellar photons shortward of the Lyman limit.\ 
A systematic decline of the ratio of the integrated H$\alpha$-to-FUV flux in low luminosity dwarf galaxies has been shown (e.g.\ Sullivan et al.\ 2004; Meurer et al.\ 2009; Lee et al.\ 2009a). Three interpretations have been suggested in the literature, namely, the leakage of ionizing photons (e.g.\ Hunter et al.\ 2010), the stochasticity of massive SF at low SFR (e.g.\ Cervino \& Luridiana 2004), and non-universal stellar IMF (especially the upper end; e.g.\ Pflamm-Altenburg, Weidner \& Kroupa 2009).\ 
These possibilities all imply that the observed H$\alpha$ emission only gives a lower limit to the current SFR, under the assumption of Case B recombination and a well-populated universal IMF, which is the case for our modeling.\ 
Therefore, our H$\alpha$ data are used to set a lower limit on the H$\alpha$ flux from the models.\ Considering the uncertainties in the observed H$\alpha$ flux (HE04), we set the lower limit to be 0.8$\times$$L_{\tt H\alpha, obs}$.\ 

We use the two-phase dust attenuation recipes developed by Charlot \& Fall (2000) as our standard dust extinction law.\ 
The effective absorption is proportional to $\lambda$$^{-0.7}$.\ 
Charlot \& Fall (2000) found a typical value of $\sim$3 for the ratio $\mu$ of the extinction experienced by the young stellar populations ($\le$ 10 Myr) to that by the older stellar populations.\ 
We adopt this value when creating our SFH library.\ The {\it V}-band dust extinction affecting the young stellar populations is  allowed to vary linearly between 0.0 and 0.5 magnitude.\ 
The gray extinction curve derived by Charlot \& Fall (2000) is in agreement with recent results for a large sample of nearby galaxies (Johnson et al.\ 2007).\

\subsection{Determining the physical parameters}
\label{subsec: meth}
Our final library consists of $\sim$ 4 $\times$10$^{6}$ different SFHs, which are created by allowing all the relevant parameters (i.e.\ dust extinction, metallicity, and relative SFR among different age bins) to vary uniformly (either logarithmically or linearly) among physically reasonable ranges.\ 
For each SFH in the library, we integrate respectively back to the past 0.03 Gyr, 0.1 Gyr, 1 Gyr, and the galaxy lifetime (here we adopt a Hubble time of 13.7 Gyr) in order to get the SFR averaged over these different timescales.\ 
In the following we denote the SFR averaged over the past 0.03 Gyr, 0.1 Gyr, 1 Gyr, and the whole lifetime as SFR$_{0.03}$,  SFR$_{0.1}$,  SFR$_{1}$, and SFR$_{13.7}$, respectively.\
The accumulated present-day stellar mass is also derived.\ 
From our experiments, the broadband SED modeling is not expected to be very sensitive to the SFR averaged over the past 0.03 Gyr (see the Appendix) and that averaged over the past 4 Gyr.\ So we concentrate on the interpretation of \sfr, \sfg, \sfl\, and the stellar mass in this work.\ 
However, as mentioned above, our choice of six age bins guarantees that we have a relatively complete SFH library adequate for broadband SED modeling.\  

We use the Bayesian technique (e.g.\ Kauffmann et al.\ 2003; Kong et al.\ 2004) to find the most probable estimates of parameters related to each of the observed broadband SEDs.\ 
Specifically, given an observed SED, we construct the probability density function (PDF) and the related cumulative distribution function (CDF) for each parameter defining our SFH library by calculating the likelihood exp(-$\chi$$^{2}$/2) that an observed SED corresponds to each SFH model in our library.\ 
As described in the previous subsection, all the models with the emergent H$\alpha$ luminosity (scaled by a factor determined from the weighted least squares) smaller than 0.8$\times$$L_{\tt H\alpha, obs}$ are excluded from the final determination of the parameters.\  
The most probable value for a specific parameter is taken as the median of the corresponding PDF.\ The confidence interval is defined as the central 68\% of the CDF.\ 
Simply adopting the single best-fitting value as the most probable estimate for the parameter is blind to degeneracies among relevant parameters.\ Our method is very robust, and the possible degeneracies are factored into the related confidence intervals.\ 
In Appendix A, we thoroughly discuss the validity of these parameter estimates from the broadband SED modeling.\ 

\subsection{Correcting the $GALEX$ NUV data for Galactic extinction}
\label{subsec: cmw}
Before feeding the data into the SED modeling, we correct the data for the foreground MW extinction (Schlegel, Finkbeiner \& Davis 1998), adopting the Cardelli et al.\ (1989) extinction curve with $R_{V} = A_{V}/E(B-V) = 3.1$.\ 
While there are nearly fixed, linear relations between $A_{\lambda}$ and $E(B-V)$ for most bands, this is not the case for $GALEX$ NUV (e.g.\ Wyder et al.\ 2007), due to the finite bandwidth ($\sim$ 1000 \AA) and the presence of the  2175 \AA\ bump in the MW extinction curve.\ The ratio $A_{\tt NUV}/E(B-V)$ is sensitive to both the emergent extra-galactic SED and the foreground MW reddening $E(B-V)$.\ For example, the SED emergent from a constant SF and the SED from a 7 Gyr old single stellar population have $A_{\tt NUV}/E(B-V)$ $\sim$ 8.2  and $\sim$ 7.5 respectively for 0.05 mag of MW reddening.\ Also, for the SED emergent from a constant SF, $A_{\tt NUV}/E(B-V)$ would be 7.9 if the Galactic reddening was 0.5 mag.\ 
Therefore, unlike the other bands, the NUV data are not corrected for MW extinction with a fixed ratio of $A_{\tt NUV}/E(B-V)$.\ Before comparing the data with each model in the library, the NUV is corrected for the Galactic extinction using the ratio of $A_{\tt NUV}/E(B-V)$ predetermined from the emergent SED of the model and the Galactic $E(B-V)$ towards the specific galaxy.\ In other words, the MW extinction correction for NUV is consistently applied before comparing each model with a given galaxy.\ 

\section{Results of SED modeling}
\begin{figure}
\centering
\includegraphics[width=0.9\textwidth]{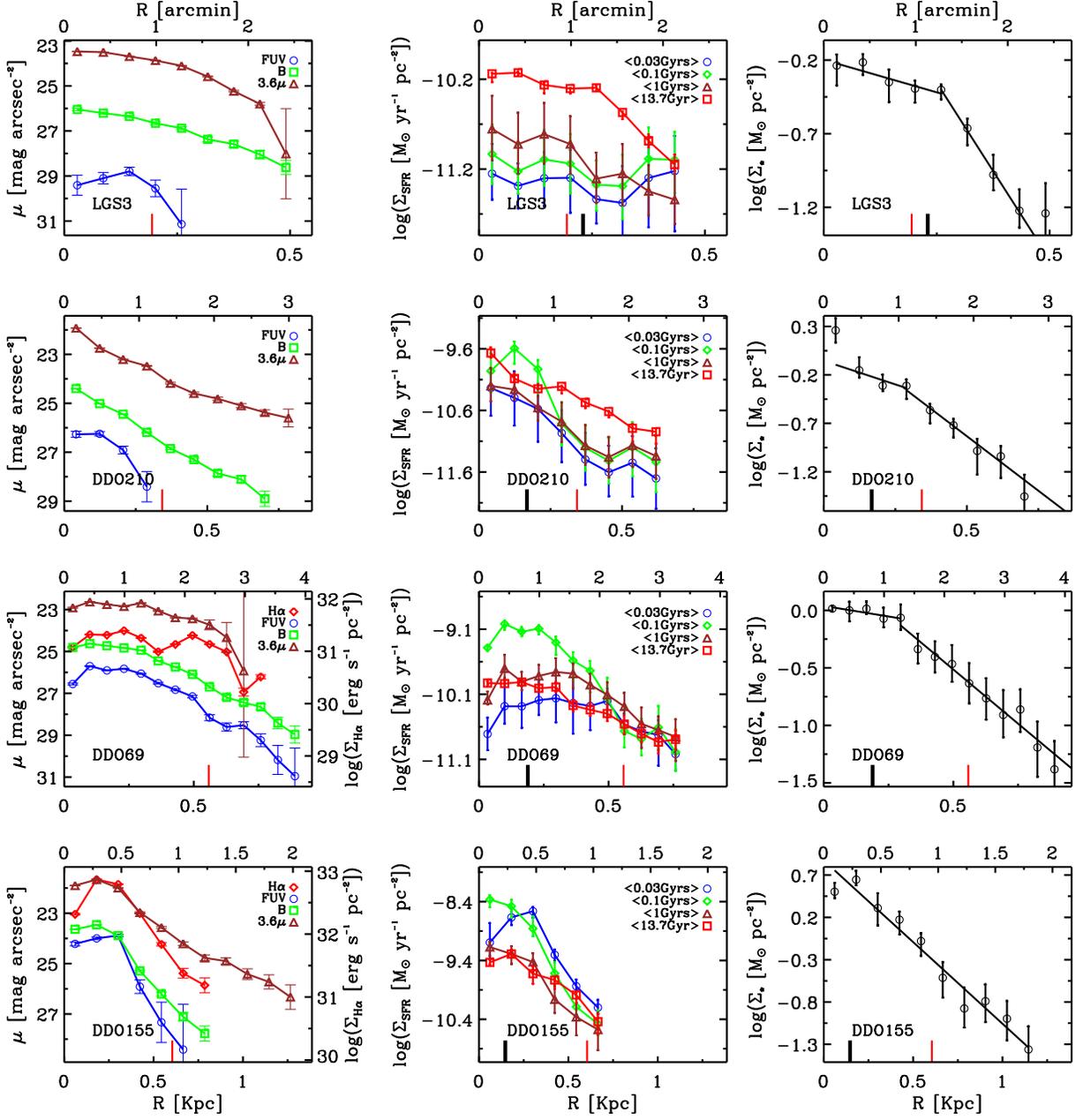}
\caption{{\it First column}: Azimuthally-averaged surface brightness profiles of the H$\alpha$, FUV, {\it B} and 3.6 $\mu$m images.\ The NUV or {\it U}-band profiles are plotted instead of FUV if there are no FUV observations. {\it Second column}: Radial variations of the SFR averaged over different timescales obtained from our multi-band SED modeling.\ The averaging timescales are the most recent 0.03 Gyr, the most recent 0.1 Gyr,  the past 1 Gyr, and the Hubble time. {\it Third column}: Inclination-corrected stellar mass surface density profiles from the SED modeling.\ The fitted exponential profiles are overplotted as {\it solid} lines.\ The {\it thin red} line marks the Holmberg radius, and the {\it thick black} vertical line marks the {\it V}-band scale length.\ 
[A color version of this figure is available in the electronic journal.] \label{fig1}}
\end{figure}

\begin{figure}
\centering
\epsscale{0.9}
\plotone{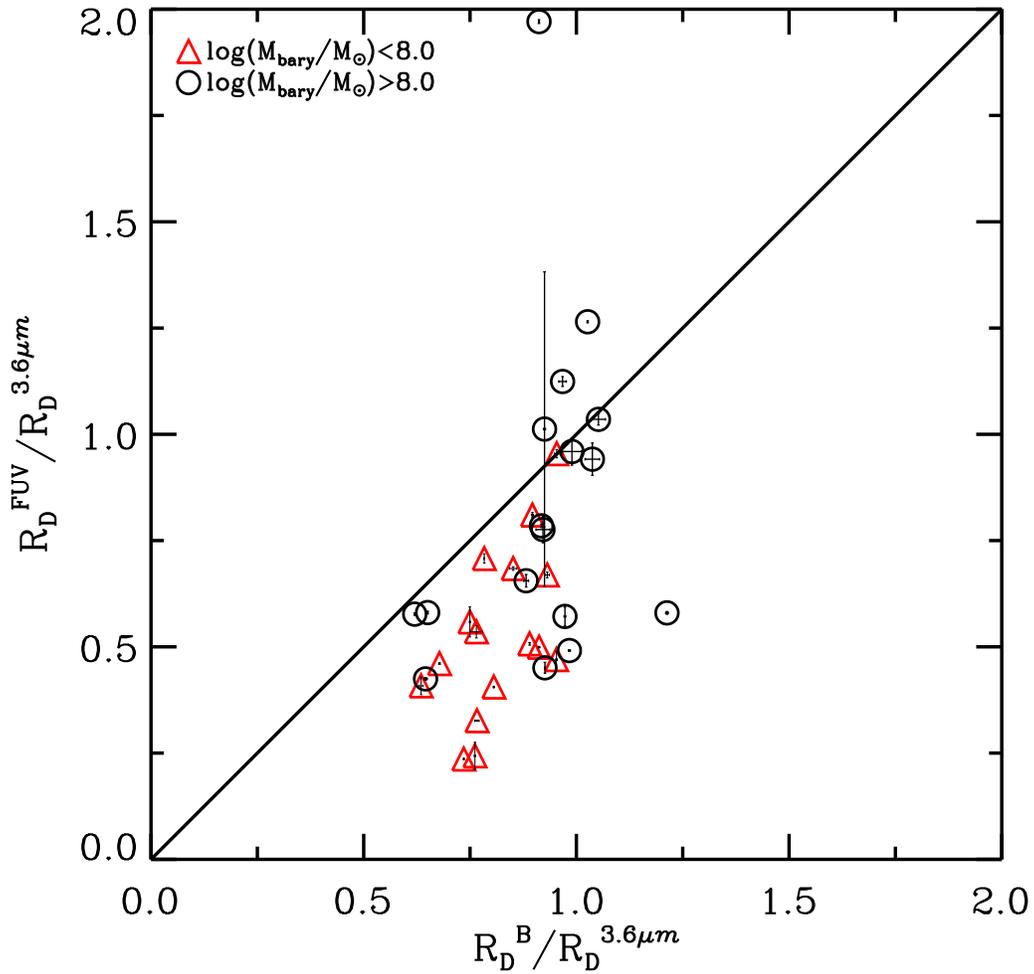}
\caption{Ratios of disk scale lengths $R_{D}^{\tt FUV}$/$R_{D}^{\tt 3.6 \mu m}$ plotted against $R_{D}^{\tt {\it B}}$/$R_{D}^{\tt 3.6 \mu m}$ for our sample of galaxies.\ $R_{D}^{\tt FUV}$, $R_{D}^{\tt {\it B}}$, and $R_{D}^{\tt 3.6 \mu m}$ denote the disk scale lengths of the FUV, {\it B}, and 3.6 $\mu$m passbands, respectively.\ The disk scale length measured at the outer disk is plotted for galaxies with broken surface brightness profiles.\ The galaxies with baryonic mass larger and smaller than 10$^{8}$ M$_{\odot}$ are denoted as {\it black open circles} and {\it red open triangles}, respectively.\ 
The ({\it solid}) line of equality is shown to guide the eye,
and galaxies below the line have larger $R_{D}^{\tt {\it B}}$ than $R_{D}^{\tt FUV}$.\
[See the electronic journal for a color version of this figure.]
\label{fig2}}
\end{figure}

\begin{figure}
\centering
\epsscale{0.9}
\plotone{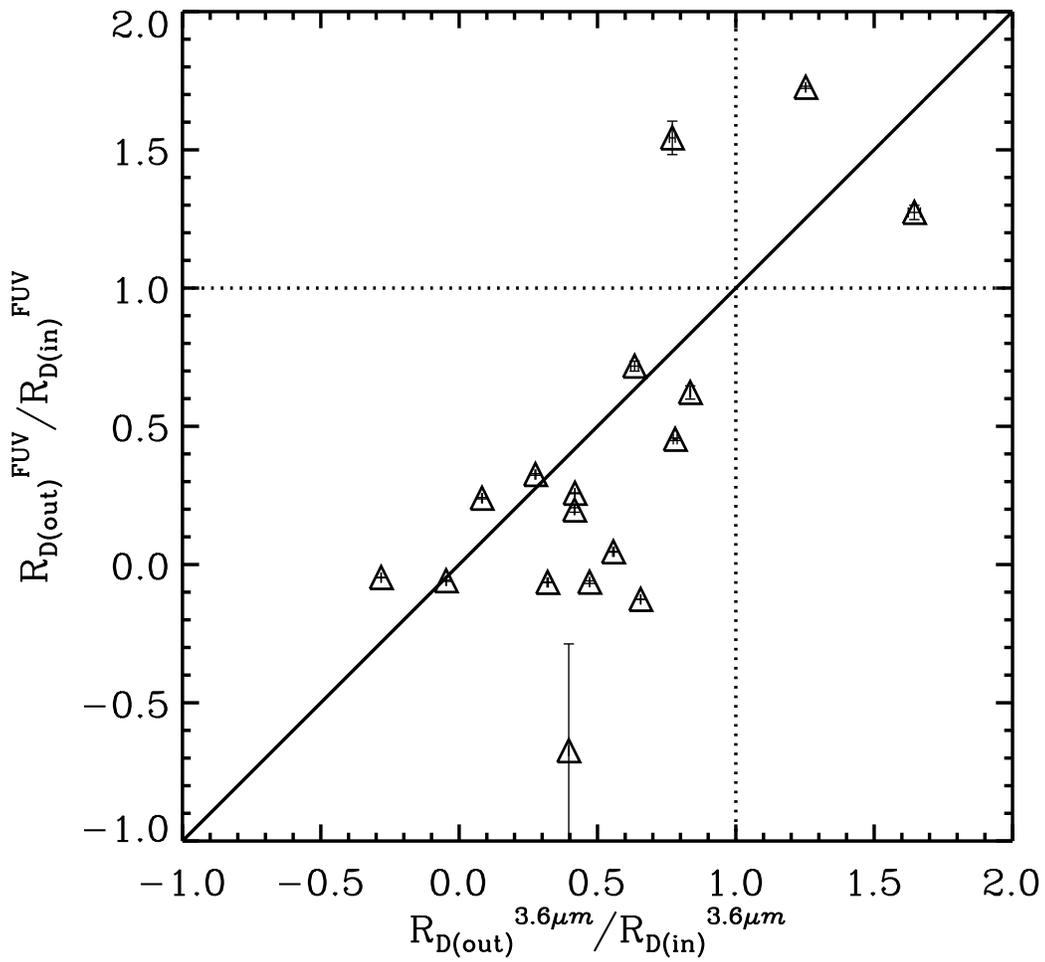}
\caption{
The disk scale length ratios of the outer-to-inner disks in the FUV plotted against the disk scale length ratios of the outer-to-inner disks in the 3.6 $\mu$m.\ 
The {\it solid } line is the relationship of equality.\ 
The {\it dotted} lines mark values for {\it non-broken} profiles. 
\label{fig3}}

\end{figure}

The SFH variations as a function of radius within the galaxies, determined from our SED modeling,
are shown for each galaxy in the middle column of Figure \ref{fig1}.\ Specifically, the SFR averaged over the past 0.03 Gyr, 0.1 Gyr, 1 Gyr and the Hubble time are shown, respectively, as {\it blue circles}, {\it green squares}, {\it brown triangles} and {\it red diamonds}.\ 
Stellar mass surface density profiles $\Sigma_*$ are shown in the third column of Figure \ref{fig1}.\
Table 3 gives the globally integrated stellar mass,
lifetime averaged SFR$_{13.7}$, and ratios of SFR$_{\tt 0.1}$ to SFR$_{\tt 13.7}$ and SFR$\tt _1$ to SFR$_{\tt 13.7}$ for each galaxy.\ 
All the photometry with uncertainties $\sigma > 0.3$ mag are excluded from the SED modeling.\ 
The global results are derived by integrating the relevant radial profiles across the whole disk.\

\subsection{Comparison with the stellar CMD-based SFHs}
\begin{figure}
\centering
\epsscale{0.9}
\plotone{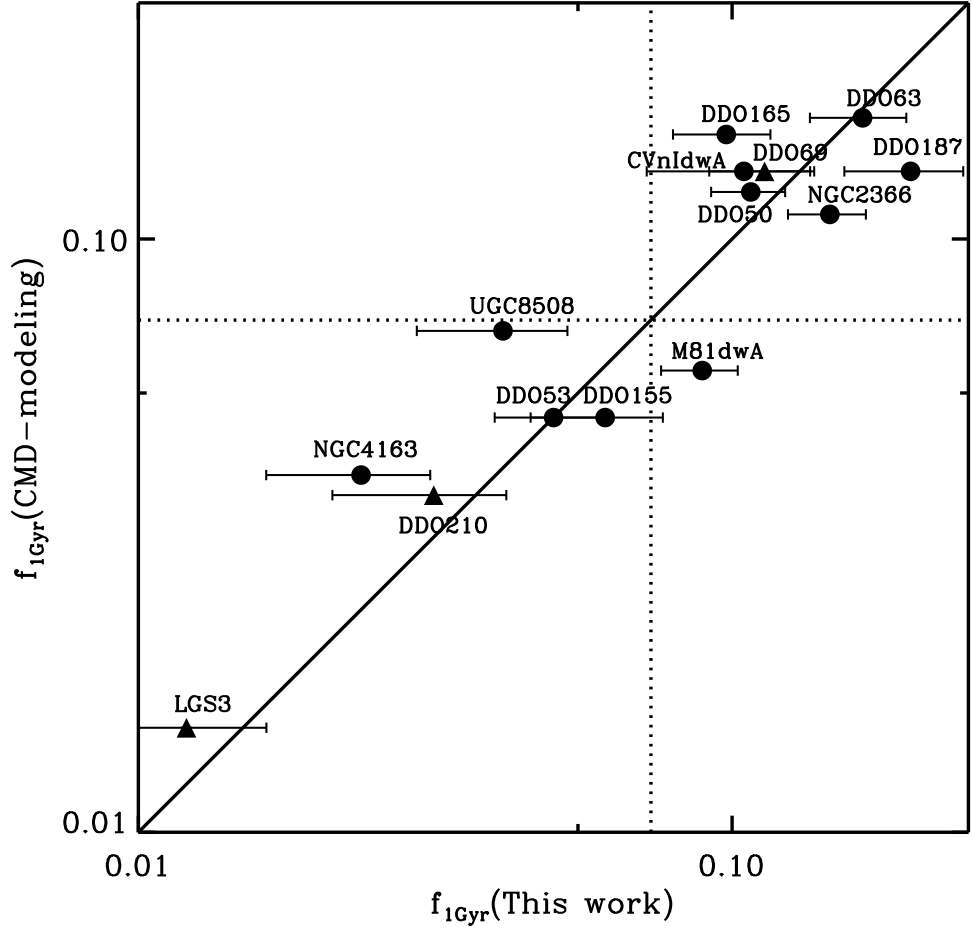}
\caption{Fractions of the total stellar mass formed during the past 1 Gyr f$_{\tt 1Gyr}$ derived from our SED-modeling are compared to those determined from stellar CMD-based analysis.\ The galaxies with CMD analysis from Weisz et al.\  (2011) and from Orban et al.\ (2008) are denoted as {\it filled circles} and {\it filled triangles}, respectively.\  
The ({\it solid}) line of equality is plotted to guide the eye.\
The {\it dotted} lines mark the expected value for a constant SFH over the Hubble time. 
\label{fig4}}

\end{figure}

Before proceeding to the interpretation of our SED modeling results, we compare our results with those derived from existing stellar CMD-based analysis.\ 
Half of our galaxies have stellar CMD-based SFHs obtained from {\it HST} data and available in the literature.\ 
Among these galaxies, 14 galaxies have CMD analysis for a significant part (the observing field covers more than 20\% of the area enclosed by the Holmberg radius) of the stellar disk.\ 
The availability of {\it GALEX} FUV and {\it Spitzer} 3.6 $\mu$m data makes our results for SFR$_{\tt 0.1}$ and M$_{\tt \star}$ particularly reliable.\ Therefore, we compare our results of the fraction of the SF accumulated during the recent 1 Gyr  f$_{\tt 1Gyr}$ ((SFR$_{\tt 1}$/\sfl)/13.7) with that derived from CMD modeling.\
Among the 14 galaxies, the f$_{\tt 1Gyr}$ for 11 galaxies (denoted as {\it black circles} in Figure \ref{fig4}) are from Weisz et al.\  (2011), and those of the other 3 galaxies (denoted as {\it triangles} in Figure \ref{fig3}) are from Orban et al.\ (2008).\ 
From the comparison of these 14 galaxies (Figure \ref{fig4}), our globally integrated results from the SED-modeling are broadly consistent with those from the CMD-based analysis.\ 

Generally speaking, studying the resolved stellar populations is the best way to derive a detailed SFH, 
provided the data are deep enough.\ 
We note that most of the above CMD analyses are carried out on relatively shallow data sets ($M_{I}$ $\lesssim$ 0).\ This means that the SFH at older times ($\gtrsim$ 2 Gyr) is mainly constrained by the age-insensitive red giant branch (RGB).\ The RGB alone at best could provide a good estimate of the total SF, from its formation to a few Gyr ago (Greggio 2002) to within a factor of $\sim$ 2.\ 
Nevertheless, the results from the current CMD-fitting are also sensitive to the modeling techniques used by different works.\ For example, using the same $HST$ data sets of WLM, Weisz et al.\ (2008) derived a two times higher  f$_{\tt 1Gyr}$ than Dolphin et al.\ (2005).\ 
Even more, adopting different stellar evolution libraries can lead to differences of at least a factor of 2 in the derived SFR (Gallart et al.\ 2005).\ Our library of SFHs are created with stellar evolution tracks combining those of Padova 1994 (see Bruzual \& Charlot 2003) and Marigo et al.\ (2008) for the evolution of AGB, whereas the above CMD-modeling adopted the tracks of Padova 2000 (Girardi et al.\ 2002) and Marigo et al.\ (2008).\ This undoubtedly leads to some differences in the SFHs.\ We also noticed that the CMD-modeling works listed above adopted different IMFs from ours.\ Finally, the CMD-modeling often covers a smaller spatial area on the galaxy than our SED fitting.\ Given these uncertainties and differences, the results from our SED-modeling are in reasonable agreement with those from CMD-modeling.\ 

SFHs derived from broadband SED-modeling have been thought in the past to be biased towards younger, luminous populations.\ The above comparison shows that the average SFR during the past $\sim$ Gyr is well constrained by modeling the broadband SED with a relatively complete library of SFHs.\ 
We point out that the availability of {\it U} and {\it B}-band data is essential in SED-modeling because the two bands straddle the age-sensitive 4000 \AA\  break.\

\subsection{Globally Integrated SFHs}
By integrating the relevant surface density profiles times the circumference over radius, we derived (in Table 3) the asymptotic total stellar masses $M_{\star}$ (See \S~5.4),  SFR$_{13.7}$, $b_{0.1Gyr}$ (\sfr/\sfl), and $b_{1Gyr}$ (\sfg/\sfl).\ 
The galaxies are listed in order of increasing total baryonic mass in Table 3.\ 
A factor of 2-3 (e.g.\ Hunter \& Gallagher 1986; Kennicutt et al.\ 2005; Lee et al.\ 2009b) enhancement of current SF compared the averaged past rate has been used as one way to define a starburst in the literature.\  Compared to SFR$_{13.7}$,
38\% (13) of the sample galaxies have \sfr\ enhanced at least by a factor of 2, and 47\% (16) have \sfr\ consistent with \sfl\ within a factor of 2.\ 
For \sfg,\  12\% (4 galaxies) have been enhanced at least by a factor of 2, and 71\% (24 galaxies) have \sfg\ consistent with \sfl\ within a factor of 2.\ 
Among the 5 galaxies with \sfr/\sfl\ $<$ 0.5, 4 (LGS 3, DDO 216, DDO 101, WLM) have baryonic mass (here approximated to be the stellar mass plus atomic gas mass, see below) smaller than 10$^{8}$ M$_{\odot}$.\  
Similarly, all 6 galaxies with \sfg/\sfl\ $<$ 0.5 have baryonic mass smaller than 10$^{8}$ M$_{\odot}$.\ 
The total atomic gas mass was derived by multiplying the \hi\ gas mass by 1.34 to account for He.\ 
The \hi\ masses used here were all from single dish \hi\ emission line observations reported in the literature (see HE04 for the references), and the total stellar mass is derived from our SED modeling (see \S~\ref{sec: masden}).\

\begin{figure}
\centering
\epsscale{0.9}
\plotone{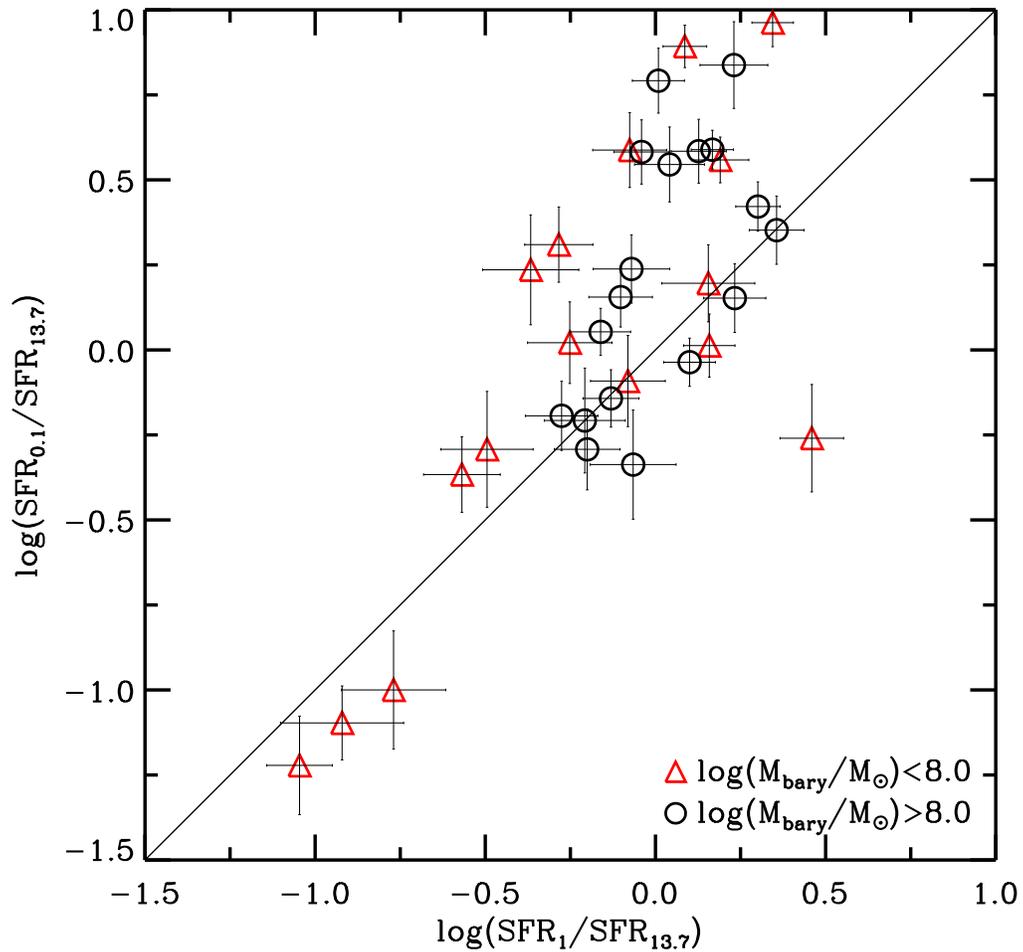}
\caption{Ratios of globally integrated \sfr/\sfl\ vs. \sfg/\sfl\ for the galaxy sample.\ The ({\it solid}) line of equality is plotted to guide the eye.\ The galaxies are divided into two groups according to total baryonic mass.\ Those with masses less than $10^8 $M$_{\odot}$ are denoted with {\it red triangles}, and those with masses greater than this are plotted as {\it black circles}.\
[See the electronic journal for a color version of this figure.]
\label{fig5}} 

\end{figure}

We plot b$_{0.1Gyr}$ and b$_{1Gyr}$ for the galaxies (Table 3) in Figure \ref{fig5}.\
The galaxies with baryonic mass smaller and higher than 10$^{8}$ M$_{\odot}$ are denoted as different symbols.\ A mass of 10$^{8}$ M$_{\odot}$ is about the median baryonic mass of our whole sample.\ There is almost no relationship between SFHs and total baryonic mass for our sample, except that all galaxies that exhibit extremely declining SFHs are relatively low mass systems.\ Among the whole sample, 
five galaxies (LGS 3, DDO 216, NGC 4163, DDO 101 and WLM) have experienced a significant ($>$30\%) decline in SF activity (both \sfr\ and \sfg) over the past $\sim$ Gyr.\ We note that all of these five galaxies have baryonic masses smaller than 10$^{8}$ M$_{\odot}$, and, except for WLM,  the (atomic) gas to stellar mass ratios are smaller than 1.

The tidal indices (Table 1) of LGS 3 and DDO 216 are larger than 1, which indicates that their evolution has likely been significantly influenced by environment.\ Most of the gas in these two galaxies could have been removed by either ram pressure stripping or tidal disturbance.\  
In particular, based on the morphology of the \hi\ distribution, McConnachie et al.\ (2007) concluded that DDO 216 is currently undergoing ram pressure stripping caused by the intragroup/intracluster medium.\ 
The tidal index (Karachentsev et al.\ 2004) of a galaxy is defined as the maximum density enhancement caused by all neighboring galaxies, and higher values correspond to the likelihood of more significant tidal interaction with neighboring galaxies.\ 
This is consistent with the expectation that, lower mass systems (below a few times 10$^{8}$ M$_{\odot}$) are more susceptible to significant influence of both the environment (Gunn \& Gott 1972) and stellar feedback (e.g.\ winds and supernova explosions (SNe)), which could even induce blow-out of interstellar gas from the galaxy (e.g.\ Mac Low \& Ferrara 1999).\ 

A few galaxies deserve special mention.\ M81dwA, DDO 75, Haro 29 and NGC 1569 hold the largest ratio ($>$ 5) of \sfr/\sfl\ among our sample.\ 
The recent starburst in NGC 1569, which has the highest current SF intensity in our sample, may be ascribed to an interaction of some kind (e.g.\ Stil \& Israel 1998; Muhle et al. 2005; Johnson et al.\ 2011).\ 
Similarly, the recent rise of the SFR in M81dwA may be due to the recent interaction within the M81 group, which has several dramatic galaxy-galaxy interactions in progress.\ 
Haro 29 is a typical BCD galaxy undergoing a intense burst of SF in the central regions (Thuan \& Martin 1981).\ 
DDO 75 and Haro 29 both have negative tidal indices (-0.6 for DDO 75, -1 for Haro 29), indicating that the recent enhancement of SF may be ascribed to internal processes.\ 
For DDO 154, Kennicutt \& Skillman (2001) reported a factor of $\sim$ 2-4 times lower recent SF compared to the past based on H$\alpha$ imaging, which is inconsistent with our results for \sfr.\ We interpret the disagreement as being due to the fact that H$\alpha$ is not a robust SFR indicator in LSB dwarf galaxies (e.g.\ Meurer et al.\ 2009; Lee et al.\ 2009a; Hunter et al.\ 2010).\ 

\subsection{Azimuthally-averaged Radial SFHs}
The sample as a whole exhibits a diversity of radial variations of SFHs (Figure \ref{fig1}).\ 
However, some systematic trends do exist.\ 

\begin{figure}
\centering
\epsscale{0.9}
\plotone{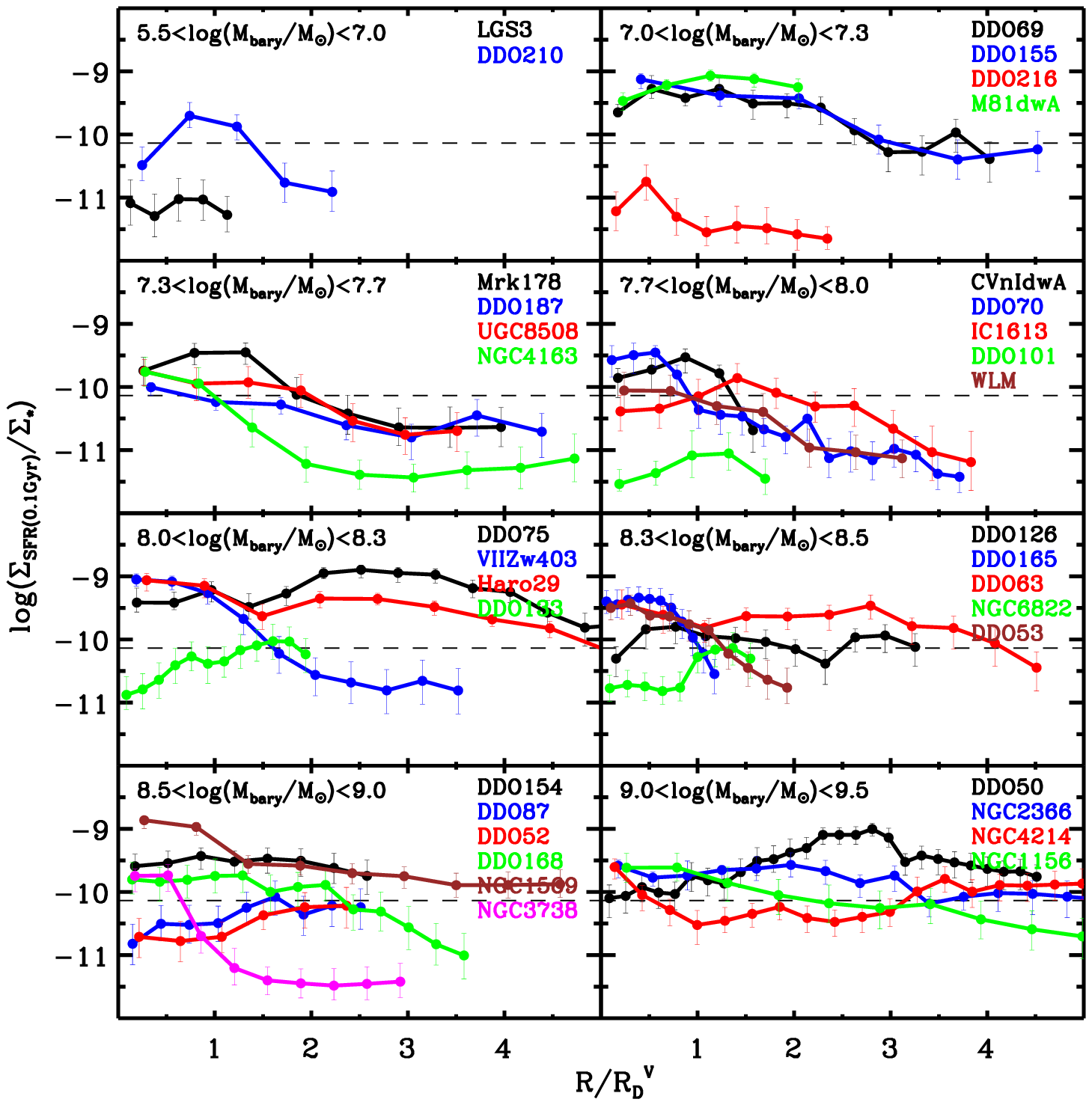}
\caption{log($\Sigma_{\tt SFR_{0.1}}$/$\Sigma_{\star}$) is plotted as a function of radius, which is normalized by the {\it V}-band disk scale length.\ The galaxies are plotted in order of total baryonic mass from the upper left to the lower right, and within each panel galaxy names are listed in order of increasing baryonic mass.\
The {\it dashed} line in each panel marks a constant SFH over a Hubble time.
\label{fig6}}

\end{figure}

\begin{figure}
\centering
\epsscale{0.9}
\plotone{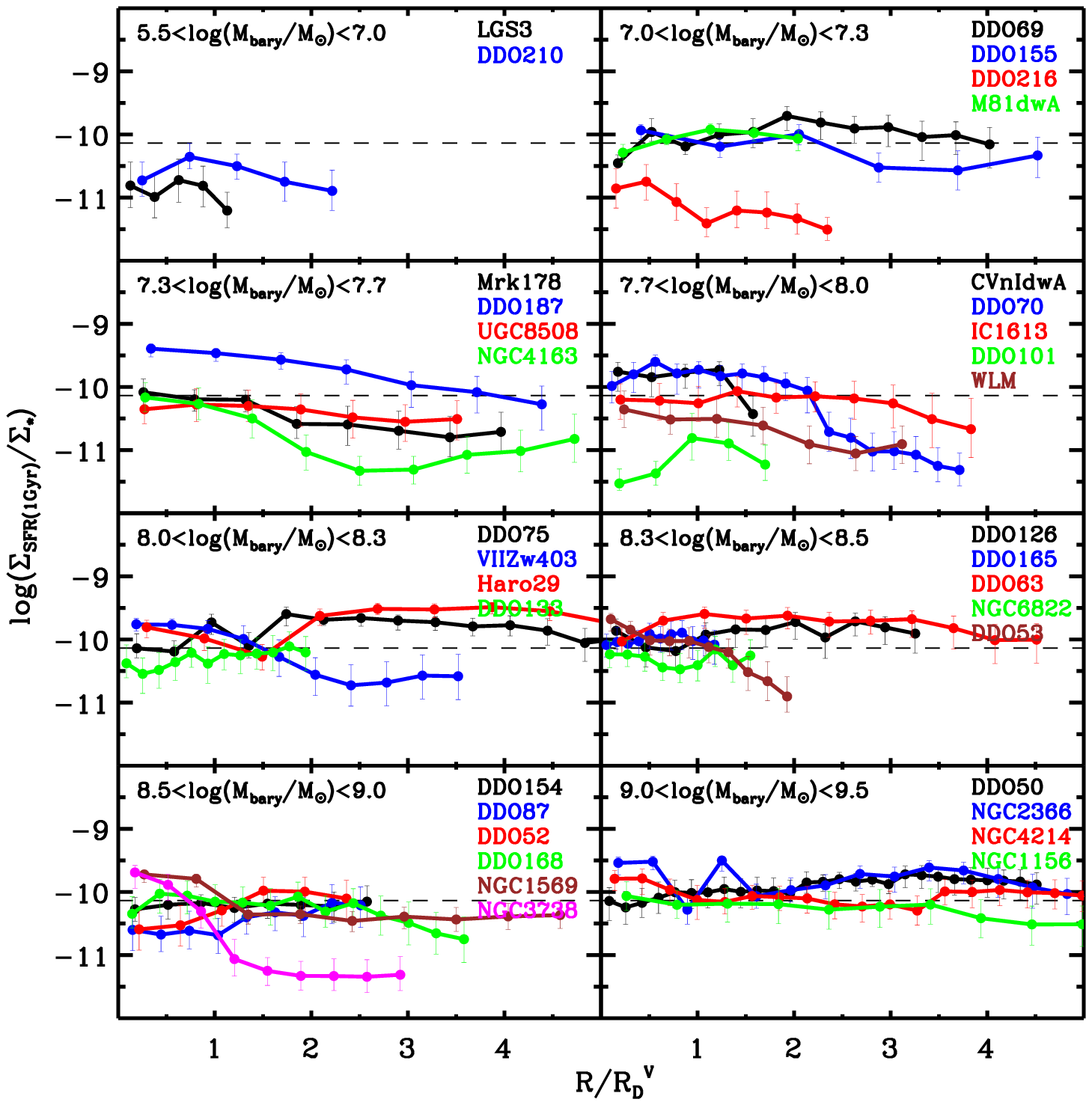}
\caption{log($\Sigma_{\tt SFR_{1}}$/$\Sigma_{\star}$) plotted as a function of radius, which is normalized by the {\it V}-band disk scale length.\ The galaxies are plotted in order of total baryonic mass from the upper left to the lower right, and within each panel galaxy names are listed in order of increasing baryonic mass.\
The {\it dashed} line in each panel marks a constant SFH over a Hubble time.
\label{fig7}}

\end{figure}

Figures \ref{fig6} and \ref{fig7} show the radial variations of log($\Sigma_{\tt SFR_{0.1}}$/$\Sigma_{\star}$) and of log($\Sigma_{\tt SFR_{1}}$/$\Sigma_{\star}$).\ 
The radius has been normalized by the {\it V}-band disk scale length.\ 
The galaxies in Figures \ref{fig6} and \ref{fig7} are plotted in order of increasing total baryonic mass from the upper left panel to the lower right panel, and the galaxy names are listed in order of increasing baryonic mass in each panel.\  
The $dashed$ line in each panel marks a constant SFH over a Hubble time.\ 
Deep stellar CMD analysis of nearby dIrr galaxies (e.g.\ IC 1613: Skillman et al.\ 2003; Leo A: Cole et al.\ 2007) suggests that most of the SF in dIrr galaxies may take place at intermediate ages, rather than persisting constantly over a Hubble time. In constrast, LGS 3 (Hidalgo et al.\ 2011) consists of mostly old populations.\  
Therefore, the data points lying above the $dashed$ line in Figures \ref{fig6} and \ref{fig7} do not necessarily mean elevated SFR compared to the past.\ 

\begin{figure}
\centering
\epsscale{1}
\plotone{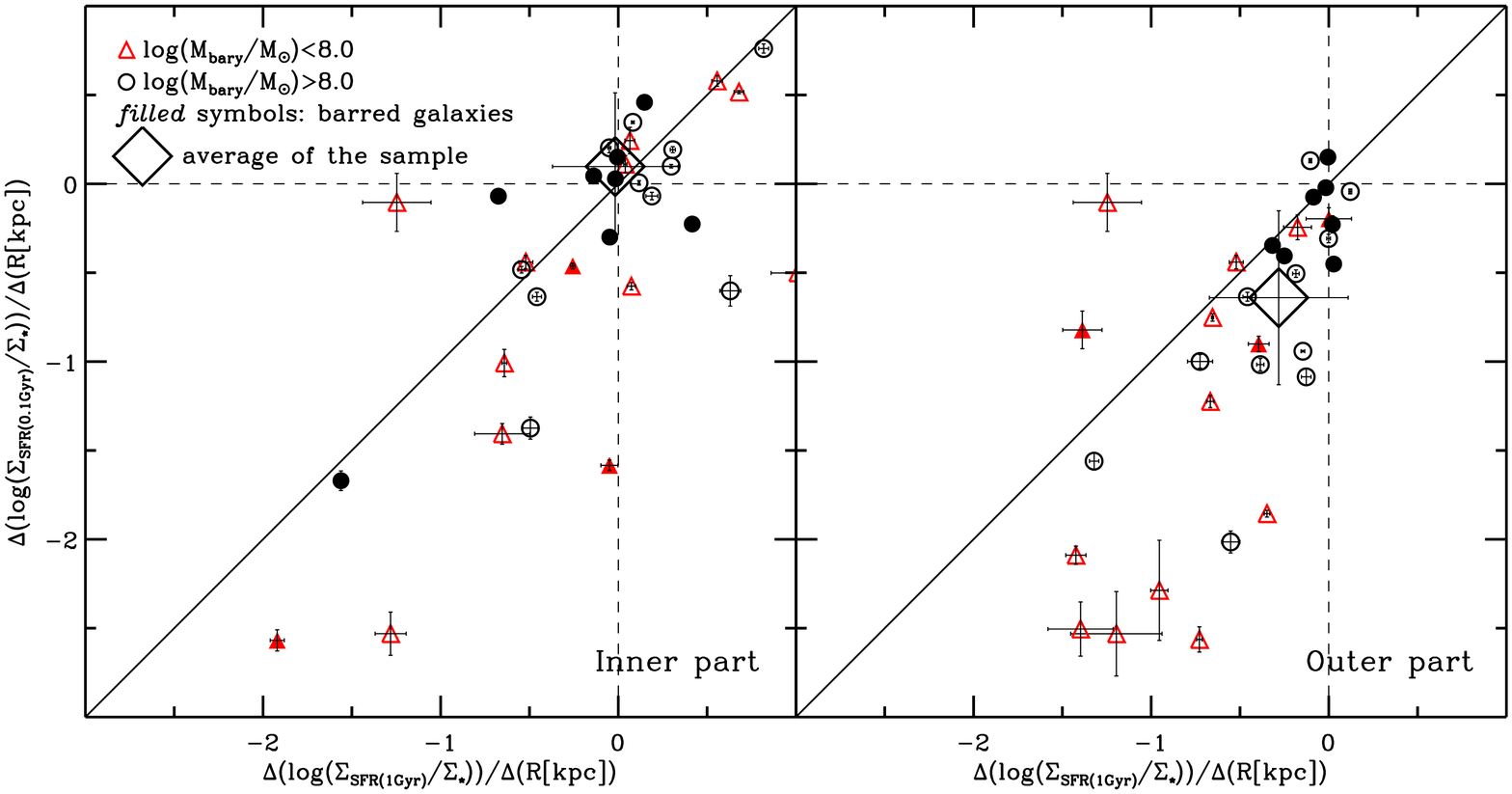}
\caption{Fitted slope of the radial variations of log($\Sigma_{\tt SFR_{0.1}}$/$\Sigma_{\star}$) plotted against the slope of the radial variations of log($\Sigma_{\tt SFR_{1}}$/$\Sigma_{\star}$).\ The $left~panel$ shows the results for the inner disks, and the $right~panel$ shows the results for the outer disks.\ 
The galaxies with baryonic mass larger and smaller than 10$^{8}$ M$_{\odot}$ are denoted as {\it black open circles} and {\it red open triangles}, respectively.\ The $large~diamonds$ represent the averages among the whole sample galaxies.\
The ({\it solid}) line of equality is plotted to guide the eye.\
The {\it dashed} lines mark a flat gradient slope.\ 
Here the slope fitting was done on the physical (kpc) scales.\
[See the electronic journal for a color version of this figure.]
\label{fig8}}

\end{figure}

\begin{figure}
\centering
\epsscale{1}
\plotone{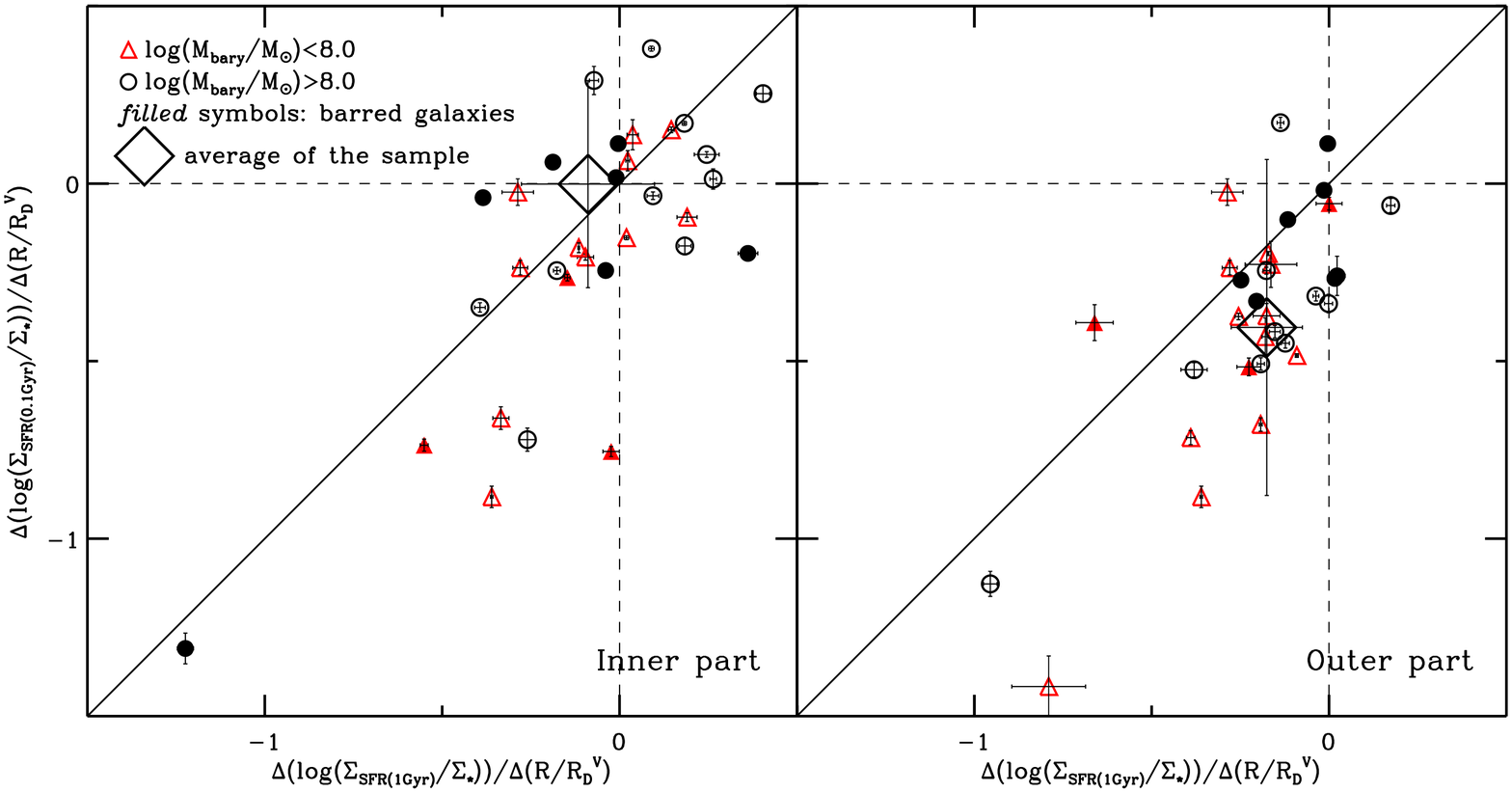}
\caption{Fitted slope of the radial variations of log($\Sigma_{\tt SFR_{0.1}}$/$\Sigma_{\star}$) plotted against the slope of the radial variations of log($\Sigma_{\tt SFR_{1}}$/$\Sigma_{\star}$).\ The $left~panel$ shows the results for the inner disks, and the $right~panel$ shows the results for the outer disks.\ 
The galaxies with baryonic mass larger and smaller than 10$^{8}$ M$_{\odot}$ are denoted as {\it black open circles} and {\it red open triangles}, respectively.\ The $large~diamonds$ represent the averages among the whole sample galaxies.\ 
The ({\it solid}) line of equality is plotted to guide the eye.\
The {\it dashed} lines mark a flat gradient slope.\ 
Here the slope fitting was done on the radius normalized to the {\it V}-band disk scale length.\
[See the electronic journal for a color version of this figure.]
\label{fig9}}

\end{figure}

To quantify the radial trends of the SFH variation, we did linear least-squares fitting to the radial variations of log($\Sigma_{\tt SFR_{0.1}}$/$\Sigma_{\star}$) and log($\Sigma_{\tt SFR_{1}}$/$\Sigma_{\star}$).\ If there is an obvious break in the stellar mass surface density profiles (Figure \ref{fig1} and Table 3), we fit the slope for the inner and the outer parts separately.\ 
The {\it GALEX} data (thus the measure of recent SFR) of several galaxies (LGS 3, DDO 210, DDO 216 and NGC 1569) have high S/N ratios only in the inner disk, in this case we only fit the inner disk.\ We also fit the relevant ratios as a function of normalized (by the {\it V}-band scale length) radius.\ The slopes (or gradients) from the fitting are listed in Table 4.\
The weighted averages of the relevant slopes for the galaxies below and above a baryonic mass 10$^{8}$ M$_{\odot}$ are also listed in Table 4.\ 
The slopes are further plotted in Figures \ref{fig8} and \ref{fig9}.\ Also shown in Figures \ref{fig8} and \ref{fig9} are the weighted averages ({\it large diamonds}) of the whole sample.\ Negative slope indicates that the SFR averaged over the relevant timescale becomes more centrally concentrated compared to the past.\ As is shown, on average, our sample galaxies have more negative radial variations of log($\Sigma_{\tt SFR_{0.1}}$/$\Sigma_{\star}$) and log($\Sigma_{\tt SFR_{1}}$/$\Sigma_{\star}$) in the outer disks than the inner disks, and log($\Sigma_{\tt SFR_{0.1}}$/$\Sigma_{\star}$) has steeper radial declining than  log($\Sigma_{\tt SFR_{1}}$/$\Sigma_{\star}$).\ 
In the inner disks, 44\% (15 galaxies) of the sample show negative radial slopes for both log($\Sigma_{\tt SFR_{0.1}}$/$\Sigma_{\star}$) and log($\Sigma_{\tt SFR_{1}}$/$\Sigma_{\star}$), and 9 of the 15 galaxies have M$_{bary}$ $<$ 10$^{8}$ M$_{\odot}$.\ 
A few galaxies with M$_{bary}$ $>$ 10$^{8}$ M$_{\odot}$ (i.e.\ DDO 165, DDO 63, DDO 154, DDO 168, NGC 2366 and NGC 4214) have almost flat slopes ($|$slope$|$ $<$ 0.2) in the inner disks.\
In the outer disks, 80\% (27 galaxies) have negative slopes for both log($\Sigma_{\tt SFR_{0.1}}$/$\Sigma_{\star}$) and log($\Sigma_{\tt SFR_{1}}$/$\Sigma_{\star}$), and 14 of the 27 galaxies have M$_{bary}$ $<$ 10$^{8}$ M$_{\odot}$.\ Particularly, 35\% (12 galaxies, 7 with M$_{bary}$ $\lesssim$ 10$^{8}$ M$_{\odot}$) of the galaxies exhibit negative slopes across the observed disk.\  
In NGC 4163 and NGC 3738, the recent SF has been suppressed by more than an order of magnitude in the outer regions, even though these two galaxies have almost constant ratios of log($\Sigma_{\tt SFR_{0.1}}$/$\Sigma_{\star}$) and log($\Sigma_{\tt SFR_{1}}$/$\Sigma_{\star}$) at large radii.\ These results are in line with Figure \ref{fig2}, which shows the comparison between the scale length of FUV, {\it B} and 3.6 $\mu$m. 
 
Among the whole sample, the weighted averages of the slopes of $\Delta({\rm log}(\Sigma_{\tt SFR_{0.1}}/\Sigma_{\star}))/\Delta({\rm R})$ are 0.10$\pm$0.42 and -0.64$\pm$0.48 for the inner and the outer disks, respectively. The weighted averages of $\Delta({\rm log}(\Sigma_{\tt SFR_{1}}/\Sigma_{\star}))/\Delta({\rm R})$ are -0.05$\pm$0.37 and -0.28$\pm$0.39 for the inner and the outer disks, respectively. 
Furthermore, as listed in Table 4, the inner disks tend to have much shallower radial slopes of $\Delta({\rm log}(\Sigma_{\tt SFR}/\Sigma_{\star}))/\Delta({\rm R})$ than the outer disks. 
The SF activity in dIrr galaxies was suggested to be a random percolating process across the stellar disk (e.g.\ van Zee 2001). According to our results, {\it this randomly percolating scenario may be only appropriate for the inner disks of some galaxies (especially those relatively massive galaxies).  
As far as the whole disk of the low mass dIrr galaxy is concerned, the star-forming disk has been shrinking from the outer part.}

\subsection{Stellar Mass Surface Density Profiles}
\label{sec: masden}

The stellar mass surface density $\Sigma$$_{\star}$ profiles of the galaxies are shown in the last column of Figure 1.\ 
The stellar mass surface density profiles are well fitted with a piece-wise exponential function, which is overplotted on the profiles.\ 
For some relatively faint galaxies, the high quality IRAC 3.6 $\mu$m data reaches further radii than the other bands.\ In this case, considering the relative insensitivity of the $M_{\star}/L_{\tt NIR}$ to the underlying stellar populations (Bell \& de Jong 2001), we extend stellar mass surface density profiles derived from our multi-band fitting to the radius reached by the 3.6 $\mu$m surface photometry, adopting the $M_{\star}/L_{3.6 \mu m}$ ratio from the radius where our multi-band SED fitting ends.\ 
The 3.6 $\mu$m photometry for some galaxies with large angular size does not reach as far as the other bands due to the limited field of view of the {\it Spitzer} observations.\ In this case, we constrained the stellar mass profiles at the outer part by modeling the {\it UBV} (and {\it GALEX} FUV/NUV if available) data.\   
We also derived the asymptotic total stellar mass by applying the ``growth curve" method to the accumulated stellar mass surface density profiles (e.g.\ Mu\~noz-Mateos et al.\ 2009).\ 
Briefly, we obtained the radial gradient of the accumulated stellar mass surface density at each observed radius, 
then the accumulated stellar mass surface density was linearly fit as a function of the gradient and the y-intercept (zero gradient) of the fit was adopted as the asymptotic stellar mass.\ 
Table 3 lists the central stellar mass density $\Sigma_{\star,center}$ extrapolated from the inner exponential disk, stellar mass concentration index C$_{31}$, the inner/outer scale lengths of the stellar mass density profiles, and the asymptotic total stellar mass for each galaxy.\ C$_{31}$ is defined here as the ratio of the radii that encompass 75\% and 25\% of the total stellar mass (de Vaucouleurs 1977).

Like the radial light profiles, the stellar mass surface density profiles of dIrr galaxies usually can be described by either a single or piece-wise exponential profile over most of the disk.\ The most striking feature about our galaxies is that most of them have {\it broken} stellar mass surface density profiles.\ Specifically, $\sim$ 80\% (27) of the galaxies exhibit obvious down-bending profiles, in the sense that the outer disks have a steeper mass profile than the inner disks.\ Whereas three (DDO 70, Haro 29, NGC 1569) galaxies have obvious up-bending (flatter outer disk) profiles, in the sense that the outer disks exhibit shallower profiles than the inner disks.\ 
Both Haro 29 and NGC 1569 have experienced intense, extended starbursts in their inner regions.\  
It has been shown that down-bending surface brightness profiles are very common in spiral galaxies (e.g.\ Ferguson \& Clarke 2001; Pohlen \& Trujillo 2006).\ Nevertheless, Bakos, Trujillo \& Pohlen (2008) found a pure exponential stellar mass profile for the spiral galaxies with down-bending surface brightness profiles.\ This is different from our dIrr galaxies which tend to have both down-bending stellar mass profiles and down-bending surface brightness profiles.\ Amor\'in et al.\ (2007) found that the LSB stellar hosts in BCD galaxies have near-exponential profiles.\ 
From our results for BCD galaxies, not only the LSB stellar hosts, but also the inner starburst regions have exponential stellar mass surface density profiles, although the scale length is different for the inner starburst region and the LSB stellar host (Figure \ref{fig1}). 

The present-day stellar mass surface density profile of dIrr galaxies should be primarily determined (see \S~6.1.1) by the radial variations of SFH.\ As is shown above (Table 4), on average, the SF in the outer disks has been decreasing more significantly both over time (Figures \ref{fig6} and \ref{fig7}) and over radius (Figures \ref{fig8} and \ref{fig9}) compared to the inner disks, which would naturally lead to down-bending stellar mass surface density profile, seen in most of our galaxies.\ This scenario of outside-in depression of SF is different from what may be happening in typical BCD (e.g.\ Haro 29, NGC 1569, see Figure \ref{fig1}) galaxies.\ The central starburst of BCD galaxies results in faster buildup of the inner stellar disk than the outer part, which could lead to up-bending stellar mass surface density profiles.

As can be seen, a local enhancement of the recent SF may affect the stellar mass density profile in these low mass dIrr galaxies.\ For example, in NGC 2366, the supergiant \hii\ complex (NGC 2363, Youngblood \& Hunter 1999) $\sim$ 1 kpc away from the center corresponds to the remarkable spike in the stellar mass surface density profile.\ The concentration index of the stellar mass surface density profile is generally close to, but lower than, that of a single exponential (C$_{31}$ = 2.81). 

Lower mass systems tend to have lower stellar mass concentrations (Table 3).\ 
However, the BCD galaxy Haro 29 is as centrally concentrated (C$_{31}$ = 5.3) as a de Vaucouleurs r$^{1/4}$ bulge profile, which suggests that the centralized starburst has lasted an extended period of time.\ Assuming the \sfr\ in the central regions is representative of the recent episode of central starburst, and the central stellar mass is dominantly contributed by the burst, the starburst would have lasted more than 0.6 Gyr.

\section{Discussions}

\subsection{The disk assembly mode of dIrr galaxies: outside-in}
The ``inside-out"  growth mode has long been suggested for the formation of galaxy disks (e.g.\ Larson 1976; Chiappini et al.\ 1997; Mo, Mao \& White 1998; Naab \& Ostriker 2006). The ``inside-out" scenario reproduces many observations of spiral galaxies, such as the radial gradients of both broad-band colors (bluer outward) and metallicity (lower outward) (de Jong 1996; Bell \& de Jong 2000; MacArthur et al.\ 2004; Wang et al.\ 2011), and the extended UV emission discovered in outer spiral disks (e.g.\ Thilker et al.\ 2007; Boissier et al.\ 2008).\ Mu\~noz-Mateos et al.\ (2007) found a moderate inside-out disk formation by studying the radial profiles of {\it GALEX} FUV and 2MASS {\it Ks} for a sample of relatively face-on nearby spiral galaxies.\ Recently, the ``inside-out" growth mode has been confirmed for M 33 (Williams et al.\ 2009) and NGC 300 (Gogarten et al.\ 2010) from analysis of CMDs obtained with $HST$.\ For M 33, it was shown (Williams et al.\ 2009; Barker et al.\ 2011) that the ``inside-out" scenario only applies to the inner disk, and the region beyond the surface brightness break exhibits positive age gradient. 

However, unlike luminous spiral galaxies, late-type dIrr galaxies exhibit a variety of behaviors in terms of color profiles (Kormendy \& Djorgovski 1989; Tully et al.\ 1996; Jansen et al.\ 2000; Taylor et al.\ 2005; HE06).\ 
Nevertheless, Tully et al.\ (1996) found that galaxies fainter than $M_{B}$ $\sim$ -17 become redder with radius in the Ursa Major cluster.\ 
Later, with a larger sample, Jansen et al.\ (2000) found that the reddening (with radius) trend may occur at a fainter absolute magnitude than that found by Tully et al.\ (1996).\ 
Of a sample of 94 dIrr galaxies studied by HE06, 64\% of the galaxies with a gradient in $B-V$ become redder with radius.\ In particular, all the galaxies with $M_{B}$ $>$ -14 become redder at least in one color (either $B-V$ or $U-B$) at outer radii.\ 
Recently, Tortora et al.\ (2010) analyzed the optical color gradients of $\sim$ 50,000 nearby Sloan Digital Sky Survey galaxies.\ 
They found a good correspondence between the color gradients and the stellar mass of the galaxies.\ In particular, below a stellar mass $M_{\star}$ $\lesssim$ 10$^{8.7}$ M$_{\odot}$, the color gradient slopes become positive, which were mainly attributed to the positive metallicity gradients.\ The lowest mass galaxies in the sample of Tortora et al.\ (2010) have $M_{\star}$ $\sim$ 10$^{8.2}$ M$_{\odot}$, compared to the median stellar mass $\sim$ 10$^{7.2}$ M$_{\odot}$ of our sample.\
All the above studies are based on optical broadband data alone, which makes the interpretation of the radial color gradients ambiguous, due to the degeneracy between age, metallicity, and extinction.

Additionally, from stellar CMD analysis of two different (HST/WFPC2) fields in IC1613, Skillman et al.\ (2003) (also Bernard et al.\ 2007) found that the SF activity in the outer field has been significantly depressed during the last Gyr, which is in contrast to the ``inside-out" growth scenario.\ Similarly, by analyzing resolved stellar populations, the characteristics of more spatially extended older populations have been reported both for transition-type dwarf galaxies (e.g.\ DDO 210: McConnachie et al.\ 2006; Phoenix: Hidalgo et al.\ 2009) and dwarf spheroidal (dSph) galaxies (e.g.\ Sculptor dSph: Tolstoy et al.\ 2004; Fornax dSph: Battaglia et al.\ 2006).\ 
A clear trend of ``outside-in" quenching of recent SF was also found (Gallart et al.\ 2008; Indu \& Subramaniam 2011) in the 
Large Magellanic Cloud, which has a mass comparable to that of the most massive galaxy (i.e.\ NGC 1156) in our sample.  

These studies are in agreement with our findings for our sample of dIrr galaxies.\ 
We point out that our analysis here is only sensitive to the SF during the recent $\sim$ Gyr and the SF averaged over the whole lifetime.\ The radial variations of SF at the intermediate ages could be different from the recent past.\ For instance, in the late-type spiral galaxy NGC 2976, Williams et al.\ (2010) found a deficit of stellar populations younger than $\sim$ Gyr beyond the break of the disk surface brightness profile, with similar ancient stellar populations at all radii.\ Nevertheless, since the suppression of SF in the outer disk was found in the majority of our sample galaxies, the trend we found here must reflect the evolutionary process of the stellar disks of low-mass dIrr galaxies in general.\ In the following, we discuss possible interpretations of the observed radial variations of the stellar populations in dIrr galaxies.

\subsubsection{{\it In situ} SF vs. Secular redistribution?}
The present-day radial distribution of the stellar populations could be ascribed either to {\it in situ} SF or to a stellar redistribution process.\ 
For example, disk non-axisymmetric instabilities (e.g.\ spiral arms, bars) are capable of a large-scale redistribution of stars.\
In spiral galaxies, resonant scattering with transient spiral arms can lead to substantial radial stellar migration (e.g.\ Sellwood \& Binney 2002; Roskar et al.\ 2008), 
which complicates determination of the {\it in situ} SFH.\ Likewise, the existence of bars can, besides inducing strong gas inflows, drive substantial redistribution of the stellar disk, even increasing the disk scale length  (e.g.\ Hohl 1971; Debattista et al.\ 2006).\ However, dIrr galaxies usually lack large-scale instabilities (H98), which is a requisite of generating spiral arms and (probably) bar instabilities (e.g.\ Lin \& Pringle 1987; Mihos, McGaugh, \& de Blok 1997).

Nevertheless, it has been shown that a slowly rising rotation curve naturally leads to the alignment of elongated orbits, and thus is more prone to bar formation (Lynden-Bell 1979).\ Even more, once the bar is made, it may exist for many dynamical times in a situation of solid body rotation, which is the case over a large part of the disk in most dIrr galaxies.\ Nevertheless, the bar growth of late-type galaxies may stop at a very early stage because the co-rotation radius moves out of the main disk (Combes \& Elmegreen 1993).\ There exists a strong correlation between the strength of the bar and the central density (e.g.\ Elmegreen et al.\  2007; Sheth et al.\ 2008).\ dIrr galaxies usually lack a central stellar mass excess, which implies that bar instabilities, if they exist, are very inefficient in transferring angular momentum across the disk of dIrr galaxies.\
Of our sample, 11 galaxies (32\%) show evidence for a bar based on ellipse fitting of optical isophotes (HE06).\ Figures \ref{fig8} and \ref{fig9} suggest that there is no preference for barred galaxies exhibiting larger or smaller radial gradients of log($\Sigma_{\tt SFR}$/$\Sigma_{\star}$).\ 
We note that identifying bars in dIrr galaxies is not as straightforward as in luminous spiral galaxies.

\subsubsection{External influences}
\begin{figure}
\centering
\epsscale{0.9}
\plotone{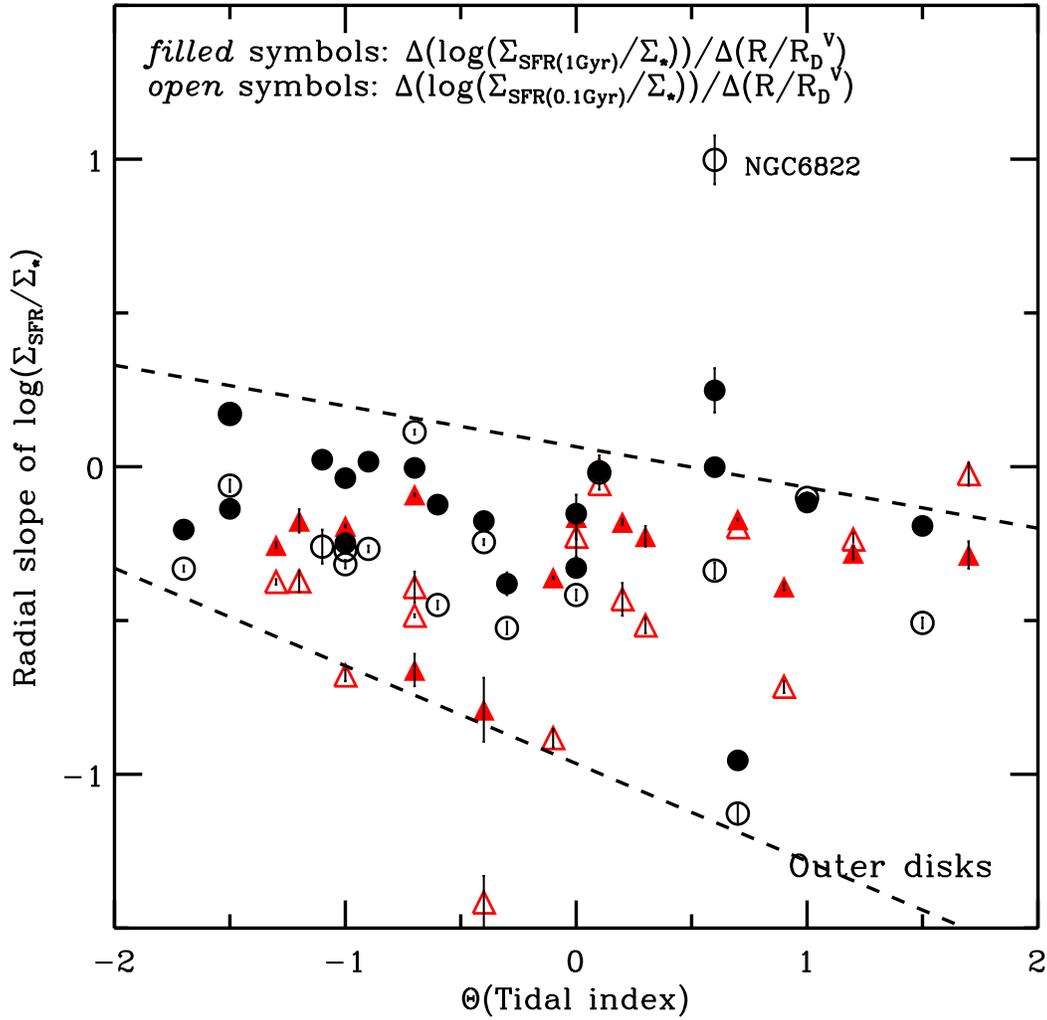}
\caption{Fitted slope of the radial variations of log($\Sigma_{\tt SFR_{0.1}}$/$\Sigma_{\star}$) and log($\Sigma_{\tt SFR_{1}}$/$\Sigma_{\star}$) in the outer disks plotted as a function of the tidal index $\Theta$.\ The {\it black circles} and the {\it red triangles} denote galaxies with baryonic mass larger and smaller than 10$^{8}$ $M_{\odot}$, respectively.\ The {\it open} and the {\it filled} symbols denote the slope of  log($\Sigma_{\tt SFR_{0.1}}$/$\Sigma_{\star}$) and the slope of log($\Sigma_{\tt SFR_{1}}$/$\Sigma_{\star}$), respectively.\
The {\it dashed} lines mark the upper and the lower envelopes of the distribution of most data points in the plot.\ 
[See the electronic journal for a color version of this figure.]
\label{fig10}}

\end{figure}

External factors, such as interactions (both with the intragroup/intracluster medium and with neighboring galaxies) and the cosmic UV background (e.g.\ Gnedin 2000), can also lead to more centrally concentrated recent SF or depressed SF in outer disks.\ The tidal index (Table 1) suggests that less than 5 of the galaxies studied here could be noticeably affected ($\Theta \gtrsim 1$) by neighboring galaxies.\
The slopes of the radial variations of log($\Sigma_{\tt SFR_{0.1}}$/$\Sigma_{\star}$) and log($\Sigma_{\tt SFR_{1}}$/$\Sigma_{\star}$) for the outer disks are plotted against the tidal index in Figure \ref{fig10} (There is no calculation of $\Theta$ for DDO 101 in Karachentsev et al.\ (2004), so we arbitrarily set $\Theta$ as 0 when creating the plot).\
There is no obvious correlation between the radial variations of the SFHs and the environment.\
However, most galaxies fall between an upper envelope and a lower envelope, as indicated by the {\it dashed} lines in the plot.\
NGC 6822 has such a large angular size projected on the sky that the observations only covered the central regions.\ 
As mentioned above, the SF of dIrr galaxies usually peaks in the circumnuclear regions, 
therefore the slopes measured here for NGC 6822 do not reflect the SFH variations across the main star-forming disk.\ 
In fact, based on $HST$ images of five fields in NGC 6822, Wyder (2001) found a higher ratio of the recent SFR to the average past rate in the inner bar regions than the outer regions, consistent with the trend we found in this study.\
The galaxies with larger $\Theta$ prefer lower (zero or negative) slopes of the radial variations of log($\Sigma_{\tt SFR_{0.1}}$/$\Sigma_{\star}$) and log($\Sigma_{\tt SFR_{1}}$/$\Sigma_{\star}$).

Two of the most obvious environmental effects studied in the literature are ram pressure stripping of the gas component caused by a hot gaseous intragroup/intracluster medium (Gunn \& Gott 1972; Lin \& Faber 1983) and tidal disturbance from neighboring galaxies (e.g.\ Mayer et al.\ 2001).\
By studying the dynamical properties of the dwarf galaxies in the Fornax cluster, Drinkwater et al.\ (2001) show that the dwarf galaxies are still falling into the cluster, and the fraction of dwarfs with active SF drops rapidly towards the cluster center.\ This is a clear indication of the environmental effect on the evolution of dwarf galaxies.\
Similarly, Pustilnik et al.\ (2002) found that the blue compact galaxies in higher density environment have on average less \hi\, which could be attributed to either ram pressure stripping of the gas or tidal interactions. 

Presumably, both ram pressure stripping and tidal disturbance should be more effective on lower mass galaxies.\
The outer disk, especially in relatively low mass systems where the gravitational well and the gas surface density are relatively low, is more prone to gas removal due to ram pressure.\
Furthermore, significant tidal forces from neighboring galaxies could induce a bar instability and ensuing gas inflows towards the inner regions (e.g.\ Mayer et al.\ 2001), which leads to a more centrally concentrated SF.\
In our sample, lower mass galaxies tend to have more centrally concentrated SF than relatively higher mass galaxies, which is in general agreement with the expectation of the above two environmental effects. 

By studying the \hi\ distribution of spiral galaxies in the Virgo cluster, Cayatte et al.\ (1994) found that some galaxies close to the cluster core have normal gas densities (compared to field counterparts) in the inner disks but show a strong decline or cutoff of the \hi\ intensity starting within the half light radius of the stellar disk.\ These galaxies should have been undergoing remarkable ram pressure stripping.\
Actually, the radial \hi\ distribution (Hunter et al.\ 2011, in preparation) for some of our galaxies do show a strong decline starting well within the stellar disk.\
Numerical simulations (e.g.\ Vollmer 2003; Kapferer et al.\ 2008) suggest that both ram pressure stripping and tidal disturbance are needed to reproduce the observations.\
Figure \ref{fig10} suggests that the tidal disturbance from neighboring galaxies should play some role in the inward movement of SF activity. 

The presence of a UV background can heat the gas in low mass halos, which prevents SF at least in the outer disk
where the gas is not dense enough for self-shielding (Susa \& Umemura 2004).\
By solving the radiative transfer equation for the diffuse UV background in a pre-galactic cloud, Tajiri \& Umemura (1998) showed that, above a critical number density $n_{\tt crit}$ $\sim$ 1.4 $\times$ 10$^{-2}$ cm$^{-3}$ (N$_{\tt H_{I}}$ $\sim$ 1.3 $\times$ 10$^{19}$ for a gas disk $\sim$ 0.3 kpc thick), which is almost independent of the UV background intensity, self-shielding of the ISM against the UV background is prominent.\ Based on our \hi\ emission line maps of these galaxies, and according to this critical density, the typical gas column density ($\sim$ 10$^{20}$ cm$^{-2}$) across the observed stellar disks of our galaxies should be sufficient for them to be self-shielded.\ Therefore, the cosmic UV background is unlikely to be the main driver of the centrally concentrated SF. 

\subsubsection{Regulation of the SFR through stellar-feedback}
Depressed SF in outer dwarf disks could be attributed to self-regulation due to stellar feedback.\
Feedback from both SNe and stellar winds has been shown to be a major factor shaping the evolution of dwarf galaxies due to their shallow potential wells (e.g.\ Dekel \& Silk 1986; Mori, Ferrara \& Madau 2002; Mashchenko, Wadsley, \& Couchman 2008).\ From the numerical simulations, the SNe/stellar winds-regulated evolution of isolated dwarf galaxies tends to be characterized by an episodic SFH (e.g.\ Chiosi \& Carraro 2002; Stinson et al.\ 2007).\ Such SFHs have been suggested to explain the anomalously blue colors and low metallicities of dIrr galaxies (e.g.\ Searle, Sargent, \& Bagnuolo 1973).

In this SNe/stellar winds-regulated evolution scenario, the SN explosions and stellar winds from massive stars following one episode of SF trigger gas heating and expansion, and some gas may be blown away from the galaxy.\ Subsequent cooling allows the remaining gas to sink deeper into the center and start another cycle of SF.\
The outside-in formation scenario has been suggested for low mass elliptical galaxies (Martinelli, Matteucci \& Colafrancesco 1998; Pipino \& Matteucci 2004) to explain the positive gradients of [Mg/Fe] and the blue colors in the center of dwarf elliptical galaxies.\ The outer regions, where the gravitational potential well is relatively low, are more susceptible to the galactic wind (thus blow-out/blow-away) compared to the inner regions.\  
The blow-out/blow-away of the high angular momentum gas (compared to the inner regions) in the outer part of the dwarf disk results in a remarkable net loss of angular momentum, and thus the shrinking of the gas disk.\
This leads to a depression of SF in the outskirts and more prolonged SF in the central regions.\
Recent simulations (Valcke et al.\ 2008) indeed found that, for a self-regulated, SNe-driven evolution of the low mass (M$_{\tt bary}$ $\lesssim$ 10$^{8}$ M$_{\odot}$) dwarf galaxies, the SF becomes more and more centrally concentrated with time, which is in agreement with what we see here.\
We point out that this scenario of outside-in shrinking may not apply to the early universe when the galaxy just started assembling and the potential was shallow (susceptible to stellar feedback) even in the inner region.\
Governato et al.\ (2010) found that the outflows resulted from stellar feedback may selectively remove low angular momentum gas, especially at z $>$ 1.

Recent high-resolution hydrodynamical simulations including cooling, SN feedback and the UV background radiation (Sawala et al.\ 2010) suggest that, while the combined effects of SN feedback and UV heating are necessary to reproduce the observations, SN feedback is the key factor in determining the evolution of low mass dwarf galaxies.\ The cosmic UV radiation has almost no effect on galaxy evolution if the SN feedback was ignored in the simulations. 

\subsubsection{Pressure support in the gas disk}
Low-mass dIrr galaxies are dynamically hot systems compared to luminous spiral galaxies.\
The (turbulent) gas velocity dispersion within dwarf galaxies can be comparable to the ordered rotational velocity, which means the gas pressure 
may provide significant radial support to the rotating gas disk (Stinson et al.\ 2009; Dalcanton \& Stilp 2010).\
Since the gas density usually declines with radius, the pressure gradient, together with the centrifugal acceleration, will counteract the gravity.\
A higher efficiency of SF in the inner regions then leads to quicker consumption of gas, which results in an initial reduction of the gas density there.\
After the enhanced turbulence from SF subsides, the pressure in the center will drop too, and more gas will flow in so that 
pressure gradients and centrifugal force balance gravity again.\ Thus some shrinking of the gas disk might be possible, 
within the limits of angular momentum conservation.\
In their simulations of isolated low mass dwarf galaxies, Stinson et al.\ (2009) found a positive age gradient with radius after gas contraction, 
stellar migration, and supernova-triggered SF, with the latter two effects being more important for lower mass galaxies.\
Dalcanton \& Stilp (2010) further demonstrated that the impact of pressure support alone on the evolution of gas disk may be weak. 

\subsection{The relationship between $\Sigma_{\tt SFR}$, $\Sigma_{\star}$ and SFH}
\label{subsec: sfrm}

H98 found that the radial profiles of current SF activity correlate better with those of the older stellar emission (e.g.\ {\it V}-band) than with the atomic gas in dIrr galaxies.\
Since star-forming molecular clouds are forming from the atomic gas, this lack of a correlation between SF and gas is unexpected.\
Similarly, Ryder \& Dopita (1994; see also James, Bretherton, \& Knapen 2009) discovered a good correlation between $I$-band surface brightness  (a proxy for the mass dominant old stellar populations) and H$\alpha$ for a sample of nearby spiral galaxies.\ Recently, a good correspondence between the surface brightness in SDSS $r$-band and in FUV was also shown for a sample of LSB galaxies (Wyder et al.\ 2009).\
If the optical broadband emission traces the underlying mass-dominant old stellar populations, the above correlations would be equivalent to a relationship between SF and the stellar mass surface density.\
Is this correlation between $\Sigma_{\tt SFR}$ and $\Sigma_{\star}$ causal or casual?

\begin{figure}
\centering
\epsscale{0.9}
\plotone{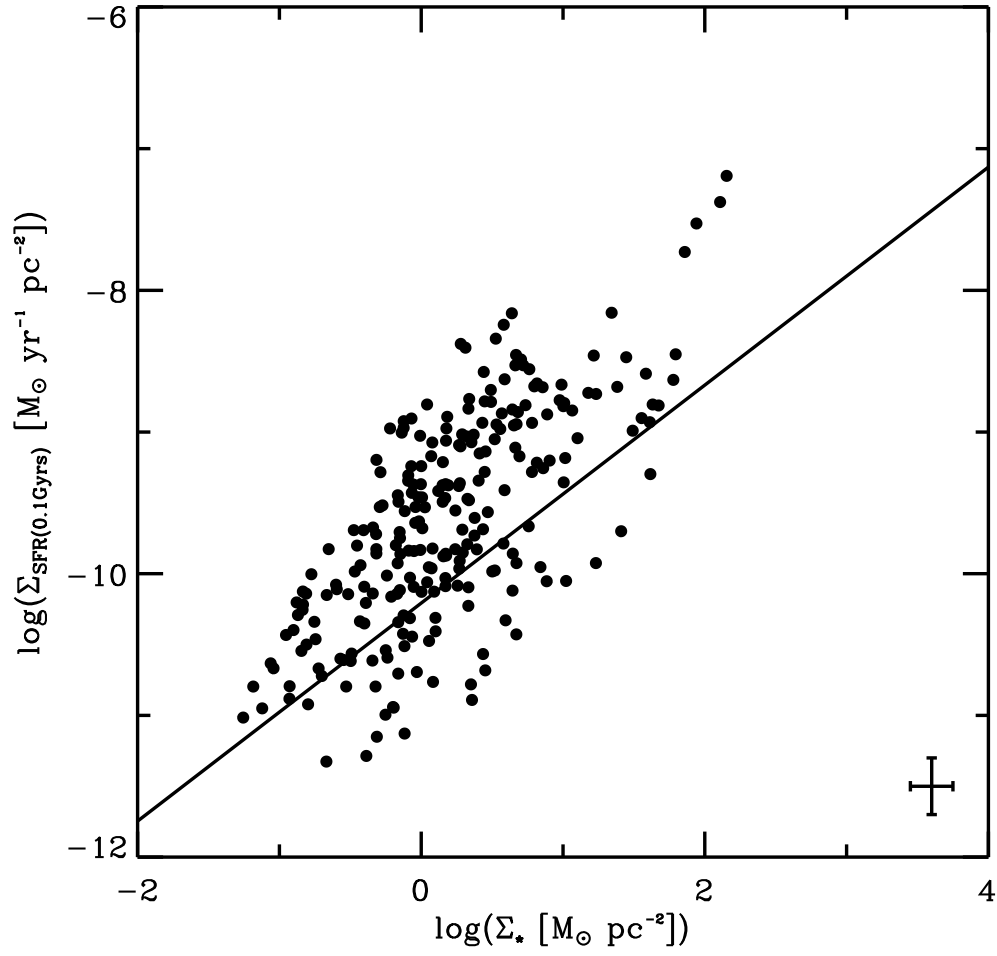}
\caption{$\Sigma$$_{\tt SFR_{0.1}}$ plotted against $\Sigma$$_{\star}$ for the galaxy sample.\
The data points are from the azimuthal averages of the relevant surface density profiles, so each galaxy is represented by many points.\
The {\it solid line} in the plot indicates the expected relationship between $\Sigma$$_{\tt SFR_{0.1}}$ and $\Sigma$$_{\star}$ for a constant SFR over a Hubble time.
\label{fig11}}

\end{figure}

In spiral galaxies, the stellar component usually dominates the inner disk in terms of both surface density and volume density, so we might expect that the stellar component plays an important role in regulating the ISM and thus the SF process (e.g.\ Wong \& Blitz; Blitz \& Rosolowsky 2006; Shi et al.\ 2011).\ However, that is not necessarily the case in dIrr galaxies, where the atomic gas usually dominates the baryonic component. 
In Figure \ref{fig11}, $\Sigma$$_{\tt SFR_{0.1}}$ is plotted against $\Sigma_{\star}$ for the whole sample.\ The data points are extracted from the azimuthal averages of the relevant surface density profiles.\
Although there is rough correspondence between  $\Sigma$$_{\tt SFR_{0.1}}$ and $\Sigma_{\star}$, the scatter is substantial.\
The Spearman rank correlation coefficient between $\Sigma$$_{\star}$ and $\Sigma$$_{\tt SFR_{0.1}}$ is $\sim$ 0.6.

As is shown in Figure \ref{fig6}, the inner regions of our galaxies tend to have higher ratios of $\Sigma_{\tt SFR_{0.1}}$ to $\Sigma_{\star}$ than outer parts, which is especially significant for relatively lower mass systems.\ In other words, $\Sigma_{\tt SFR_{0.1}}$ usually has a steeper radial dependence than $\Sigma_{\star}$.\
This is consistent with our finding that shorter wavelengths tend to have shorter scale lengths than longer wavelengths (Figure \ref{fig2}).\
Within the inner disks, our sample galaxies, on average, have almost zero radial gradients of $\Sigma$$_{\tt SFR}$/$\Sigma_{\star}$ (Figures \ref{fig8} and \ref{fig9}).\
Therefore, the correlation between $\Sigma_{\tt SFR}$ and $\Sigma_{\star}$ is only good within the inner disk.

Recently, Leroy et al.\ (2008) found that, in dIrr galaxies and outer disks of spiral galaxies, the ratio of SFR surface density to atomic gas surface density exhibits an almost linear relationship to the stellar mass surface density.\
Given the narrow range of the \hi\ surface density (5-10 M$_{\odot}$ pc$^{-2}$, Leroy et al.\ 2008) within the optical disk, the above correlation actually reflects the correlation between the SFR and the stellar mass surface density.\
We note that all of the dIrr galaxies studied by Leroy et al.\ (2008) have baryonic masses well above 10$^{8}$ M$_{\odot}$.\
For the four galaxies in common with the sample of Leroy et al.\ (i.e.\ DDO 154, DDO 63, DDO 50 and NGC 4214), 
except for DDO 50, all the other three have shallow ($|$$\Delta({\rm log}(\Sigma_{\tt SFR_{0.1}}/\Sigma_{\star}))/\Delta({\rm R})|$ $<$ 0.2, Table 4) radial slopes of log($\Sigma_{\tt SFR_{0.1}}$/$\Sigma_{\star}$) across the inner disks, which extend to more than $\sim$ 3 disk scale lengths,  in agreement with Leroy et al.\ (2008).\
Nevertheless, some galaxies (DDO 52, NGC 1569, NGC 3738,  and NGC 1156) in our sample with baryonic masses larger than 10$^{8}$ M$_{\odot}$ show obvious gradients (either positive or negative, see Table 4) of log($\Sigma_{\tt SFR_{0.1}}$/$\Sigma_{\star}$) across the whole disks.

The SFE in dIrr galaxies generally is very low (e.g.\ van Zee 2001; Lee et al.\ 2011), as is evidenced by high gas-to-stellar mass ratios (e.g.\ see Table 1).\ Therefore, the stellar disk may have always been a small perturbation on the gas-dominated baryonic disk.\
The constant average ratios of $\Sigma_{\tt SFR_{0.1}}$ to $\Sigma_{\star}$ and $\Sigma_{\tt SFR_{1}}$ to $\Sigma_{\star}$ seen in the inner disks of dIrr galaxies could imply that the SFE at a given radius in the inner disk may have been nearly the same for most of the galaxy's lifetime.\ This would naturally lead to similar radial profiles of the SFR to the accumulated stellar mass.\
In agreement with this evolutionary picture, 
Elmegreen \& Hunter (2006) considered multi-component SF triggering processes, including turbulent compression, to interpret SF in outer spiral disks and in dIrr galaxies.\ They suggested that the SF profile should be about the same as the accumulated stellar mass profile when the baryonic component of the disk is still dominated by atomic gas.\ In this scenario, the stellar mass distribution naturally follows the inefficient, patchy SF.\
On the other hand, for the outer disks, the SF process has been significantly affected by stellar feedback (e.g.\ SNe) and the external disturbance.\ As discussed in the previous section, feedback-regulated evolution or environmental effects may result in shrinking of the star-forming gas disk toward the inner part, and thus more and more centrally concentrated SF, without a remarkable correlation between SFR and stellar mass.\ We noticed that ``inner disk"  of galaxies with $M_{bary}$ $>$10$^{8}$ M$_{\odot}$ could extend over more than $\sim$ 3 disk scale lengths.

\section{Summary}
We present multi-band ({\it GALEX} FUV/NUV, {\it UBV}, H$\alpha$ and {\it Spitzer} 3.6 $\mu$m) surface brightness profiles for a representative sample of nearby dIrr galaxies.\
The median of the stellar mass and of the total baryonic mass of our sample galaxies are $\sim$ 10$^{7.2}$ M$_{\odot}$ and 10$^{8}$ M$_{\odot}$, respectively.\
By modeling the multi-band SEDs with a relatively complete library of SFHs,  we constrained: the radial variations of the SFR averaged over the recent 0.1 Gyr, \sfr, the SFR averaged over the recent 1 Gyr, \sfg, and the stellar mass surface density profiles. To summarize the results: 
\begin{enumerate}
\item
We show that, with a relatively complete library of model SFHs, the averaged SFR during the past 1 Gyr can be well constrained by modeling integrated multi-band data.\ The exponential-plus-burst SFH library commonly used for spirals tends to systematically overestimate recent SF of dIrr galaxies, which are characterized as having much more complex SFHs than the larger spiral galaxies.    
\item
For $\sim$ 80\% (27) of the dIrr galaxies studied in this work, and all (15) with baryonic mass less than $10^8$ M$_{\odot}$, surface brightness profiles determined from shorter-wavelength passbands have shorter exponential disk scale lengths than those based on passbands at longer wavelengths at least in the outer disks.\ This trend suggests that the star-forming stellar disk may be shrinking for most dIrr galaxies.\ This produces an ``outside-in" disk formation scenario for dIrr galaxies.
\item
The recent SF activity of $\sim$ 38\% of our galaxies has been enhanced by a factor of at least 2 compared to the lifetime averaged SFR.\ Even though no obvious correlation between global SFHs and total baryonic mass was found in our sample, all of the galaxies with significantly declining global SF activity have baryonic masses smaller than 10$^{8}$ M$_{\odot}$.\ This may be ascribed to the susceptibility of low mass dwarf galaxies to significant influence by both internal and external processes.
\item
Consistent with the ``outside-in" scenario, 
80\% of our sample galaxies have negative slopes of radial variations of log($\Sigma_{\tt SFR_{0.1}}$/$\Sigma_{\star}$) and of log($\Sigma_{\tt SFR_{1}}$/$\Sigma_{\star}$) at least in the outer disks (beyond the breaks of stellar mass surface density profiles).\
In the inner disks, especially for the galaxies with $M_{bary}$ $>$10$^{8}$ M$_{\odot}$, the slopes are much shallower (consistent with zero slopes on average) than that in the outer disks.\
Both internal (stellar feedback, gas pressure support) and external (ram pressure stripping of gas and tidal disturbance) processes are discussed as the possible explanation for the ``outside-in" shrinking of the star-forming disk. 
 \item
The radial distribution of the stellar mass density of dIrr galaxies is well described as a single or, for the majority, a piece-wise exponential profile.\
In particular, the majority ($\sim$ 80\%) of our sample galaxies exhibit steeper outer (down-bending) stellar mass profiles.\
Since the outer disks on average have much steeper radial declining $\Sigma_{\tt SFR}$/$\Sigma_{\star}$ than the inner regions, 
we interpret the down-bending stellar mass surface density profiles as a natural result of a gradual shrinking of the star-forming disks.\ A few galaxies exhibit much steeper stellar mass profiles in the inner disk than the outer disk, which may be caused by an intense starburst in their inner regions.\ 
Spiral galaxies with double exponential surface brightness profiles and inside-out star formation may require a different explanation for the steep outer profiles.\ In spirals, the outer disks are sometimes younger than the inner disks, while in dwarfs, the outer disks are usually older than the inner disks.\ Perhaps spirals are still accreting gas in their outer parts, whereas the dwarfs have lost the gas in their outer parts and have other dynamical effects not present in larger galaxies.
\item
The previously found correlation between surface density of the stellar mass and recent SFR is only remarkable in the inner disks, which could extend over more than 3 disk scale lengths for the relatively massive ($M_{bary}$ $\gtrsim$ 10$^{8}$ M$_{\odot}$) dIrr galaxies.\
The correlation can be explained by inefficient, constant SF at a given radius during most of the galaxy's lifetime.\ For the outer disks of most (80\% for the present sample) dIrr galaxies, on the other hand, the correlation between SFR and stellar mass surface densities is not strong, due to effects of internal feedback, turbulent gas pressure support, and external disturbances.   

\end{enumerate}

\begin{acknowledgements}
This work was funded in part by the National Science Foundation through
grants AST-0707563 and AST-0707426 to DAH and BGE.\
HZ was partly supported by NSF of China through grants \#10425313, \#10833006 and \#10621303 to YG.\
Support for AS was provided by the Deutsche Forschungsgemeinschaft (DFG) Priority Program 1177.\
We are grateful to the anonymous referee for the valuable comments, which resulted in significant improvement of the paper.\
We thank Kimberly A. Hermann for a thorough reading of the paper and the useful suggestions which have improved the paper.\
We are grateful to the LVL and the SINGS team for kindly making the {\it Spitzer} imagery data available to the public.
\end{acknowledgements}

{\it Facilities: }\facility{Lowell Observatory: Hall 42$''$ and the Perkins 72$''$}, \facility{KPNO: 2.1m}, \facility{CTIO: 0.9m}, \facility{GALEX}, \facility{Spitzer}

\clearpage

\begin{appendix}
\section{Details concerning SED modeling}
\subsection{Recovery of physical parameters}
Here we explore the reliability of multi-band SED modeling in recovering the rough SFH of dIrr galaxies.\ 
In these tests, we first logarithmically divided the Hubble time (13.7 Gyr) into 14 independent age bins, and took the SF as constant within each age bin.\ Then a series of 3000 mock SFHs were generated by randomly changing the SFR for each independent age bin.\ The internal extinction was fixed at $A_{V}=0.1$ when generating the SEDs related to each mock SFH.\ 
We then modeled the SEDs ({\it GALEX} FUV/NUV, {\it U, B, V}, 3.6 $\mu$m) related to the mock SFHs with our library of 6-component superposition (6-C) SFHs used in this work.\
The calibration uncertainties of the real data were taken as the photometric uncertainties of the SED related to each mock SFH.

\begin{figure}
\centering
\epsscale{0.9}
\plotone{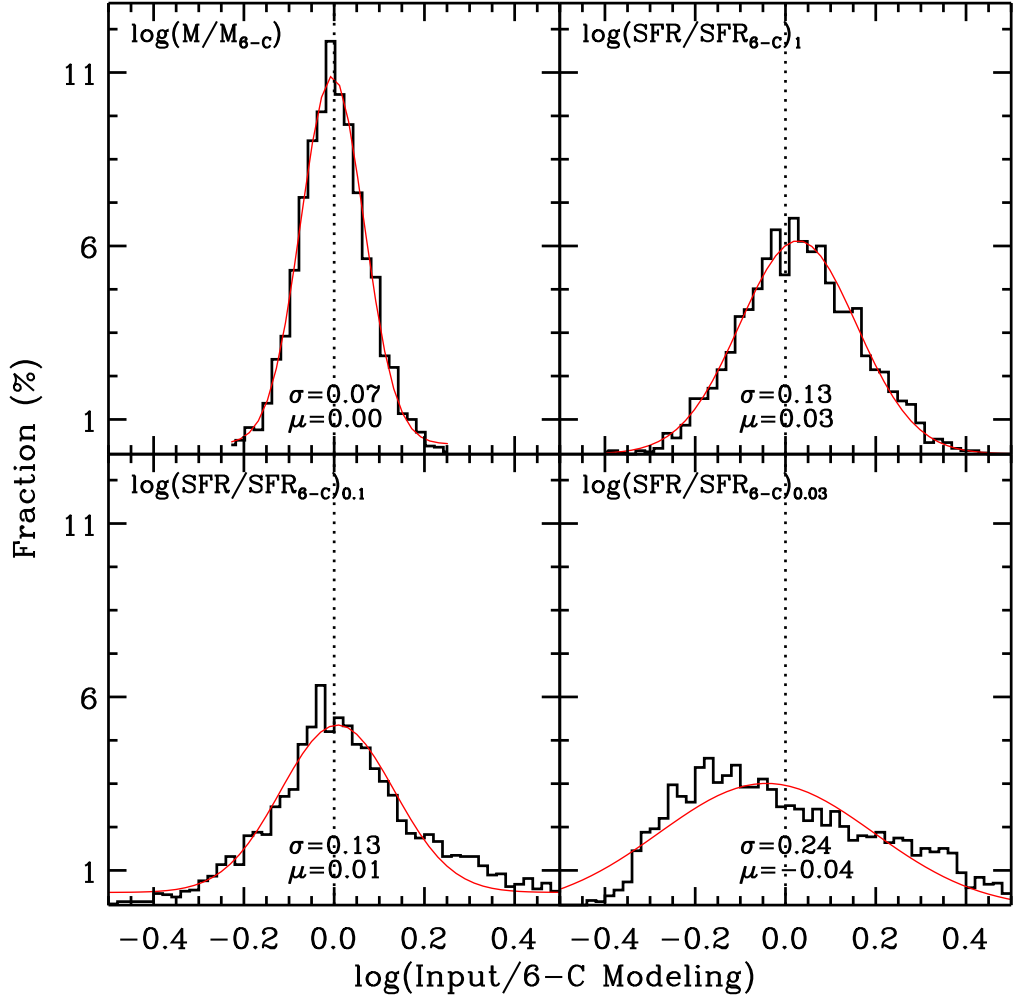}
\caption{
Histograms of the ratios of the parameters (M$_{\star}$, SFR$_{1}$, SFR$_{0.1}$ and SFR$_{0.03}$) from the input mock SFHs to the best estimates derived from the 6-component (6-C) modeling.\
The {\it abscissa} in each panel is the logarithm of the relevant ratio (indicated in each panel) of the input and of the 6-C modeling results.\ The {\it ordinate} is the per cent of the data points in each bin.\
The {\it dotted} line in each panel marks the relationship of equality.\
The {\it red} curve in each panel is the gaussian fit to the relevant histogram. $\sigma$ and $\mu$ indicate the best fitting parameters defining the gaussian curve. 
\label{figa1}} 

\end{figure}

\begin{figure}
\centering
\epsscale{0.9}
\plotone{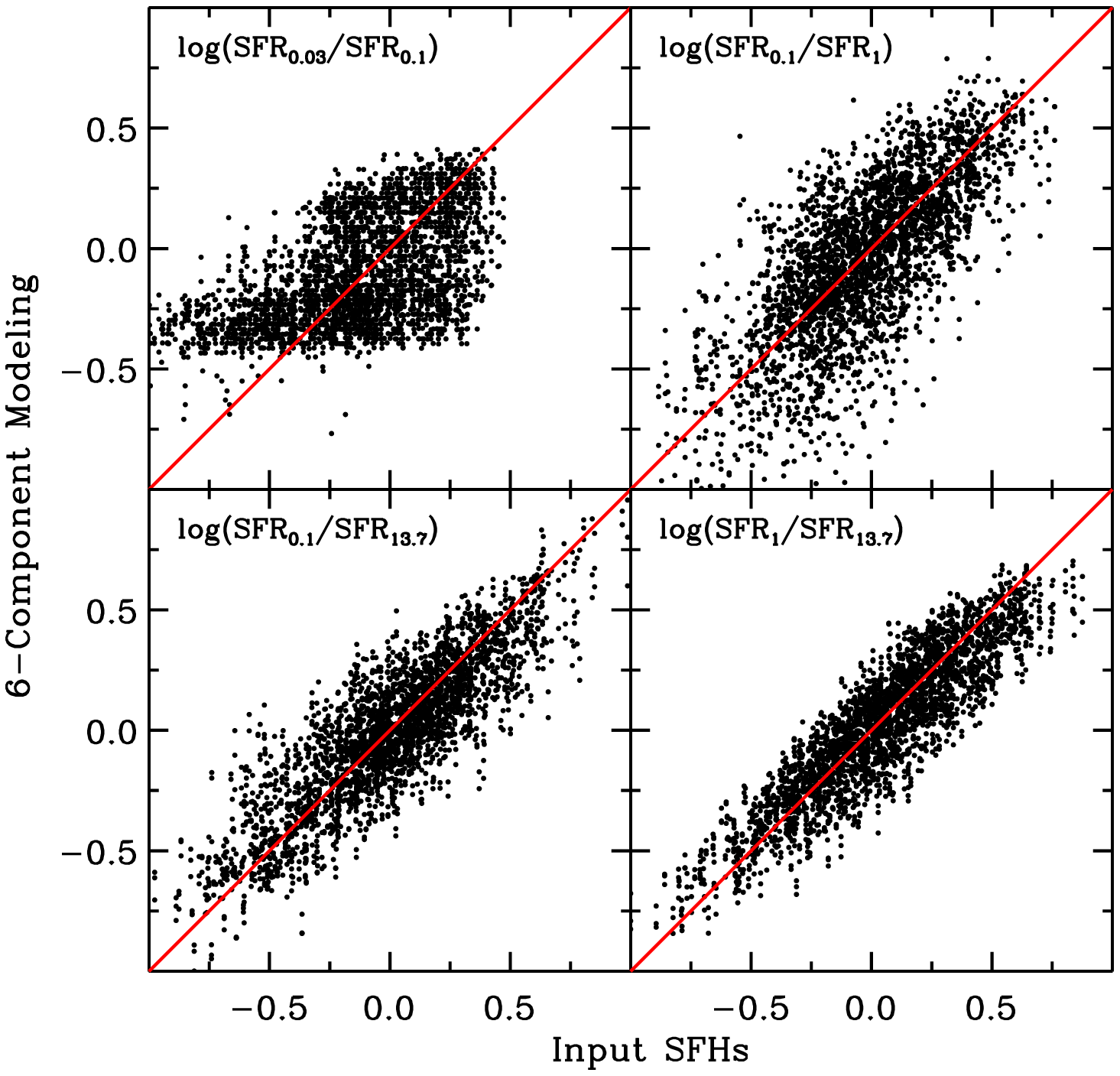}
\caption{
Comparison between the ratios of the SFR averaged over different timescales (indicated in each panel) from the input SFHs (the {\it abscissa}) and the relevant quantities from the 6-C modeling (the {\it ordinate}).\
The {\it red solid} line in each panel marks the relationship of equality.  
\label{figa2}} 

\end{figure}

Figure \ref{figa1} presents a comparison of the modeling results, including M$_{\star}$, SFR$_{1}$, SFR$_{0.1}$ and SFR$_{0.03}$, and relevant integrated quantities of the mock input SFHs.\ 
We fit each histogram in Figure \ref{figa1} with a gaussian curve, which is overplotted as a {\it red solid} line in each panel.\ 
The standard deviation ($\sigma$) and mean value ($\mu$) of the gaussian curve is also indicated.\ 
As can be seen, except for SFR$_{0.03}$, all the parameters are well recovered in our SED modeling, with acceptable uncertainties related to the varying SFHs.\  Figure \ref{figa2} further shows ratios of SFR averaged over different timescales.\ 
Again, except for SFR$_{0.03}$, the modeling with the 6-C library recovers the SFHs with negligible, if any, bias. 

\subsection{Uncertainty in converting the {\it GALEX} FUV luminosity to SFR}
FUV has been extensively used to estimate the recent ($\sim$ 0.1 Gyr) SFR in galaxies (e.g.\ Kennicutt 1998; Salim et al.\ 2007; Hunter et al.\ 2010).\ The most commonly used formula (e.g.\ Kennicutt 1998) assumes constant SF during the past 0.1 Gyr, which is comparable to a galactic dynamical timescale.\ 
The assumption of more or less constant recent SF is generally true for a high luminosity galaxy as a whole, as is evidenced by the consistency between the H$\alpha$-based SFR and the FUV-based SFR (e.g.\ Sullivan et al.\ 2004).\ 
However, this is not necessarily the case in low-luminosity dIrr galaxies that may have more fluctuating SFHs. 

\begin{figure}
\centering
\epsscale{0.9}
\plotone{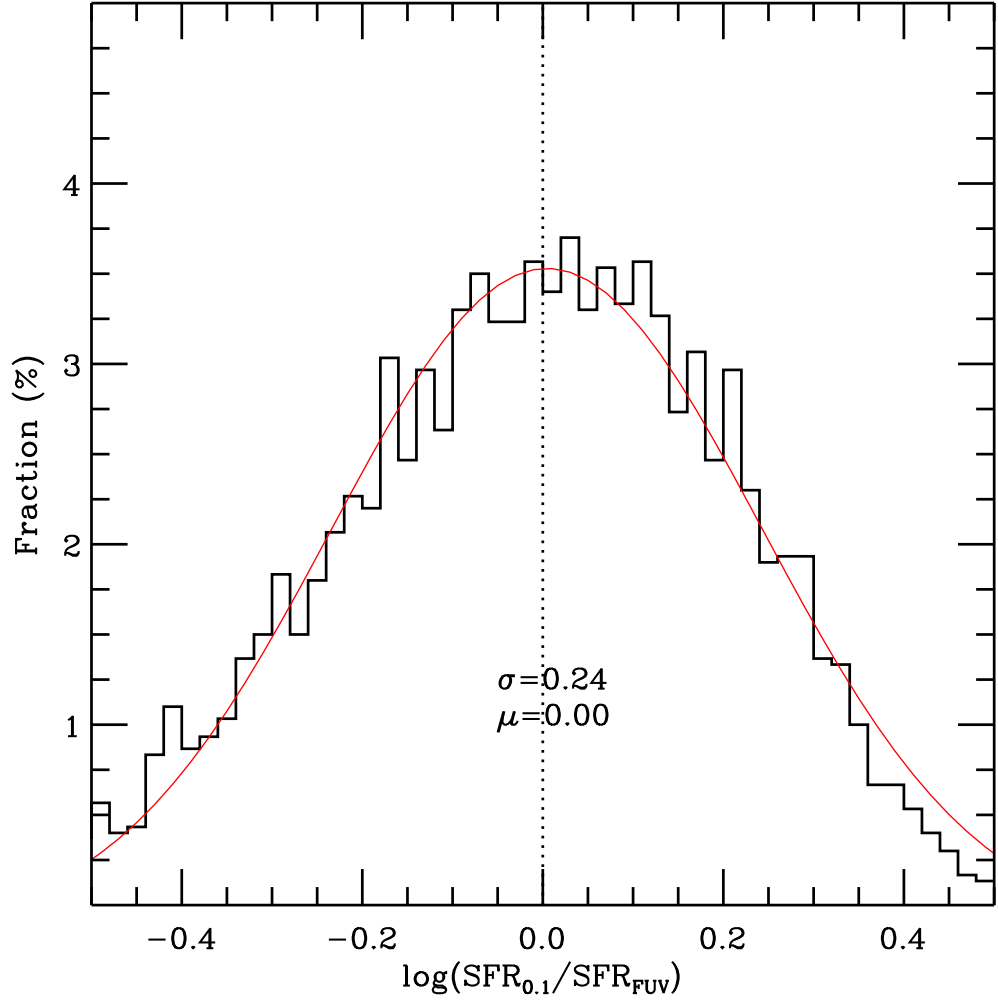}
\caption{
Histogram of the ratio of SFR$_{0.1}$ from the input mock SFHs to SFR$_{\tt FUV}$ estimated with the modified formula of Hunter et al.\ (2010).\
The {\it ordinate} is the per cent of the data points in each bin.\
The {\it red} curve is a single gaussian fitting to the histogram. $\sigma$ and $\mu$ indicate the best fitting parameters defining the gaussian curve.\
The {\it dotted} line marks the relationship of equality. 
\label{figa3}}

\end{figure}

Figure \ref{figa3} shows how well the {\it GALEX} FUV passband can recover the SFR averaged over the past 0.1 Gyr
from our mock SFHs.\ 
The formula for transforming FUV to SFR$_{\tt FUV}$ used here was from Hunter et al.\ (2010), where the Kennicutt (1998) formula was modified to account for the low metallicity of dwarf galaxies.\ Here we further divided their proportionality constant by 1.59 to account for the difference between the initial mass function (IMF) assumed here (Chabrier 2003) and the Salpeter (1955) IMF that they used.\ 
figure \ref{figa4} justifies the practice of using FUV as an unbiased estimator of the recent SFR, with an uncertainty less than a factor of 2 incurred by different SFHs.\
One should keep in mind that here we assumed low dust extinction, which is reasonable for dwarf galaxies in general.  

\subsection{Bias introduced by an incomplete SFH library}
The incompleteness of the SFH library used in the SED modeling could significantly affect or bias the estimation of physical parameters (e.g.\ Kauffmann et al.\ 2003).\ 
To demonstrate this, we create another SFH library by assuming the real SFH can be approximated as an underlying component that varies exponentially with time plus a single random burst of finite length.\ Here we denote this SFH library as `E+B'.  E+B is now the most commonly used SFH library in the literature.\
The exponential component is described with two parameters: the SF timescale $\tau$ and the age.\ The burst component is described with three parameters:  the age, strength, and length of the burst.\ 
The Charlot \& Fall (2000) extinction recipe was adopted here.\ 
All the parameters, including the extinction, are allowed to vary uniformly within physically reasonable ranges when creating the library.\ 
We model the same SEDs as related to the mock SFHs generated above.\  
We obtained the most probable physical parameters, including M$_{\star}$, SFR$_{1}$, SFR$_{0.1}$ and SFR$_{0.03}$, with the same method as described in \S~\ref{subsec: meth}. 

\begin{figure}
\centering
\epsscale{0.9}
\plotone{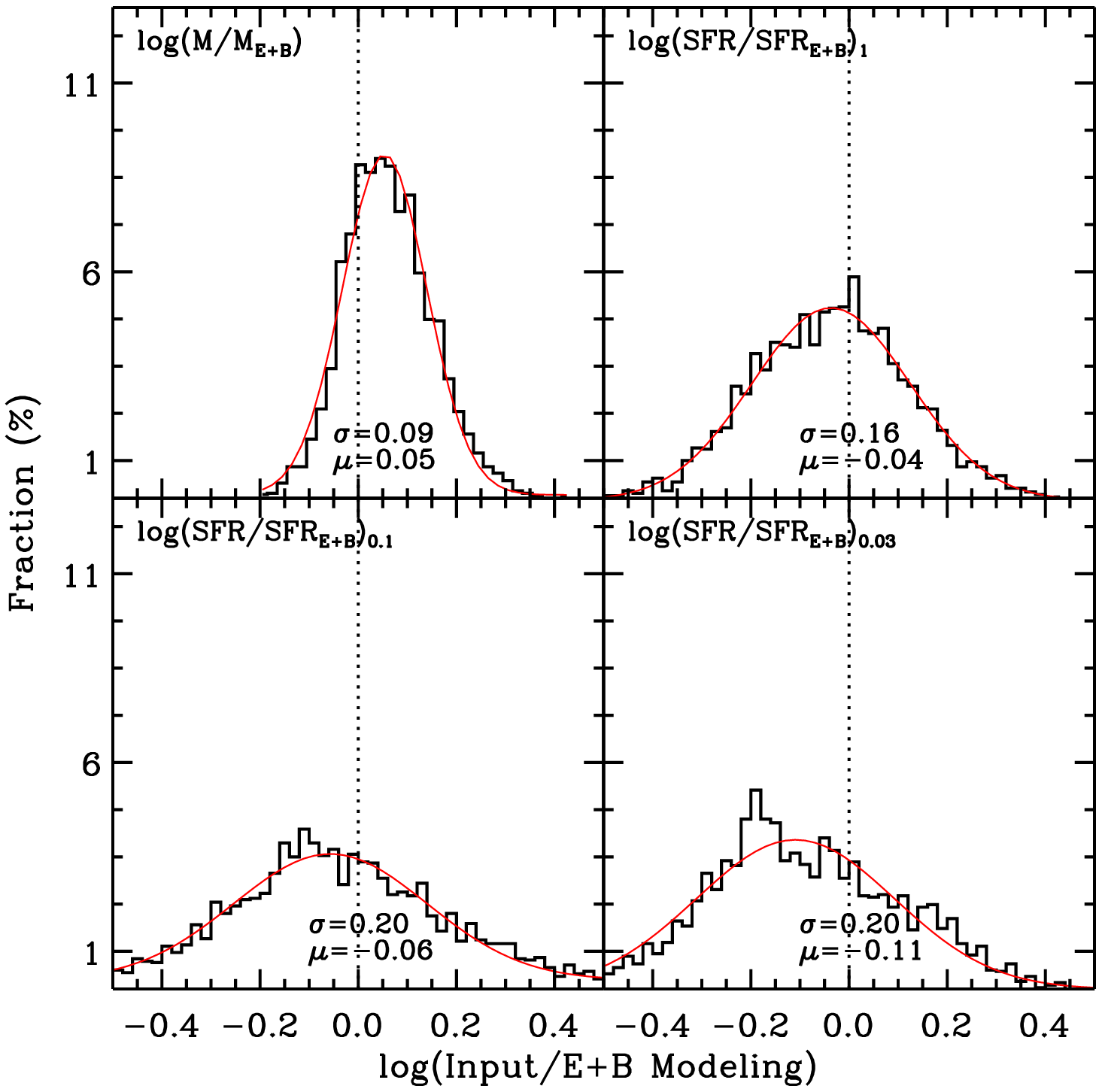}
\caption{
Histograms of the ratios of the parameters (M$_{\star}$, SFR$_{1}$, SFR$_{0.1}$ and SFR$_{0.03}$) from the input mock SFHs to the best estimates from the exponential plus burst (E+B) modeling.\
The {\it abscissa} in each panel is the logarithm of the relevant ratio (indicated in each panel) of the input and of the E+B modeling results.\ The {\it ordinate} is the per cent of the data points in each bin.\
The {\it red} curve in each panel is the gaussian fitting to the relevant histogram. $\sigma$ and $\mu$ indicate the best fitting parameters defining the gaussian curve.\
The {\it dotted} line in each panel marks the relationship of equality. 
\label{figa4}}

\end{figure}

\begin{figure}
\centering
\epsscale{0.9}
\plotone{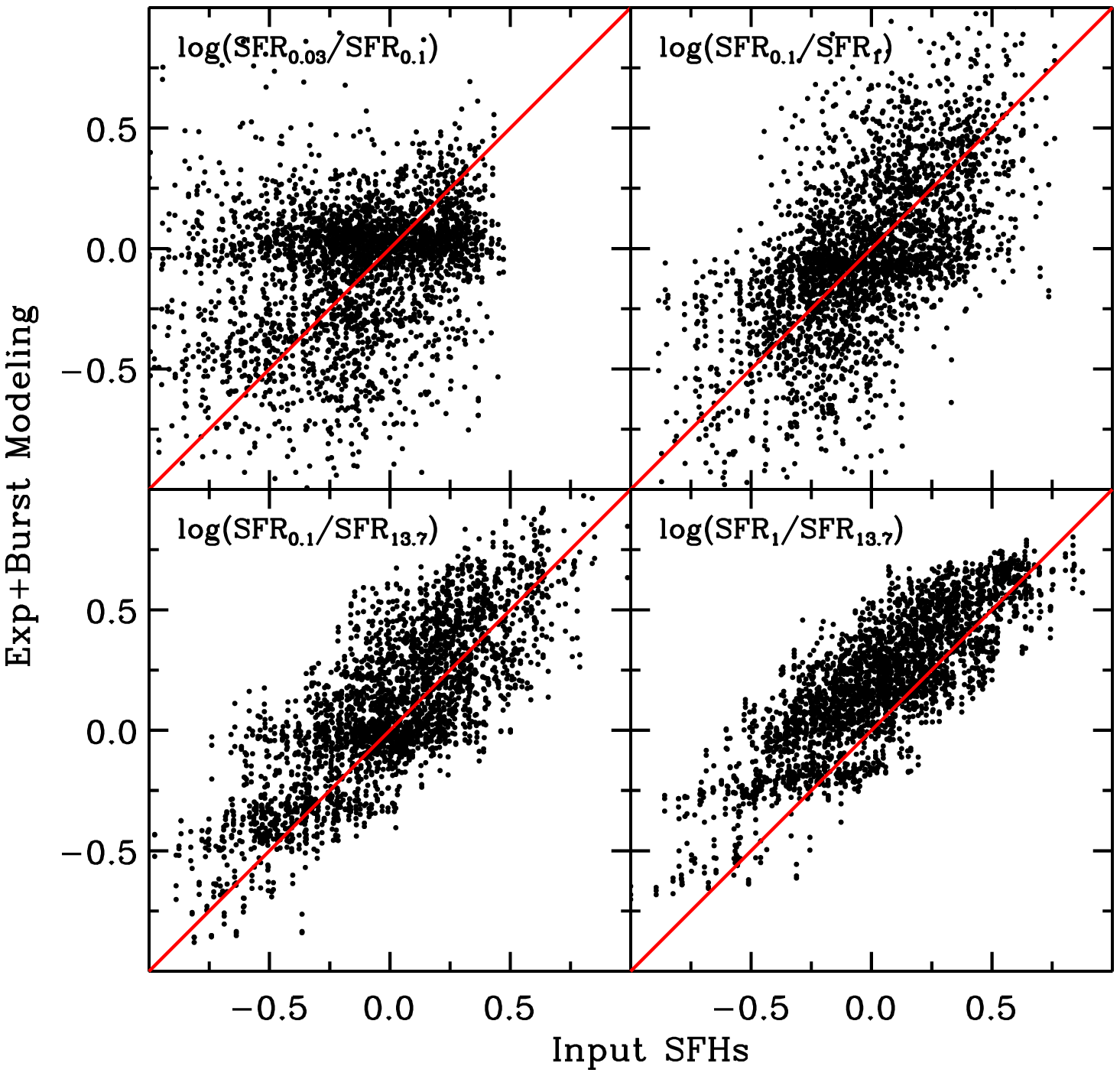}
\caption{
Comparison between the ratios of the SFR averaged over different timescales (indicated in each panel) from the input SFHs (the {\it abscissa}) and the relevant quantities from the E+B modeling (the {\it ordinate}).\
The {\it red solid} line in each panel marks the relationship of equality.  
\label{figa5}}

\end{figure}

Figures \ref{figa4} and \ref{figa5} show the modeling results of using the E+B library.\ 
Compared to the modeling results from our standard 6-C superposition method, 
the estimate of the relevant parameters exhibits a larger uncertainty and a considerable bias.\ For instance, 
the stellar mass M$_{\star}$ has a tendency to be slightly underestimated, and the recent SFR, i.e.\ SFR$_{0.1}$ and \sfg,\ tend to be overestimated.\  
The problem of using the E+B library in modeling can be more clearly seen in Figure \ref{figa5}.\ 
The E+B library tends to overestimate the SFR averaged over recent times, e.g.\ SFR$_{0.1}$, SFR$_{1}$.\ 
As is expected for the more or less continuous nature of the SFHs in the E+B library, there is a very strong bias towards almost constant recent SF.   \
The above bias could be ascribed to the fact that, while the SF averaged over a timescale of $\sim$ Gyr may be relevant to the overall evolution histories of the galaxies, the SF over shorter timescales (i.e.\ 0.1 Gyr) is not necessarily as important. 

\end{appendix}

                                                                                  
%
                                                                                                                   
\begin{deluxetable}{lccccccccr} 
\tabletypesize{\scriptsize}                                                
\tablenum{1}                                                               
\tablecolumns{10}                                                           
\tablewidth{0pt}                                                           
\tablecaption{Galaxy Sample}
\tablehead{                                                                                                                
\colhead{Galaxy}           
& \colhead{Other Names}  
& \colhead{{\it D}}  
& \colhead{{\it b/a}}  
& \colhead{{\it E(B-V)$_{f}$}}
& \colhead{{\it M$_{B}$}}
& \colhead{log($\Sigma_{\tt SFR(H\alpha)}$)} 
& \colhead{{\it M$_{bary}$}}  
& \colhead{{\it M$_{gas}$/M$_{\star}$}}  
& \colhead{$\Theta$} \\
\colhead{}
& \colhead{}
& \colhead{(Mpc)}     
& \colhead{}
& \colhead{}
& \colhead{}
& \colhead{($M_{\odot}$ yr$^{-1}$ kpc$^{-2}$)}  
& \colhead{($M_{\odot}$)}  
& \colhead{} 
& \colhead{} \\                                                                 
\colhead{(1)}                                                                             
& \colhead{(2)}                                                                             
& \colhead{(3)}                                                                             
& \colhead{(4)}                                                                             
& \colhead{(5)}                                                                             
& \colhead{(6)}                                                                             
& \colhead{(7)}   
& \colhead{(8)}
& \colhead{(9)}
& \colhead{(10)}                                                                                                                                                                 
}                                                                          
\startdata                                                                                                               

LGS 3 \dotfill & Pisces Dwarf & 0.7 & 0.51 & 0.041 & -9.11 & \nodata& 5.68 & 0.76 & 1.7 \\
DDO 210 \dotfill & Aquarius Dwarf & 0.9 & 0.48 & 0.051 & -10.38 & \nodata & 6.60 & 4.90 & -0.1 \\
DDO 69 \dotfill & UGC 5364, Leo A & 0.8 & 0.56 & 0.021 & -11.38 & -3.28 & 7.16 &13.19 & 0.2 \\
DDO 155 \dotfill  & UGC 8091, GR 8 & 2.2 & 0.71 & 0.026 & -12.25 & -1.50 & 7.22 & 4.54 & -1.2 \\
DDO 216 \dotfill & UGC 12613, Pegasus Dwarf & 1.1 & 0.45 & 0.066 & -13.06 & -4.15 & 7.22 & 0.10 & 1.2 \\
M81dwA \dotfill & KDG 052 & 3.6 & 0.73 & 0.021 & -11.46 & \nodata & 7.29 & 13.17 & 0.7 \\
Mrk 178 \dotfill & UGC 6541 & 3.9 & 0.46 & 0.018 & -13.76 & -1.53 & 7.39 & 1.21 & -0.7 \\
DDO 187 \dotfill & UGC 9128 & 2.2 & 0.80 & 0.024 & -12.38 & -2.64 & 7.43 & 6.78 & -1.3 \\
UGC 8508 \dotfill  & IZw 60 & 2.6 & 0.54 & 0.015 & -13.24 & -2.12 & 7.63 & 4.54 & -1.0 \\
NGC 4163 \dotfill & NGC 4167, UGC 7199 & 2.9 & 0.64 & 0.020 & -13.95 & -2.43 & 7.69 & 0.81 & 0.1 \\
CVnIdwA \dotfill  & UGCA 292 & 3.6 & 0.78 & 0.016 & -12.16 & -2.64 & 7.83 & 15.66 & -0.4 \\
DDO 70 \dotfill  & UGC 5373, Sextans B & 1.3 & 0.59 & 0.032 & -13.74 & -2.86 & 7.86 & 2.68 & -0.7 \\
IC 1613 \dotfill & UGC 668, DDO 8 & 0.7 & 0.81 & 0.025 & -14.17 & -2.64 & 7.87 & 1.57 & 0.9 \\
DDO 101 \dotfill & UGC 6900 & 6.4 & 0.69 & 0.022 & -14.40 & -2.99 & 7.89 & 0.19 & \nodata \\
WLM \dotfill & UGCA 444, DDO 221 & 1.0 & 0.44 & 0.037 & -13.98 & -2.85 & 7.98 & 4.88 & 0.3 \\
DDO75 \dotfill  & UGCA 205, Sextans A & 1.3 & 0.85 & 0.044 & -13.72 & -1.40 & 8.04 & 13.05 & -0.6 \\
VIIZW 403 \dotfill & UGC 6456, VV 574 & 4.4 & 0.85 & 0.036 & -13.99 & -1.82 & 8.05 & 5.37 & -0.3 \\
Haro 29 \dotfill & UGCA 281, Mrk 209, IZw 36 & 5.9 & 0.58 & 0.015 & -14.39 & -0.82 & 8.06 & 7.04 & -1.0 \\ 
DDO 133 \dotfill & UGC 7698 & 3.5 & 0.69 & 0.016 & -14.36 & -2.93 & 8.30 & 5.55 & -1.1 \\
DDO 126 \dotfill & UGC 7559 & 4.9 & 0.47 & 0.014 & -14.56 & -2.45 & 8.31 & 11.60 & 0.1 \\
DDO 165 \dotfill & UGC 8201, IIZw 499 & 4.6 & 0.54 & 0.024 & -15.38 & -3.52 & 8.35 & 5.56 & 0.0 \\
DDO 63 \dotfill  & UGC 5139, Holmberg I & 3.9 & 1.00 & 0.048 & -14.58 & -3.44 & 8.40 & 6.45 & 1.5 \\
NGC 6822 \dotfill & IC 4895, DDO 209 & 0.5 & 0.79 & 0.240 & -14.76 & -1.96 & 8.40 & 2.28 & 0.6 \\
DDO 53 \dotfill & UGC 4459, VIIZw 238 & 3.6 & 0.51 & 0.037 & -13.43 & -2.50 & 8.41 & 25.32 & 0.7 \\
DDO 154 \dotfill & UGC 8024, NGC 4789A & 3.7 & 0.50 & 0.009 & -13.88 & -2.60 & 8.50 & 36.97 & -0.9 \\
DDO 87 \dotfill & UGC 5918, KDG 072, VIIZw 347 & 7.7 & 0.58 & 0.011 & -14.52 & -3.16 & 8.53 & 9.37 & -1.5 \\
DDO 52 \dotfill & UGC 4426 & 10.3 & 0.67 & 0.037 & -15.05 & -3.27 & 8.63 & 7.04 & -1.5 \\
DDO 168 \dotfill & UGC 8320 & 4.3 & 0.63 & 0.015 & -15.35 & -2.33 & 8.71 & 7.77 & 0.0 \\
NGC 1569 \dotfill & UGC 3056, Arp 210, VIIZw 16 & 3.4 & 0.55 & 0.604\tablenotemark{a} & -17.35 & 0.11 & 8.78 & 0.66 & -0.4 \\
NGC 3738 \dotfill & UGC 6565, Arp 234 & 4.9 & 1.00 & 0.010 & -16.70 & -1.72 & 8.83 & 0.45 & -1.0 \\
DDO 50 \dotfill  & UGC 4305, Holmberg II & 3.4 & 0.72 & 0.032 & -16.39 & -1.83 & 9.03 & 9.11 & 0.6 \\
NGC 2366 \dotfill  & UGC 3851, DDO 42 & 3.4 & 0.42 & 0.036 & -16.49 & -1.73 & 9.04 & 14.81 & 1.0 \\
NGC 4214 \dotfill  & UGC 7278 & 3.0 & 0.91 & 0.022 & -17.26 & -1.10 & 9.11 & 1.69 & -0.7 \\
NGC 1156 \dotfill & UGC 2455, VV 531 & 7.8 & 0.86 & 0.224 & -18.29 & -0.87 & 9.42 & 1.19 & -1.7 \\

\enddata

\tablecomments{
(1) Galaxy names adopted in this work.\ 
(2) The other commonly used names in the literature.\ 
(3) Distance from Hunter et al.\ , in prep and references therein.\ 
(4) Minor-to-major axis ratio measured on the {\it V}-band images.\
(5) The foreground reddening from Schlegel et al.\ (1998).\
(6) {\it B}-band absolute magnitude.\
(7) Integrated star formation rate (SFR) normalized to the area of the galaxy within one {\it V}-band disk scale length.\ The SFR is derived from H$\alpha$.\
(8) The baryonic (stellar plus atomic gas) mass.\ The stellar mass is derived from our SED modeling in this work.\ The atomic gas (1.34$\times$$M_{H_{\tt I}}$) mass is collected from single dish observations in the literature (see Hunter \& Elmegreen (2004) for the references).\
(9) The atomic gas-to-stellar mass ratio.\
(10) Tidal index $\Theta$ from Karachentsev et al.\ (2004).\ $\Theta$ quantifies the collective gravitational disturbance from neighboring galaxies.\ The larger the value of $\Theta$ is, the stronger gravitational disturbance exerted by the neighboring galaxies.\ Galaxies with zero or negative values of $\Theta$ could be considered as isolated objects.}                                                                   
\tablenotetext{a}{The average of the {\it E(B-V)} derived by Burstein \& Heiles (1982) (0.508) and by Schlegel et al.\ (1998) (0.700).\ See Johnson et al.\  in prep.}           
\end{deluxetable}                                                          

\clearpage

\clearpage
                                                                           
%
                                                                           
                                   
\begin{deluxetable}{lccccccccccccr} 
\tabletypesize{\scriptsize}         
\tablenum{2}                                                               
\tablecolumns{14}                                                           
\tablewidth{0pt}                                                           
\tablecaption{Multi-band Disk Scale Length}
\tablehead{                                                                                                                
\colhead{Galaxy}
& \multicolumn{6}{c}{Inner} 
& \colhead{}
& \multicolumn{6}{c}{Outer} \\
\cline{2-7}
\cline{9-14}
\colhead{}
& \colhead{$R_{D}^{\tt FUV}$}
& \colhead{$R_{D}^{\tt NUV}$}
& \colhead{$R_{D}^{\tt {\it U}}$}  
& \colhead{$R_{D}^{\tt {\it B}}$} 
& \colhead{$R_{D}^{\tt {\it V}}$} 
& \colhead{$R_{D}^{\tt 3.6 \mu m}$} 
& \colhead{}
& \colhead{$R_{D}^{\tt FUV}$}
& \colhead{$R_{D}^{\tt NUV}$}
& \colhead{$R_{D}^{\tt {\it U}}$}  
& \colhead{$R_{D}^{\tt {\it B}}$} 
& \colhead{$R_{D}^{\tt {\it V}}$} 
& \colhead{$R_{D}^{\tt 3.6 \mu m}$}  \\
\colhead{}
& \colhead{(kpc)}                                                                                                                                                                                                                                 
& \colhead{(kpc)}                                                                             
& \colhead{(kpc)} 
& \colhead{(kpc)}                                                                                                                                                                                                                                 
& \colhead{(kpc)}                                                                             
& \colhead{(kpc)} 
& \colhead{}
& \colhead{(kpc)}                                                                                                                                                                                                                                 
& \colhead{(kpc)}                                                                             
& \colhead{(kpc)} 
& \colhead{(kpc)}                                                                                                                                                                                                                                 
& \colhead{(kpc)}                                                                             
& \colhead{(kpc)}  \\                                                                        
\colhead{(1)}                                                                             
& \colhead{(2)}                                                                             
& \colhead{(3)}                                                                             
& \colhead{(4)} 
& \colhead{(5)}                                                                             
& \colhead{(6)}                                                                             
& \colhead{(7)} 
& \colhead{}         
& \colhead{(8)}                                                                             
& \colhead{(9)}                                                                             
& \colhead{(10)}      
& \colhead{(11)}                                                                             
& \colhead{(12)}                                                                             
& \colhead{(13)}                                                                                                                                                             
}                                                                          
\startdata                                                                                                               

LGS 3          \dotfill & 0.07$\pm$0.01 & 0.15$\pm$0.00 & 0.20$\pm$0.00 & 0.23$\pm$0.00 & 0.35$\pm$0.00 & 0.31$\pm$0.00 & & \nodata & \nodata & \nodata & \nodata & \nodata & \nodata  \\
DDO 210    \dotfill &  0.12$\pm$0.01 & 0.14$\pm$0.00 &  0.13$\pm$0.00 & 0.16$\pm$0.00 &  0.17$\pm$0.00 & 0.21$\pm$0.00 & & \nodata & \nodata & \nodata & \nodata & \nodata & \nodata  \\
DDO 69      \dotfill & 0.72$\pm$0.00 &  0.59$\pm$0.00 &   0.65$\pm$0.00 &   0.69$\pm$0.00 &   0.83$\pm$0.00 &  3.02$\pm$0.05 & & 0.17$\pm$0.00 &  0.17$\pm$0.00 &  0.17$\pm$0.00 &   0.17$\pm$0.00 &   0.18$\pm$0.00 &   0.25$\pm$0.00 \\
DDO 155    \dotfill  & 0.07$\pm$0.00 &  0.08$\pm$0.00 &  0.09$\pm$0.00 &  0.11$\pm$0.00 &   0.12$\pm$0.00 &   0.16$\pm$0.00 & & \nodata & \nodata & \nodata \\
DDO 216    \dotfill & 0.16$\pm$0.00 &  0.24$\pm$0.00 &   0.42$\pm$0.00 &  0.48$\pm$0.00 &   0.56$\pm$0.00 &   0.64$\pm$0.00 & & 0.48$\pm$1.90 & 0.32$\pm$0.01 &  0.61$\pm$0.01 &   0.50$\pm$0.00 &   0.50$\pm$0.00 &  0.50$\pm$0.00  \\
M81dwA     \dotfill & 0.20$\pm$0.00 & 0.25$\pm$0.00 &  0.24$\pm$0.00 &   0.25$\pm$0.00 &  0.27$\pm$0.00 &   0.30$\pm$0.00  & & \nodata & \nodata & \nodata \\\
Mrk 178      \dotfill & 0.01$\pm$0.00 &  0.01$\pm$0.00 &   0.18$\pm$0.00 &   0.19$\pm$0.00 &  0.21$\pm$0.00 &   0.29$\pm$0.00 & & 0.36$\pm$0.00 &  0.36$\pm$0.00 &  \nodata &   \nodata &  \nodata &   0.83 \\
DDO 187    \dotfill & 0.34$\pm$0.00 & 0.32$\pm$0.00 &   0.32$\pm$0.00 &    0.34$\pm$0.00 &   0.38$\pm$0.00 &   0.56$\pm$0.00  & & 0.11$\pm$0.00 &   0.12$\pm$0.00 &  0.14$\pm$0.00 &  0.15$\pm$0.00 &  0.17$\pm$0.00 &  0.15$\pm$0.00 \\
UGC 8508  \dotfill  & \nodata & \nodata & 0.21$\pm$0.00 & 0.24$\pm$0.00  & 0.26$\pm$0.00 & 0.30$\pm$0.00  & & \nodata & \nodata & \nodata & \nodata & \nodata & \nodata  \\
NGC 4163  \dotfill & 0.11$\pm$0.00 & 0.13$\pm$0.00 &  0.19$\pm$0.00 &   0.22$\pm$0.00 &   0.24$\pm$0.00 &  0.27$\pm$0.00 & &\nodata & \nodata & \nodata & \nodata & \nodata & \nodata  \\
CVnIdwA   \dotfill  & 0.12$\pm$0.00 & 0.14$\pm$0.00 &   0.20$\pm$0.00 &  0.28$\pm$0.00 &  0.44$\pm$0.01 &   0.37$\pm$0.00 & & \nodata & \nodata & \nodata & \nodata & \nodata & \nodata  \\ 
DDO 70      \dotfill  & 0.20$\pm$0.00 &   0.23$\pm$0.00 &   0.30$\pm$0.00 &   0.33$\pm$0.00 &   0.37$\pm$0.00 &   0.42$\pm$0.00 & & 0.25$\pm$0.00 &  0.36$\pm$0.01 &  0.28$\pm$0.00 &   0.36$\pm$0.00 &   0.47$\pm$0.00 &   0.69$\pm$0.01 \\
IC 1613      \dotfill & -2.40$\pm$0.00 &   -7.05$\pm$0.01 &  2.46$\pm$0.01 &  1.26$\pm$0.00 &  1.12$\pm$0.00 &   0.92$\pm$0.00 & & 0.30$\pm$0.00 & 0.36$\pm$0.00 & 0.29$\pm$0.00 &   0.55$\pm$0.00 &  0.61$\pm$0.00 &  \nodata \\
DDO 101    \dotfill & -3.5$\pm$0.22 &   6.4$\pm$0.13 &   1.30$\pm$0.00 &   1.06$\pm$0.00 &  0.94$\pm$0.00 &  0.93$\pm$0.00 & & 0.22$\pm$0.01 & 0.26$\pm$0.00 &  0.42$\pm$0.00 &   0.45$\pm$0.00 &  0.47$\pm$0.00 &  0.44$\pm$0.00  \\
WLM           \dotfill & 1.03$\pm$0.00 &  0.95$\pm$0.00  &   0.90$\pm$0.00 &  1.23$\pm$0.00 &  1.18$\pm$0.00 &  1.27$\pm$0.00  & & 0.26$\pm$0.00 &  0.29$\pm$0.00 &  0.40$\pm$0.00 &  0.46$\pm$0.00 &   0.52$\pm$0.00 &   0.53$\pm$0.00 \\
DDO 75      \dotfill  & -3.48$\pm$0.08 &   -5.75$\pm$0.09 &   -6.17$\pm$0.14 &  15.08$\pm$0.74 &  9.86$\pm$0.25 &  -1.15$\pm$0.00 & & 0.16$\pm$0.00 &  0.16$\pm$0.00 &   0.18$\pm$0.00 &   0.18$\pm$0.00 &   0.20$\pm$0.00 & 0.32$\pm$0.00 \\
VIIZW 403  \dotfill & 0.18$\pm$0.00 &   0.20$\pm$0.00 &  0.23$\pm$0.00 &   0.27$\pm$0.00 &  0.32$\pm$0.00 &   0.42$\pm$0.00 & & \nodata & \nodata & \nodata & \nodata & \nodata & \nodata  \\
Haro 29      \dotfill & 0.16$\pm$0.00 & 0.18$\pm$0.00 &   0.16$\pm$0.00 &   0.17$\pm$0.00 &  0.17$\pm$0.00 &  0.28$\pm$0.00 & & 0.54$\pm$0.03 &  0.54$\pm$0.02 &  0.34$\pm$0.00 &   0.50$\pm$0.00 &  0.32$\pm$0.00 &   1.78$\pm$0.01 \\ 
DDO 133   \dotfill & 6.79$\pm$0.08 &  3.70$\pm$0.01 &  1.83$\pm$0.01 &  1.46$\pm$0.00 &  1.34$\pm$0.00 &  1.16$\pm$0.00 & & 0.31$\pm$0.01 & 0.37$\pm$0.00 &  0.38$\pm$0.00 &   0.42$\pm$0.00 &   0.47$\pm$0.00 &    0.65$\pm$0.00  \\
DDO 126    \dotfill & 0.72$\pm$0.00 & 0.75$\pm$0.00 &   0.74$\pm$0.00 &   0.81$\pm$0.00 &   0.82$\pm$0.00 &  1.25$\pm$0.00  & & \nodata & \nodata & \nodata & \nodata & \nodata & \nodata  \\
DDO 165   \dotfill & \nodata  &   -214.11$\pm$28.16 &  3.55$\pm$0.01 &   2.50$\pm$0.00 &  2.37$\pm$0.00 &  2.40$\pm$0.01  & & \nodata & 0.46$\pm$0.00 &   0.60$\pm$0.00 &   0.65$\pm$0.00 &  0.71$\pm$0.00 &  0.95$\pm$0.00 \\
DDO 63      \dotfill  & -4.87$\pm$0.00 &  -31.25$\pm$0.10 &   6.71$\pm$0.11 &   2.84$\pm$0.01 &   2.65$\pm$0.01 &  2.14$\pm$0.01  & & 0.31$\pm$0.00 &  0.38$\pm$0.00 &   0.46$\pm$0.00 &   0.63$\pm$0.00 &   0.64$\pm$0.00 &   0.68$\pm$0.00 \\
NGC 6822 \dotfill & 1.29$\pm$0.00 & 0.74$\pm$0.00 & 0.63$\pm$0.00 & 0.59$\pm$0.00 & 0.65$\pm$0.00 &  0.65$\pm$0.00 & & \nodata & \nodata & \nodata & \nodata & \nodata & \nodata  \\
DDO 53     \dotfill & 0.20$\pm$0.00 &  0.24$\pm$0.00 &   0.31$\pm$0.00 &   0.43$\pm$0.00 &  0.54$\pm$0.00 &  0.35$\pm$0.00 & & \nodata & \nodata & \nodata & \nodata & \nodata & \nodata  \\
DDO 154   \dotfill & 0.60$\pm$0.00 &   0.56$\pm$0.00 &   0.52$\pm$0.00 &  0.51$\pm$0.00 &  0.52$\pm$0.00 &   0.53$\pm$0.00 & & \nodata & \nodata & \nodata & \nodata & \nodata & \nodata  \\
DDO 87     \dotfill & 0.56$\pm$0.01 & 0.71$\pm$0.01 & 1.94$\pm$0.01 & 1.84$\pm$0.01 &  1.73$\pm$0.00 & 1.75$\pm$0.01 & & 0.87$\pm$0.03 & 1.40$\pm$0.02 &  0.81$\pm$0.01 &  1.03$\pm$0.01 &  1.10$\pm$0.00 &  1.35$\pm$0.02 \\
DDO 52      \dotfill & 4.30$\pm$0.03 &   3.07$\pm$0.03 &  1.39$\pm$0.00 & 1.45$\pm$0.00 & 1.33$\pm$0.00 &  2.09$\pm$0.00  & & 0.85$\pm$0.03 &  0.89$\pm$0.01 &  0.72$\pm$0.07 &   0.93$\pm$0.02 &  0.90$\pm$0.01 &  0.87$\pm$0.00 \\
DDO 168   \dotfill & -10.15$\pm$0.36 & -15.31$\pm$0.17 &  174.25$\pm$19.25 &  -58.18$\pm$0.89 & 82.50$\pm$0.98 &  -17.72$\pm$0.34  & & 0.60$\pm$0.00 & 0.60$\pm$0.00 &  0.66$\pm$0.00 &  0.69$\pm$0.00 & 0.74$\pm$0.00 & 0.83$\pm$0.00 \\
NGC 1569 \dotfill & 0.22$\pm$0.00 & 0.23$\pm$0.00 &  0.22$\pm$0.00 &   0.22$\pm$0.00 &  0.24$\pm$0.00 &  0.31$\pm$0.00 & & 0.05$\pm$0.17 &  0.16$\pm$0.58 &  0.60$\pm$0.12 &   1.99$\pm$0.54 &   0.66$\pm$0.02 &   0.83$\pm$0.00 \\
NGC 3738 \dotfill & 0.21$\pm$0.01 & 0.24$\pm$0.00 &  0.36$\pm$0.00 &  0.42$\pm$0.00 &  0.45$\pm$0.00 &  0.45$\pm$0.00 & & \nodata & \nodata & \nodata & \nodata & \nodata & \nodata  \\
DDO 50      \dotfill  & 1.78$\pm$0.01 & 1.67$\pm$0.00 &   1.37$\pm$0.00 &  1.40$\pm$0.00 & 1.27$\pm$0.00 &   0.99$\pm$0.00  & & 0.80$\pm$0.01 &  0.75$\pm$0.01 &   0.76$\pm$0.01 &  0.82$\pm$0.01 &  0.78$\pm$0.01 &  \nodata \\
NGC 2366 \dotfill  &  1.49$\pm$0.01 & 1.39$\pm$0.00 & 1.50$\pm$0.00 & 1.58$\pm$0.00 &  1.58$\pm$0.00 &  1.83$\pm$0.00  & & 1.07$\pm$0.03 & 1.06$\pm$0.01 & 1.08$\pm$0.00 & 1.11$\pm$0.00 &  1.12$\pm$0.00 &  1.16$\pm$0.03 \\
NGC 4214 \dotfill  & 0.49$\pm$0.00 &   0.46$\pm$0.00 &   0.50$\pm$0.00 &  0.55$\pm$0.00 &  0.57$\pm$0.00 &  0.54$\pm$0.00  & & 0.85$\pm$0.00 &  0.79$\pm$0.00 &   0.71$\pm$0.00 &   0.69$\pm$0.00 &  0.67$\pm$0.00 &   \nodata \\
NGC 1156 \dotfill & 0.68$\pm$0.00 & 0.78$\pm$0.00 &  0.88$\pm$0.00 &  0.94$\pm$0.00 &  0.96$\pm$0.00 &  1.04$\pm$0.00 & & 0.42$\pm$0.02 &  0.52$\pm$0.02 &   0.62$\pm$0.00 &   0.72$\pm$0.00 &  0.74$\pm$0.00 &   0.87$\pm$0.00 \\

\enddata         
\tablecomments{
(1) Galaxy names.\
(2-7) Inner disk scale lengths measured on the surface brightness profiles of FUV, NUV, {\it U}, {\it B}, {\it V}, and 3.6 $\mu$m,  respectively.\ The typical uncertainties of the scale length are smaller than 0.01 kpc.\
(8-13) Disk scale lengths measured in the outer part of the disks, if an obvious break is present in the surface brightness profiles. 
}
\end{deluxetable}                                           

                                                                                    
%
                                                                           
\begin{deluxetable}{lcccccccccr} 
\tabletypesize{\scriptsize}                                                
\tablenum{3}                                                               
\tablecolumns{11}                                                           
\tablewidth{0pt}                                                           
\tablecaption{SED-fitting results}
\tablehead{                                                                                                                
\colhead{Galaxy}           
& \colhead{M$_{\star}$}  
& \colhead{$\langle {\tt SFR} \rangle_{\tt 13.7 Gyr}$}  
& \multicolumn{2}{c}{$R_{D}^{\star}$}
& \colhead{R$_{break,\star}$}
& \colhead{log($\Sigma_{\star, break}$)}
& \colhead{log($\Sigma_{\star,center}$)}
& \colhead{C$_{31,\star}$}                                                                                                                                                                             
& \colhead{$b_{\tt 0.1 Gyr}$}                                                    
& \colhead{$b_{\tt 1 Gyr}$} \\
\cline{4-5}
\colhead{} 
& \colhead{}                                                                             
& \colhead{}  
& \colhead{Inner}                                                                             
& \colhead{Outer}
& \colhead{}                                                                             
& \colhead{}  
& \colhead{}                                                                             
& \colhead{}  
& \colhead{}                                                                             
& \colhead{}  \\
\colhead{}                                                                             
& \colhead{[10$^{7}$ M$_{\odot}$]}                                                                             
& \colhead{ [10$^{-3}$ M$_{\odot}$yr$^{-1}$]}                                                                             
& \colhead{(kpc)}    
& \colhead{(kpc)}
& \colhead{(kpc)}
& \colhead{(log(M$_{\odot}$pc$^{-2}$))}                                                                         
& \colhead{(log(M$_{\odot}$pc$^{-2}$))}                                                                             
& \colhead{}                                                                             
& \colhead{}                                                                             
& \colhead{}   \\                                                                           
\colhead{(1)}                                                                             
& \colhead{(2)}                                                                             
& \colhead{(3)}                                                                             
& \colhead{(4)}                                                                             
& \colhead{(5)}                                                                             
& \colhead{(6)}                                                                             
& \colhead{(7)}                                                                           
& \colhead{(8)}       
& \colhead{(9)}                                                                             
& \colhead{(10)}                                                                           
& \colhead{(11)}                                                                                           
}                                                                          
\startdata                                                                                                               

LGS 3 \dotfill & 0.027 & 0.038 & 0.49 & 0.09 & 0.26 & -0.32 & -0.20 & 2.04 & 0.10$\pm$0.04 & 0.17$\pm$0.06 \\
DDO 210 \dotfill & 0.068 & 0.094 & 0.44 & 0.19 & 0.29 & -0.21 & -0.05 & 2.63 & 1.72$\pm$0.64 & 0.43$\pm$0.14 \\
DDO 69 \dotfill & 0.102 & 0.137 & 1.18 & 0.23 & 0.25 & 0.03 & 0.04 & 2.36 & 3.62$\pm$0.56 & 1.55$\pm$0.30 \\
DDO 155 \dotfill  & 0.297 & 0.404 & 0.22 & \nodata & \nodata & \nodata & 0.87 & 3.06 & 3.87$\pm$0.98 & 0.84$\pm$0.21 \\
DDO 216 \dotfill & 1.521 & 2.140 & 0.89 & 0.31 & 1.12 & 0.06 & 0.44 & 2.25 & 0.06$\pm$0.02 & 0.09$\pm$0.02 \\
M81dwA \dotfill & 0.138 & 0.182 & -14.28 & 0.30 & 0.41 & -0.12 & -0.16 & 2.02 & 7.81$\pm$1.12 & 1.22$\pm$0.18  \\
Mrk 178 \dotfill & 1.110 & 1.531 & 0.27 & 0.87 & 1.58 & -1.14  & 0.69 & 2.78 & 2.04$\pm$0.52 & 0.52$\pm$0.12 \\
DDO 187 \dotfill & 0.345 & 0.451 & 0.53 & 0.23 & 0.30 & 0.48 & 0.71 & 2.51 & 0.55$\pm$0.20 & 2.88$\pm$0.62 \\
UGC 8508 \dotfill  & 0.764 & 1.061 & 0.57 & 0.27 & 0.46 & 0.50 & 0.77 & 2.49 & 1.05$\pm$0.29 & 0.56$\pm$0.16  \\
NGC 4163 \dotfill & 2.708 & 3.807 & 0.43 & 0.29 & 0.61 & 0.83 & 1.40 & 2.62 & 0.51$\pm$0.20 & 0.32$\pm$0.10 \\
CVnIdwA \dotfill  & 0.410 & 0.532 & 0.68 & \nodata & \nodata & \nodata & 0.22 & 2.53 & 1.57$\pm$0.41 & 1.43$\pm$0.45 \\
DDO 70 \dotfill  & 1.960 & 2.668 & 0.50 & 1.59 & 0.77 & 0.14 & 0.76 & 2.77 & 1.03$\pm$0.22 & 1.44$\pm$0.25 \\
IC 1613 \dotfill & 2.920 & 4.030 & 0.74 & 0.39 & 1.98 & -0.20 & 0.96 & 2.64 & 0.81$\pm$0.25 & 0.83$\pm$0.21 \\
DDO 101 \dotfill & 6.543 & 9.325 & 0.68 & 0.41 & 1.54 & 0.39 & 1.37 & 2.52 & 0.08$\pm$0.02 & 0.12$\pm$0.05 \\
WLM \dotfill & 1.629 & 2.268 & 1.24 & 0.57 & 1.34 & -0.08 & 0.15 & 2.31 & 0.43$\pm$0.11 & 0.27$\pm$0.07 \\
DDO75 \dotfill  & 0.784 & 1.023 & 0.64 & 0.27 & 0.73 & 0.31 & 0.71 & 2.03 & 9.17$\pm$1.49 & 2.21$\pm$0.31 \\
VIIZW 403 \dotfill & 1.777 & 2.449 &  0.96 & 0.55 & 1.14 & 0.17 & 0.67 & 2.45 & 3.51$\pm$0.89 & 1.10$\pm$0.26 \\
Haro 29 \dotfill & 1.443 & 1.921 & 0.20 & 1.98 & 1.12 & -0.81 & 1.59 & 5.29 & 6.88$\pm$2.02 & 1.70$\pm$0.39 \\ 
DDO 133 \dotfill & 3.042 & 4.214 & 0.91 & 0.57 & 2.26 & -0.28 & 0.77 & 2.54 & 0.72$\pm$0.14 & 0.74$\pm$0.14 \\
DDO 126 \dotfill & 1.605 & 2.132 & 0.82 & \nodata & \nodata & \nodata & 0.37 & 2.58 & 1.42$\pm$0.33 & 1.71$\pm$0.36 \\
DDO 165 \dotfill & 3.453 & 4.631 & 1.82 & 0.92 & 1.34 & 0.22  & 0.45 & 2.30 & 3.84$\pm$0.83 & 1.34$\pm$0.25 \\
DDO 63 \dotfill  & 3.337 & 4.421 & 1.61 & 1.00 & 1.57 & 0.14 & 0.56 & 2.29 & 2.25$\pm$0.52 & 2.27$\pm$0.42 \\
NGC 6822 \dotfill & 7.630 & 10.451 & 1.47 & 0.36 & 0.38 & 1.59 & 1.66 & 2.35 & 0.51$\pm$0.14 & 0.63$\pm$0.14 \\
DDO 53 \dotfill & 0.970 & 1.337 & 20.87 & 0.79 & 0.96 & -0.06 & -0.15 & 2.10 & 1.13$\pm$0.18 & 0.69$\pm$0.14 \\
DDO 154 \dotfill & 0.835 & 1.131 & 1.00 & 0.49 & 0.68 & 0.10 & 0.28 & 2.47 & 3.82$\pm$0.83 & 0.91$\pm$0.17 \\
DDO 87 \dotfill & 3.306 & 4.589 & 2.27 & 1.06 & 1.21 & 0.15 & 0.32 & 2.69 & 0.64$\pm$0.15 & 0.53$\pm$0.13 \\
DDO 52 \dotfill & 5.273 & 7.296 & 1.05 & 0.52 & 3.15 & -0.41 & 0.85 & 2.68 & 0.46$\pm$0.17 & 0.86$\pm$0.25 \\
DDO 168 \dotfill & 5.865 & 7.953 & 32.04 & 0.99 & 0.73 & 0.65  & 0.63 & 2.64 & 1.43$\pm$0.29 & 0.79$\pm$0.17 \\
NGC 1569 \dotfill & 36.016 & 49.259 & 0.43 & 0.87 & 2.25 & 0.13 & 2.35 & 3.13 & 6.19$\pm$1.36 & 1.02$\pm$0.18 \\
NGC 3738 \dotfill & 46.571 & 65.325 & 0.68 & 0.61 & 1.59 & 1.16 & 2.20 & 2.95 & 0.62$\pm$0.22 & 0.62$\pm$0.17 \\
DDO 50 \dotfill  & 10.725 & 14.416 & 1.08  & 0.71 & 2.00 & 0.42 & 1.02 & 2.45 & 3.88$\pm$0.51 & 1.47$\pm$0.21 \\
NGC 2366 \dotfill  & 6.954 & 9.186 & 3.69 & 1.11 & 2.66 & -0.14 & -0.18 & 2.70 & 2.64$\pm$0.44 & 2.00$\pm$0.30 \\
NGC 4214 \dotfill  & 47.999 & 65.390 & 0.82 & 0.60 & 2.73 & 0.56 & 2.01 & 3.09 & 0.92$\pm$0.15 & 1.26$\pm$0.22 \\
NGC 1156 \dotfill & 121.759 & 164.914 & 0.97 & 0.64 & 3.92 & 0.55 & 2.30 & 2.64 & 1.73$\pm$0.40 & 0.85$\pm$0.22 \\

\enddata         
\tablecomments{
(1) Galaxy names.\
(2) Total stellar mass derived by applying the curve-of-growth method to the stellar mass profiles from our SED modeling.\
(3)  SFR averaged over the Hubble time (13.7 Gyr).\
(4-5)  Disk scale length measured from the surface stellar mass density profiles.\
The scale lengths of the inner and the outer parts are listed, respectively.\ If there is no obvious break in the profile, the related scale length is listed in column 4.\
(6) Radius where the break of the surface stellar mass density profile happens.\
(7) Surface stellar mass density at the broken radius.\ $\Sigma_{\star, break}$ has been inclination-corrected under the assumption that the intrinsic $(b/a)_{0}$ = 0.3.\
(8) Inclination-corrected central surface stellar mass density, extrapolated from the inner regions around the center.\
(9) Concentration index (see the text for details).\
(10) Ratio of the SFR averaged over the past 0.1 Gyr to the SFR averaged over the Hubble time.\
(11) Ratio of the SFR averaged over the past 1 Gyr to the SFR averaged over the Hubble time.
}
                                                          
\end{deluxetable}                                                          

\clearpage

                                                                           
%
                                                                           
\begin{deluxetable}{lcccccccr} 
\tabletypesize{\scriptsize}                                                
\tablenum{4}                                                               
\tablecolumns{9}                                                           
\tablewidth{0pt}                                                           
\tablecaption{Slopes of the SFH Variations with Radius}
\tablehead{                                                                                                                
\colhead{Galaxy}
& \multicolumn{2}{c}{$\frac{\Delta({\rm log}(\Sigma_{\tt SFR_{0.1}}/\Sigma_{\star}))}{\Delta({\rm R[kpc])}}$} 
& \multicolumn{2}{c}{$\frac{\Delta({\rm log}(\Sigma_{\tt SFR_{0.1}}/\Sigma_{\star}))}{\Delta({\rm R/R_{D}^{\tt {\it V}})}}$} 
& \multicolumn{2}{c}{$\frac{\Delta({\rm log}(\Sigma_{\tt SFR_{1}}/\Sigma_{\star}))}{\Delta({\rm R[kpc])}}$} 
& \multicolumn{2}{c}{$\frac{\Delta({\rm log}(\Sigma_{\tt SFR_{1}}/\Sigma_{\star}))}{\Delta({\rm R/R_{D}^{\tt {\it V}})}}$} \\
\cline{2-9}
\colhead{}
& \colhead{Inner}
& \colhead{Outer}
& \colhead{Inner}  
& \colhead{Outer} 
& \colhead{Inner} 
& \colhead{Outer} 
& \colhead{Inner} 
& \colhead{Outer} \\                                                                    
\colhead{(1)}                                                                             
& \colhead{(2)}                                                                             
& \colhead{(3)}                                                                             
& \colhead{(4)} 
& \colhead{(5)}                                                                             
& \colhead{(6)}                                                                             
& \colhead{(7)}       
& \colhead{(8)}                                                                             
& \colhead{(9)}                                                                                                                                                                                                                                      
}                                                                          
\startdata                                                                                                     
\cutinhead{Galaxies with M$_{bary}$ $<$ 10$^{8}$ M$_{\odot}$}
LGS 3          \dotfill  &  -0.10$\pm$0.16 & \nodata & -0.02$\pm$0.04 & \nodata & -1.25$\pm$0.19 & \nodata & -0.29$\pm$0.04 & \nodata \\
DDO 210    \dotfill  &  -2.59$\pm$0.59 & \nodata & -0.43$\pm$0.10 & \nodata & -0.85$\pm$0.22 & \nodata & -0.14$\pm$0.04 & \nodata \\
DDO 69      \dotfill  &  -0.50$\pm$0.06 & -2.29$\pm$0.28 & -0.09$\pm$0.01 & -0.43$\pm$0.05 & 1.01$\pm$0.15 & -0.95$\pm$0.05 & 0.19$\pm$0.03 & -0.18$\pm$0.01 \\
DDO 155    \dotfill  &  -1.41$\pm$0.06 & -2.53$\pm$0.24 & -0.21$\pm$0.01 & -0.37$\pm$0.03 & -0.65$\pm$0.15 & -1.19$\pm$0.26 & -0.10$\pm$0.02 & -0.18$\pm$0.04\\
DDO 216    \dotfill  &   -0.44$\pm$0.04 & \nodata & -0.24$\pm$0.02 & \nodata & -0.52$\pm$0.04 & \nodata & -0.28$\pm$0.02 & \nodata \\
M81dwA     \dotfill  &   0.58$\pm$0.03 & -0.75$\pm$0.02 & 0.15$\pm$0.01 & -0.20$\pm$0.01 & 0.56$\pm$0.03 & -0.65$\pm$0.01 & 0.14$\pm$0.01 & -0.17$\pm$0.01 \\
Mrk 178      \dotfill  &   -2.53$\pm$0.12 & -1.86$\pm$0.02 & -0.66$\pm$0.03 & -0.48$\pm$0.01 & -1.28$\pm$0.09 & -0.35$\pm$0.02 & -0.33$\pm$0.02 & -0.09$\pm$0.01 \\
DDO 187    \dotfill  &   -1.01$\pm$0.08 & -2.09$\pm$0.05 & -0.18$\pm$0.01 & -0.37$\pm$0.01 & -0.64$\pm$0.01 & -1.42$\pm$0.06 & -0.11$\pm$0.01 & -0.26$\pm$0.01\\
UGC 8508  \dotfill  &   -0.58$\pm$0.02 & -2.56$\pm$0.07 & -0.15$\pm$0.01 & -0.68$\pm$0.02 & 0.07$\pm$0.02 & -0.72$\pm$0.02 & 0.02$\pm$0.01 & -0.19$\pm$0.01 \\
NGC 4163  \dotfill  &   -0.74$\pm$0.02 & -0.06$\pm$0.02 & -0.74$\pm$0.02 & -0.06$\pm$0.02 & -0.55$\pm$0.01 &  0.00$\pm$0.04 & -0.55$\pm$0.01 &  0.00$\pm$0.04  \\
CVnIdwA   \dotfill   &      0.24$\pm$0.07 & -2.50$\pm$0.15 & 0.14$\pm$0.04 & -1.42$\pm$0.09 & 0.07$\pm$0.03 & -1.40$\pm$0.18 & 0.04$\pm$0.02 & -0.79$\pm$0.10 \\ 
DDO 70      \dotfill   &     -1.58$\pm$0.03 & -0.82$\pm$0.11 & -0.75$\pm$0.01 & -0.39$\pm$0.05 & 0.00$\pm$0.05 & -1.42$\pm$0.11 & 0.00$\pm$0.02 & -0.68$\pm$0.05 \\
IC 1613      \dotfill   &     0.11$\pm$0.05 & -1.22$\pm$0.03 & 0.06$\pm$0.03 & -0.72$\pm$0.02 & 0.04$\pm$0.01 & -0.67$\pm$0.02 & 0.02$\pm$0.01 & -0.39$\pm$0.01 \\
DDO 101    \dotfill   &     0.52$\pm$0.01 & -0.24$\pm$0.07 & 0.48$\pm$0.01 & -0.23$\pm$0.06 & 0.68$\pm$0.03 & -0.18$\pm$0.08 & 0.63$\pm$0.03 & -0.16$\pm$0.07  \\
WLM           \dotfill   &    -0.46$\pm$0.02 & -0.90$\pm$0.04 & -0.26$\pm$0.01 & -0.52$\pm$0.02 & -0.26$\pm$0.01 & -0.39$\pm$0.06 & -0.15$\pm$0.01 & -0.23$\pm$0.03 \\
\\
Averages  \dotfill    &    0.01$\pm$0.72 & -1.26$\pm$0.68  &  -0.10$\pm$0.31  &  -0.37$\pm$0.17  &  -0.29$\pm$0.65  &  -0.71$\pm$0.35  &  -0.14$\pm$0.16  &  -0.19$\pm$0.07 \\
\cutinhead{Galaxies with M$_{bary}$ $>$ 10$^{8}$ M$_{\odot}$} 
DDO 75      \dotfill   &     0.76$\pm$0.03 & -2.01$\pm$0.06 &  0.17$\pm$0.01 & -0.45$\pm$0.01 &  0.82$\pm$0.03 & -0.55$\pm$0.05 &  0.18$\pm$0.01 & -0.12$\pm$0.01 \\
VIIZW 403  \dotfill   &      -1.42$\pm$0.06 & -0.95$\pm$0.03 & -0.75$\pm$0.03 & -0.50$\pm$0.02 & -0.52$\pm$0.05 & -0.71$\pm$0.06 & -0.27$\pm$0.02 & -0.37$\pm$0.03 \\
Haro 29      \dotfill   &      -0.60$\pm$0.09 & -1.09$\pm$0.05 & -0.18$\pm$0.02 & -0.32$\pm$0.01 &  0.63$\pm$0.06 & -0.13$\pm$0.03 &  0.18$\pm$0.02 & -0.04$\pm$0.01 \\ 
DDO 133   \dotfill   &       0.46$\pm$0.01 & -0.23$\pm$0.05 &  0.52$\pm$0.01 & -0.26$\pm$0.06 &  0.15$\pm$0.01 &  0.02$\pm$0.02 &  0.17$\pm$0.01 &  0.02$\pm$0.02 \\
DDO 126    \dotfill   &      -0.23$\pm$0.02 & -0.02$\pm$0.02 & -0.20$\pm$0.01 & -0.02$\pm$0.02 &  0.42$\pm$0.03 & -0.02$\pm$0.01 &  0.36$\pm$0.03 & -0.01$\pm$0.01 \\
DDO 165   \dotfill    &      0.05$\pm$0.01 & -1.00$\pm$0.02 &  0.12$\pm$0.01 & -2.27$\pm$0.04 &  0.12$\pm$0.01 & -0.16$\pm$0.01 &  0.26$\pm$0.02 & -0.35$\pm$0.02 \\
DDO 63      \dotfill   &       -0.07$\pm$0.02 & -1.02$\pm$0.03 & -0.03$\pm$0.01 & -0.51$\pm$0.02 &  0.19$\pm$0.03 & -0.39$\pm$0.02 &  0.09$\pm$0.02 & -0.19$\pm$0.01 \\
NGC 6822 \dotfill   &       -0.07$\pm$0.02 &  1.75$\pm$0.14 & -0.04$\pm$0.01 &  1.00$\pm$0.08 & -0.67$\pm$0.02 &  0.44$\pm$0.13 & -0.38$\pm$0.01 &  0.25$\pm$0.07 \\
DDO 53     \dotfill   &       -0.48$\pm$0.02 & -1.56$\pm$0.05 & -0.35$\pm$0.01 & -1.13$\pm$0.04 & -0.54$\pm$0.02 & -1.32$\pm$0.03 & -0.39$\pm$0.02 & -0.95$\pm$0.02 \\
DDO 154   \dotfill   &       0.03$\pm$0.03 & -0.45$\pm$0.02 &  0.02$\pm$0.02 & -0.27$\pm$0.01 & -0.02$\pm$0.01 &  0.03$\pm$0.02 & -0.01$\pm$0.01 &  0.02$\pm$0.01 \\
DDO 87     \dotfill    &      0.20$\pm$0.03 & -0.04$\pm$0.01 &  0.29$\pm$0.04 & -0.06$\pm$0.02 & -0.05$\pm$0.01 &  0.12$\pm$0.01 & -0.07$\pm$0.01 &  0.17$\pm$0.01 \\
DDO 52      \dotfill   &     0.19$\pm$0.02 &  0.13$\pm$0.01 &  0.25$\pm$0.02 &  0.17$\pm$0.02 &  0.31$\pm$0.01 & -0.10$\pm$0.01 &  0.40$\pm$0.02 & -0.14$\pm$0.01  \\
DDO 168   \dotfill    &      -0.06$\pm$0.01 & -0.88$\pm$0.02 & -0.05$\pm$0.01 & -0.73$\pm$0.02 &  0.01$\pm$0.02 & -0.40$\pm$0.02 &  0.01$\pm$0.02 & -0.33$\pm$0.02 \\
NGC 1569 \dotfill    &      -0.64$\pm$0.03 & \nodata & -0.25$\pm$0.01 & \nodata & -0.46$\pm$0.03 & \nodata & -0.18$\pm$0.01 & \nodata \\
NGC 3738 \dotfill    &      -1.31$\pm$0.04 & -0.27$\pm$0.01 & -1.31$\pm$0.04 & -0.27$\pm$0.01 & -1.22$\pm$0.02 & -0.25$\pm$0.02 & -1.22$\pm$0.02 & -0.25$\pm$0.02 \\
DDO 50      \dotfill    &      0.35$\pm$0.01 & -0.31$\pm$0.02 &  0.38$\pm$0.01 & -0.34$\pm$0.03 &  0.08$\pm$0.01 & 0.00$\pm$0.01 &  0.09$\pm$0.01 & 0.00$\pm$0.01  \\
NGC 2366 \dotfill     &      0.04$\pm$0.01 & -0.07$\pm$0.01 &  0.06$\pm$0.01 & -0.10$\pm$0.02 & -0.14$\pm$0.01 & -0.09$\pm$0.01 & -0.19$\pm$0.02 & -0.12$\pm$0.02 \\
NGC 4214 \dotfill     &     0.15$\pm$0.01 &  \nodata &  0.11$\pm$0.01 & \nodata & -0.01$\pm$0.01 & \nodata & 0.00$\pm$0.01 & \nodata \\
NGC 1156 \dotfill     &     -0.30$\pm$0.01 & -0.41$\pm$0.01 & -0.24$\pm$0.01 & -0.33$\pm$0.01 & -0.05$\pm$0.01 & -0.25$\pm$0.02 & -0.04$\pm$0.01 & -0.20$\pm$0.02 \\
\\
Averages  \dotfill    &    0.12$\pm$0.29 & -0.59$\pm$0.44  &  0.08$\pm$0.25  &  -0.44$\pm$0.65  &  0.04$\pm$0.21  &  -0.08$\pm$0.20  &  0.02$\pm$0.20  &  -0.09$\pm$0.16 \\

\enddata         
\tablecomments{
(1) Galaxy names.\
(2-5) Slopes of the radial variations of $\Delta({\rm log}(\Sigma_{\tt SFR_{0.1}}/\Sigma_{\star}))/\Delta({\rm R[kpc]})$
and $\Delta({\rm log}(\Sigma_{\tt SFR_{0.1}}/\Sigma_{\star}))/\Delta({\rm R/R_{D}^{\tt {\it V}}})$, for the inner disks and the outer disks, respectively.\
(6-9) Slopes of the radial variations of $\Delta({\rm log}(\Sigma_{\tt SFR_{1}}/\Sigma_{\star}))/\Delta({\rm R[kpc]})$
and $\Delta({\rm log}(\Sigma_{\tt SFR_{1}}/\Sigma_{\star}))/\Delta({\rm R/R_{D}^{\tt {\it V}}})$, for the inner disks and the outer disks, respectively.\
}
\end{deluxetable}                                                          

\clearpage

\clearpage

\centering
\includegraphics[width=0.9\textwidth]{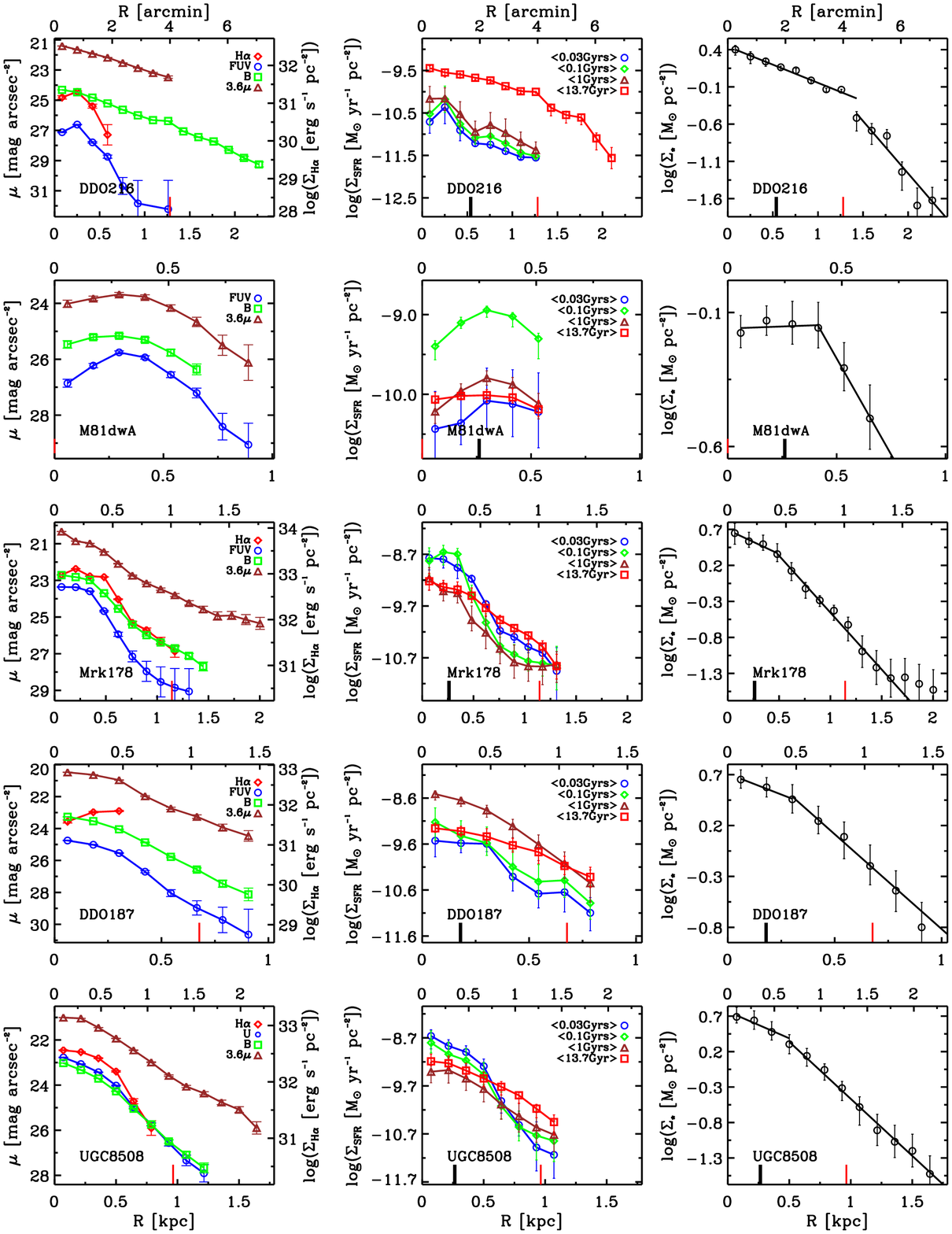}

Fig. 1--{\it Continued}

\centering
\includegraphics[width=0.9\textwidth]{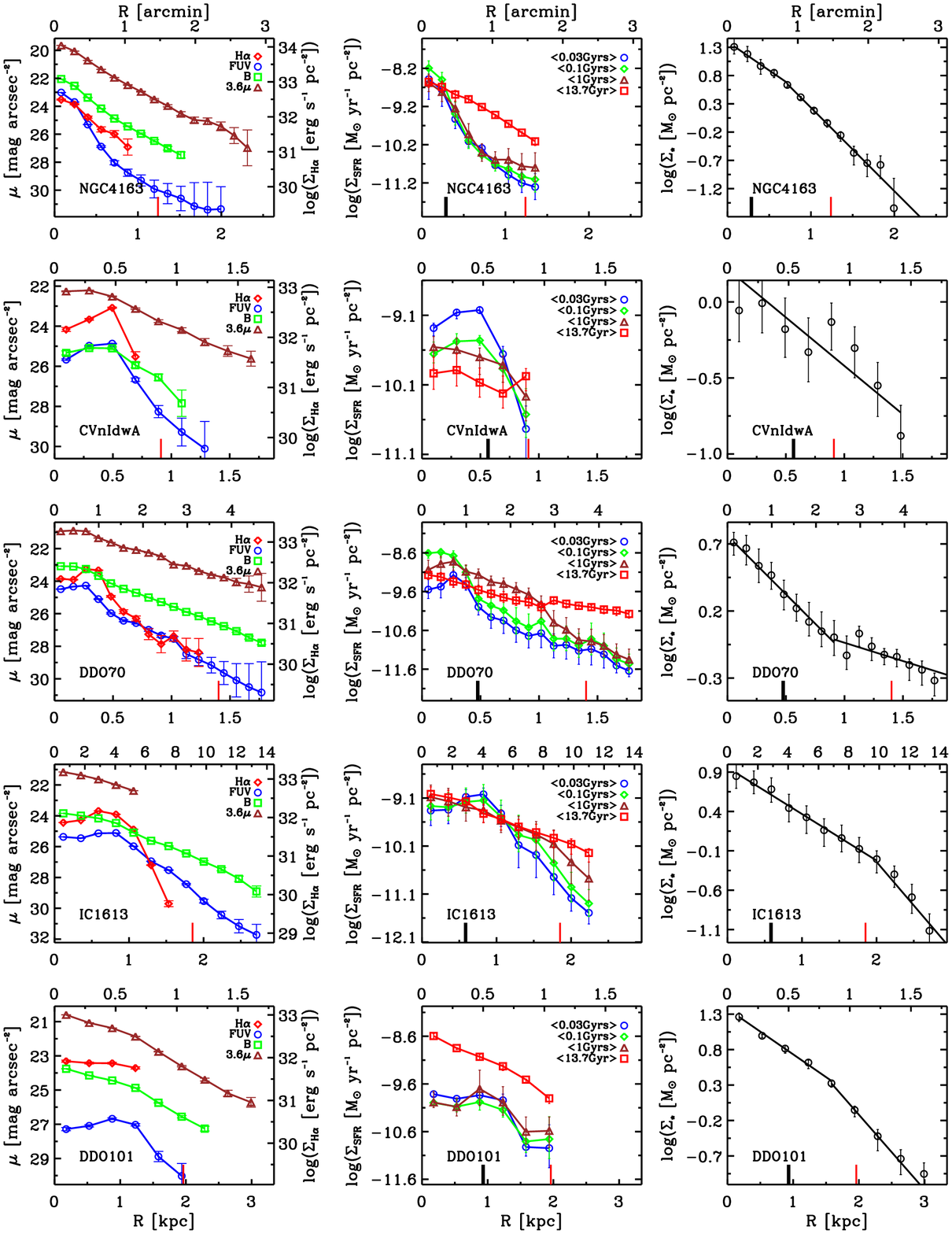}

Fig. 1--{\it Continued}
\clearpage

\centering
\includegraphics[width=0.9\textwidth]{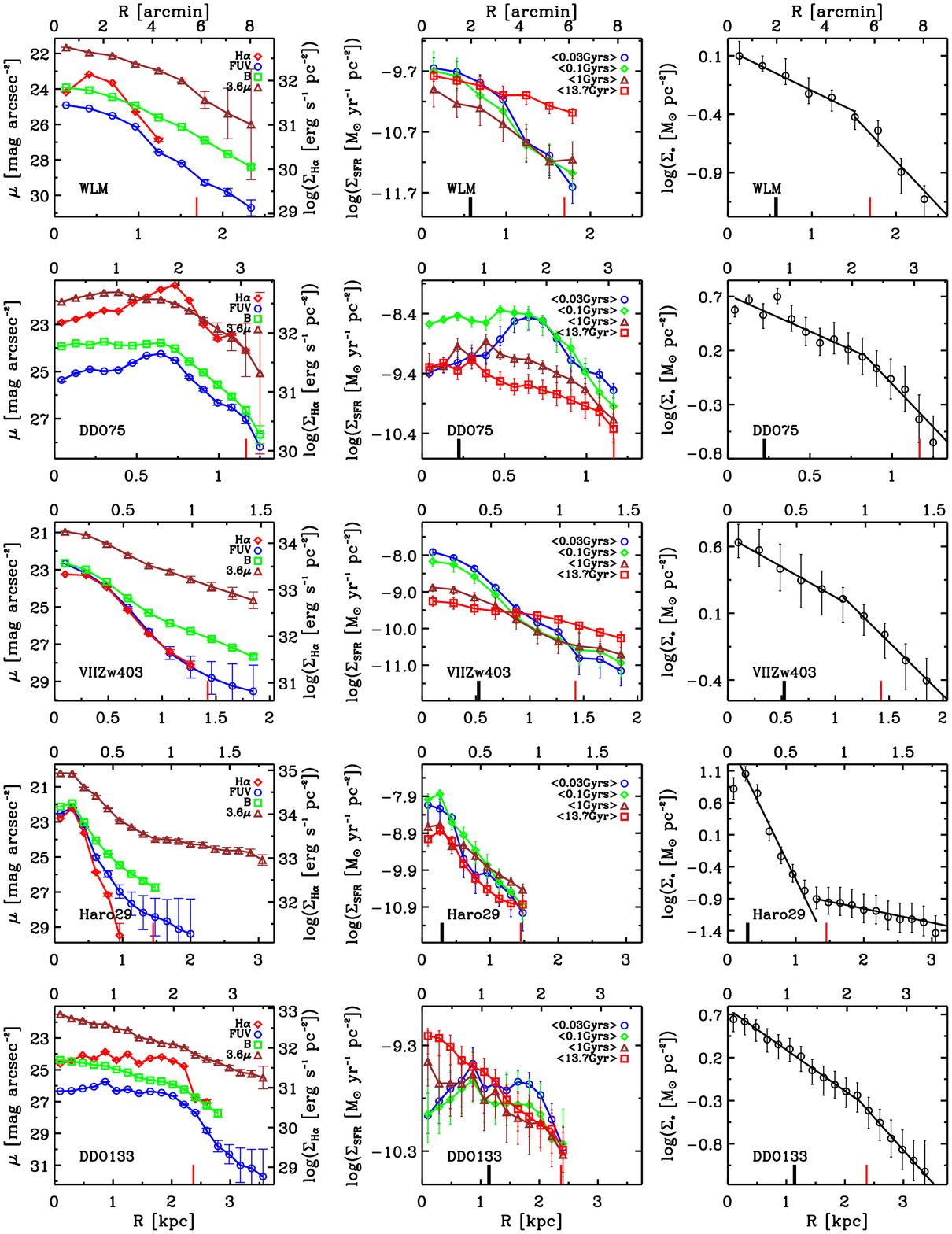}

Fig. 1--{\it Continued}
\clearpage

\centering
\includegraphics[width=0.9\textwidth]{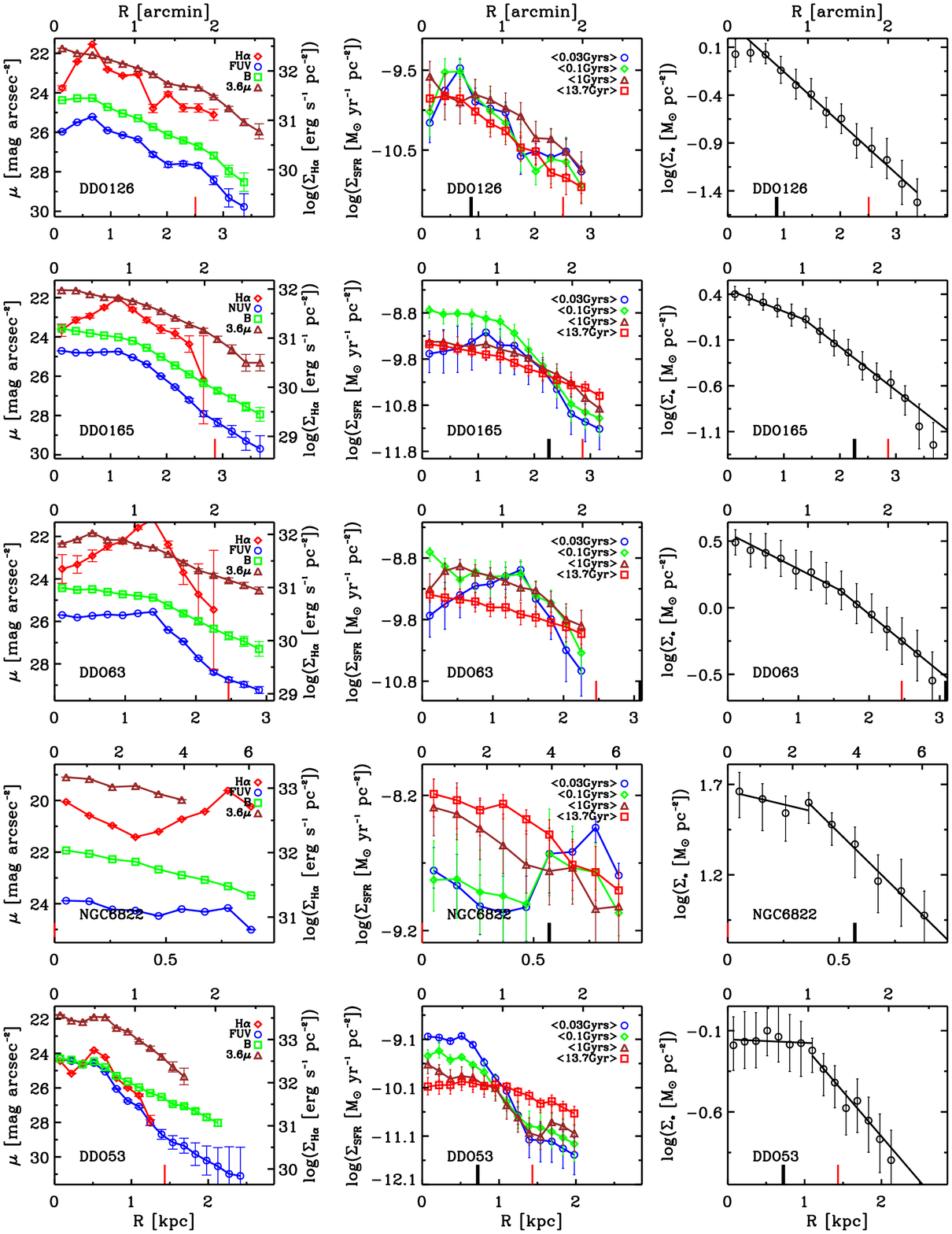}

Fig. 1--{\it Continued}
\clearpage

\centering
\includegraphics[width=0.9\textwidth]{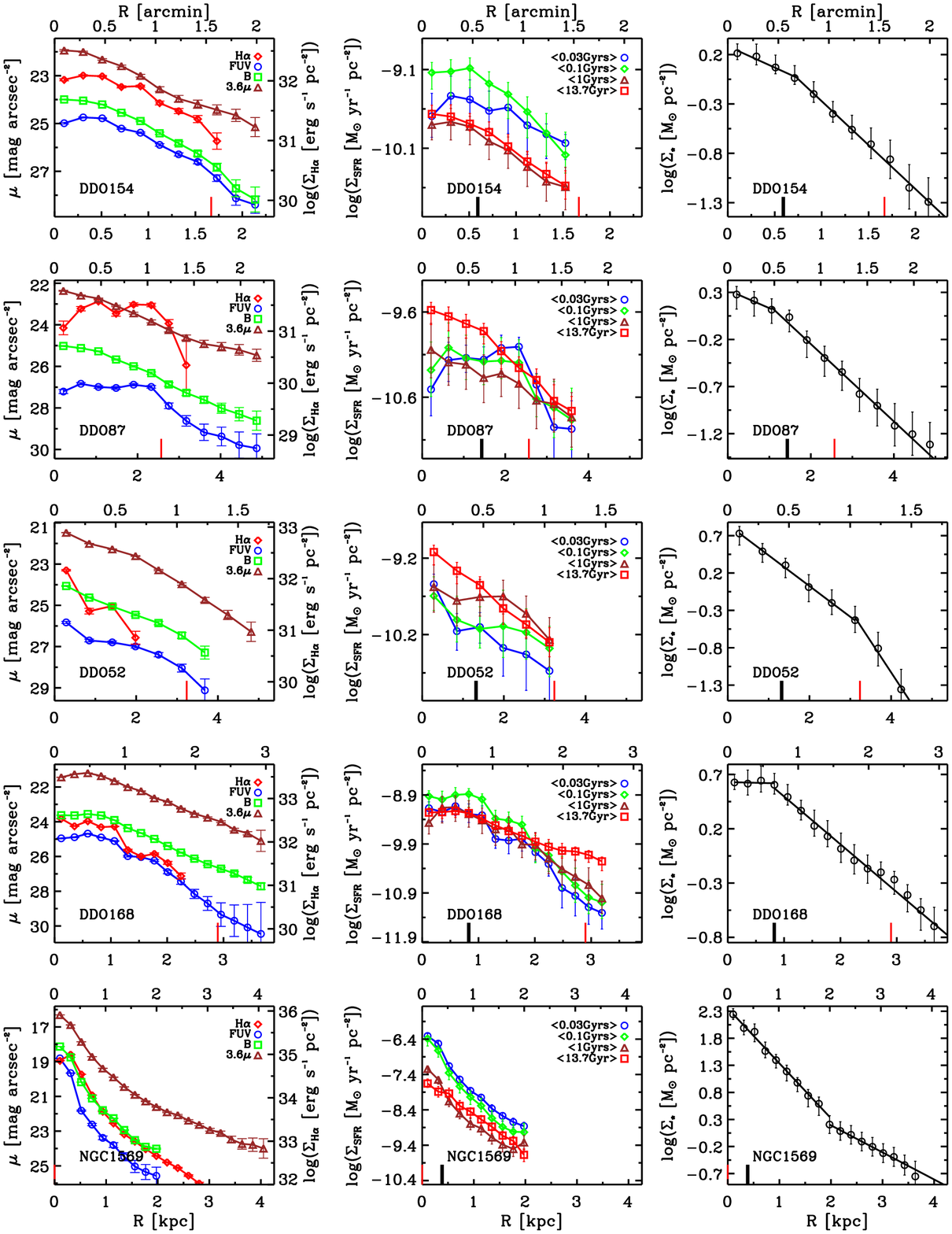}

Fig. 1--{\it Continued}
\clearpage

\centering
\includegraphics[width=0.9\textwidth]{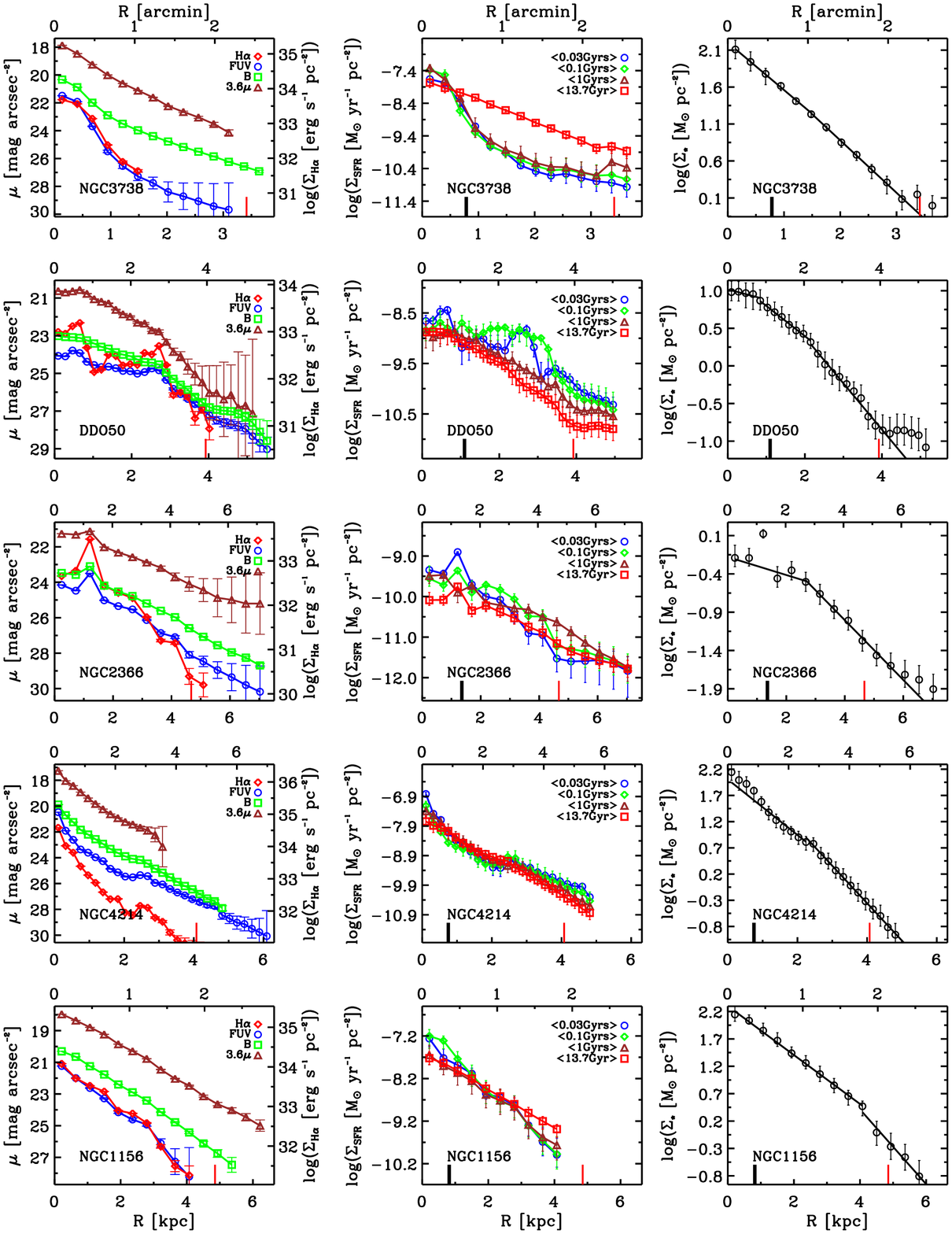}

Fig. 1--{\it Continued}
\clearpage

\clearpage

\clearpage

\clearpage

\clearpage

\clearpage

\clearpage

\clearpage

\clearpage

\clearpage

\clearpage

\clearpage

\clearpage

\clearpage

\clearpage

\clearpage

\end{document}